\documentclass[twocolumn,aps,prd,amsmath,amssymb,preprintnumbers,longbibliography]{revtex4-1}
\usepackage{amsmath} \usepackage{amsfonts} \usepackage{amssymb}
\usepackage{bbm}
\usepackage{epsfig}
\usepackage{graphics}
\usepackage{graphicx}
\usepackage{titlesec}
\usepackage{mathtools}
\usepackage{environ}
\usepackage{dsfont}

\usepackage[colorlinks=true]{hyperref}
\hypersetup{urlcolor=black,linkcolor=black,citecolor=black}
\usepackage[toc,page]{appendix}

\makeatletter
\renewcommand*\env@matrix[1][c]{\hskip -\arraycolsep
  \let\@ifnextchar\new@ifnextchar
  \array{*\c@MaxMatrixCols #1}}
\makeatother

\newcommand{\be}{\begin{equation}}
\newcommand{\ee}{\end{equation}}
\newcommand{\ba}{\begin{eqnarray}}
\newcommand{\ea}{\end{eqnarray}}
\newcommand{\nn}{\nonumber}

\newcommand{\diag}{\mathrm{diag}}

\newcommand{\te}{t~+~\epsilon}

\titleformat{\subsection}[block]{\normalfont\bfseries}{\thesubsection.}{1ex}{}
\titlespacing{\subsection}{0pt}{10pt}{1pt}[0pt]
\titleformat*{\section}{\large\bfseries}
\renewcommand{\thesubsection}{\arabic{subsection}}

\usepackage{todonotes}
\usepackage{braket}
\usepackage{bm}

\newcommand{\tr}{\mathrm{tr}}
\newcommand{\prob}[8]{#1 p_{1} #2 p_{2} #3 p_{3} #4 p_{4} #5 p_{5}#6 p_{6}#7 p_{7} #8 p_{8}}
\newcommand{\raw}{\rightarrow}
\newcommand{\ua}{\uparrow}
\newcommand{\da}{\downarrow}
\newcommand{\exval}[1]{\langle #1 \rangle}

\newcommand{\com}{\, ,\quad}
\newcommand{\mn}{\mu\nu}
\newcommand{\bC}{\mathbb{C}}
\newcommand{\bP}{\mathbb{P}}
\newcommand{\muQ}{\mu_{1}\mu_{2}\ldots\mu_{Q}}
\newcommand{\cD}{\mathcal{D}}
\newcommand{\cL}{\mathcal{L}}
\newcommand{\diff}{\text{d}}
\newcommand{\eul}{\text{e}}
\newcommand{\erf}{\text{erf}}

\definecolor{refkey}{rgb}{0,0,1}
\definecolor{labelkey}{rgb}{0,1,0}

\begin{document}

\title[ ]{Quantum computing with classical bits}

\author{C. Wetterich}
\affiliation{Institut  f\"ur Theoretische Physik\\
Universit\"at Heidelberg\\
Philosophenweg 16, D-69120 Heidelberg}

\begin{abstract}

A bit-quantum map relates probabilistic information for Ising spins or classical bits to quantum spins or qubits. Quantum systems are subsystems of classical statistical systems. The Ising spins can represent macroscopic two-level observables, and the quantum subsystem employs suitable expectation values and correlations. We discuss static memory materials based on Ising spins for which boundary information can be transported through the bulk in a generalized equilibrium state. They can realize quantum operations as the Hadamard or CNOT-gate for the quantum subsystem. Classical probabilistic systems can account for the entanglement of quantum spins. An arbitrary unitary evolution for an arbitrary number of quantum spins can be described by static memory materials for an infinite number of Ising spins which may, in turn, correspond to continuous variables or fields. We discuss discrete subsets of unitary operations realized by a finite number of Ising spins. They may be useful for new computational structures. We suggest that features of quantum computation or more general probabilistic computation may be realized by neural networks, neuromorphic computing or perhaps even the brain. This does neither require small isolated entities nor very low temperature. In these systems the processing of probabilistic information can be more general than for static memory materials. We propose a general formalism for probabilistic computing for which deterministic computing and quantum computing are special limiting cases.

\end{abstract}

\maketitle

\section{Introduction}
\label{sec:Introduction}

Quantum computing is based on a sequence of unitary operations acting on the wave function or the density matrix $\rho$ for a number of quantum spins or qubits \cite{BEN,MAN,FEY,DEU}. The unitarity of the operations is rooted in the unitary time evolution of $\rho$ for isolated quantum systems. For a quantum computation the continuous time evolution is well approximated by a series of discrete steps or operations. The actual continuity of the time evolution of $\rho$ is no longer important. Quantum gates change $\rho$ ``instantaneously'', while the evolution inbetween gates is neglected. A typical computation is a sequence of gates, ordered by some variable $t$, that transforms the initial density matrix $\rho(t_{in})$ to $\rho(t)$ at larger $t=t_{in}+n\epsilon$. Quantum computations are inherently probabilistic. The initial probabilistic information at $t_{in}$ is transformed to the probabilistic information at $t$ in the form of $\rho(t)$. This is done in discrete steps from $t$ to $t+\epsilon$. The unitary transformations $U(t)$ associated to quantum gates depend on $t$,
\begin{equation}\label{eq:I1} 
\rho(\te)=U(t)\rho(t)U^{\dagger}(t).
\end{equation}

``Classical computing'' uses bits or Ising spins that can only take two values. A computation is again a series of operations on the bits. These are generally viewed as deterministic operations, with probabilistic aspects associated to ``errors''. Still, conceptually a sequence of classical operations on bits transforms the probabilistic information about the bits at $t_{in}$ to the one at $t$, proceeding in discrete steps from $t$ to $t+\epsilon$. The individual discrete operations do not commute - the order in the sequence matters. This non-commutative aspect is similar for classical and quantum computation.

In this paper we propose that quantum operations or gates can be performed by Ising spins or classical bits, within a classical statistical setting. We also investigate first questions about possible realizations of quantum operations by classical bits. Three central issues need to be addressed. The first concerns the probabilistic nature of quantum computing which has to be implemented by classical bits. Classical systems realizing quantum computing have to be truly probabilistic. Not all macroscopic two-level observables or Ising spins can have fixed values - only probabilistic information as expectation values or correlations is available. The second issue concerns the embedding of the quantum subsystem into the classical statistical system. The choice of the bit-quantum map, which defines the quantum subsystem, is crucial for assessing which type of operations can be performed. Finally a third topic involves the practical realization of the unitary transformations needed for quantum gates.

It is not the purpose of this work to challenge the high potential of quantum computations employing isolated quantum systems. Nature offers for this purpose a powerful toolkit. It is well possible that certain quantum operations are difficult to be realized in practice by classical bits. Our investigations find, however, that there is no sharp boundary between probabilistic computation with classical bits and quantum computation with isolated qubits. For infinite resources classical bit operations can do whatever quantum operations can do -- there is no conceptual barrier. The issue for the future will be to explore how much quantum features can be effectively realized by a finite number of classical bits. This work presents first achievements, but also points to some of the potential practical difficulties, mainly in the appendices.

The physical nature of the ordering variable $t$ for the sequence of quantum gates does not matter. One obvious possibility is a sequence of time steps. As an interesting alternative, one may conceive $t$ as a spatial variable, ordering qubits on a chain. On each location $t=t_{in}+n\epsilon$ of the chain one may realize $Q$ qubits. The local probabilistic information about the qubits at $t$ is then encoded in the hermitian $2^{Q}\times 2^{Q}$-density matrix $\rho(t)$. Quantum gates relate $\rho(\te)$ to $\rho(t)$ according to eq.~\eqref{eq:I1}. They can be realized by coupling the spins at $t$ to the ones at $\te$. The relations between $\rho(\te)$ and $\rho(t)$ for characteristic gates may be realized, for example, in the ground state of the system. They do not need to involve the use of an explicit time evolution. Such a chain should be constructed such that the relations between $\rho(\te)$ and $\rho(t)$ depend on $t$, e.g. realizing different gates $U(t)$ for different $t$. A chain with several $t$-layers can perform rather complex quantum operations $U(t)U(t-\epsilon)\ldots U(t_{in}+\epsilon)U(t_{in})$, even for simple individual gates $U(t)$.

A similar setting on a chain can be realized for the gates of classical computing. At every location of the chain one places $M$ Ising spins. Gates map the probabilistic information at $t$ to the one at $t+\epsilon$.  As one possibility, this can be realized by generalized Ising models with interactions between the Ising spins at $t$ and at $\te$. Again, we are interested in different interactions for different $t$, such that a sequence of different ``classical gates'' can be realized. The transport of probabilistic information from $t_{in}$ to $t$ can be achieved in a generalized equilibrium state where the local probabilistic information at $t$ depends on the boundary condition at $t_{in}$. No explicit time dependence is needed, and the generalized Ising models can be considered as static in this respect, justifying the naming of ``static memory materials''. 

We also may associate the different locations $t$ on the chain with layers of a neural network. Generalized Ising models can be considered as models for certain aspects of neural networks, which treat the input information at $t_{in}$, computing output information at $t>t_{in}$. Neural networks may allow for a transport of probabilistic information between layers at $t$ and $t+\epsilon$ that is more general than the one allowed for the ``equilibrium states'' of generalized Ising models. We develop a formalism for probabilistic computing beyond the restriction of static memory materials. An important aspect of neural networks is learning - the parameters for the transmission of information between layers can be adapted in the learning process. We do not deal with this aspect in the present paper. 

Neural networks have been proposed for performing intermediate tasks for quantum computing \cite{LPLS,SNSSW,CARLTRO,CATRO,KIQB,JBC}, as finding the optimal decomposition of a unitary matrix in terms of some ``basis matrices'' or optimizing an ansatz for multi-qubit wave functions. Our purpose is more general. We propose that quantum operations can directly be performed by neural networks. Since neural networks operate classically, this amounts to realizing quantum statistics by classical statistical systems. Neuromorphic computing \cite{PBBSM,PJTM,JPBSM,BBNM,ASM,FSGH,DBKB}, or computational steps in our brain, may also realize quantum operations based on classical statistics. Learning by evolution, living systems may be able to perform quantum computational steps without realizing conditions that are often assumed to be necessary for quantum mechanics, as the presence of small and well isolated systems or low temperature. 

Quite generally, the Ising spins should be viewed as some macroscopic observables that can take two values. An example may be a spiking neuron either in the active or refractory state \cite{PBBSM,BBNM}, or the value of some macroscopic variable being above or below a given threshold. Generalizations to observables that can take more than two possible values, or continuous observables, are rather straightforward by composing them from several or infinitely many Ising spins. The key features of our argument can be seen with Ising spins, and we concentrate the discussion on this case. For a direct construction of quantum gates the macroscopic observables should be known. For questions of general structure, as the possibility of quantum computing by neural networks or the brain, only the existence of the macroscopic two-level observables matters. It may not always be easy to identify them explicitly in such systems. Learning may be partly associated to the identification of suitable relevant macroscopic observables by the network. 

Static generalized Ising models offer the advantage that the probabilistic aspects of a quantum computation can be implemented directly. The relevant macroscopic ``spin variables'' are identified explicitly and the probabilistic aspects follow directly from a standard equilibrium probability distribution or partition function. Generalized Ising models are therefore prime examples for understanding important aspects of probabilistic computing. For generalized Ising models the transport of information from a layer at $t$ to a neighboring layer at $\te$ proceeds according to the probabilistic rules of classical statistics for generalized equilibrium systems, while the input information at $t_{in}$ may be given in the form of probabilistic information about expectation values and correlations of Ising spins at $t_{in}$.

Generalized Ising models of this type can be useful for computations if at least part of the information at the boundary $t_{in}$ is transported effectively into the ``bulk'' at $t$ - a key issue for the understanding of memory and computing structures \cite{SHA,LTP,CWEL,NP,ALL,ICJ,KWC,LBL,MMW,BRH}. A formalism for the transport of information in classical statistical systems can be based on the stepwise information transport from $t$ to $\te$ \cite{CWIT,CWQF}. This work has proposed ``static memory materials'' \cite{CWIT,CWQF,SEW} for which part of the boundary information is indeed available in the bulk. The present work investigates the use of such static memory materials for performing steps of quantum computing. 

The appropriate theoretical framework for the transport of probabilistic information is the quantum formalism for classical statistics \cite{CWIT,CWQF}. It can be based on the ``classical density matrix'' $\rho^{\prime}(t)$, whose diagonal elements are the ``local'' probabilities $p(t)$ at $t$. In contrast to the local probabilities, the change from $\rho^{\prime}(t)$ to a neighboring layer $\rho^{\prime}(\te)$ obeys a simple linear evolution law
\begin{equation}\label{eq:I2} 
\rho^{\prime}(\te)=S(t)\, \rho^{\prime}(t)\, S^{-1}(t)\, .
\end{equation}
Here $S(t)$ is the step evolution operator at $t$, which corresponds to a particular normalization of the transfer matrix \cite{TM,MS,FU}. In the occupation number basis the step evolution operator for generalized Ising models is a real non-negative matrix \cite{CWIT,CWQF}. This restriction may be removed for more general forms of probabilistic computing. 

The structural similarity between classical gates \eqref{eq:I2} and quantum gates \eqref{eq:I1} is striking. In both cases one can realize sequences of gates by simple matrix multiplication. Both for quantum computations and classical computations the product of different matrices $U(\te)U(t)$ or $S(\te)S(t)$ is not commutative. Our basic ansatz is a ``bit-quantum map'' from the classical density matrices $\rho^{\prime}$ to quantum density matrices $\rho$. The map $\rho=f(\rho^{\prime})$ or shortly $\rho(\rho^{\prime})$ is not invertible, such that part of the probabilistic information in $\rho^{\prime}$ is lost for $\rho$. This classifies the quantum system described by $\rho$ as a subsystem of the classical statistical system described by $\rho^{\prime}$. As often for subsystems, the statistical information of the subsystem is incomplete in the sense that not all classical correlation functions for the Ising spins at $t$ are computable from $\rho(t)$. 

\begin{figure}[t!]
\includegraphics[scale=0.55]{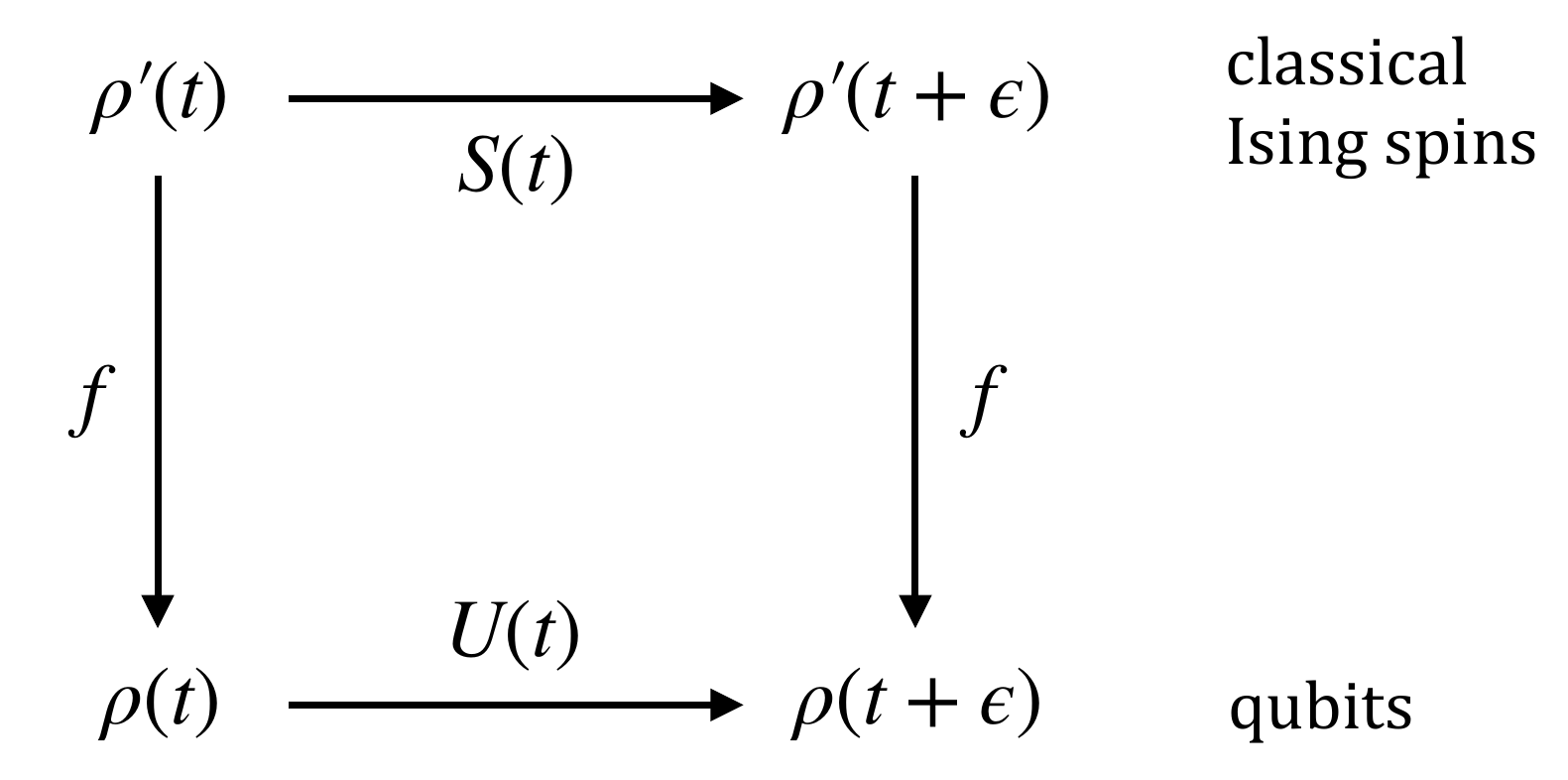}
\caption{Schematic representation of the bit-quantum map $f$.}\label{fig:1} 
\end{figure}

The map $\rho(\rho')$ induces a map from the step evolution operator $S(t)$ to a corresponding evolution operator for the quantum subsystem. This holds provided that $\rho(\te)$ can be expressed in terms of $\rho(t)$, i.e., the information in the quantum subsystem at $t$ is sufficient for the computation of the properties of the quantum subsystem at $\te$. If, in addition, pure quantum states at $t$ are mapped to pure quantum states at $\te$, the evolution operator of the quantum system is given by the unitary transformation \eqref{eq:I1}. In this case the map $\rho(\rho^{\prime})$ induces a map from $S(t)$ to $U(t)$, see Fig.~\ref{fig:1}. 

We will discuss which unitary quantum operations can be realized by the processing of probabilistic information for classical Ising spins. If one allows for arbitrary changes of classical statistical probability distributions between $t$ and $\te$, arbitrary unitary transformations can be realized for suitable quantum subsystems. This follows from the basic requirement of completeness for a quantum subsystem, namely that every possible density matrix for the subsystem can be realized by a suitable probability distribution of the classical statistical system. In general, this realization is not unique. In short, universal quantum computation can be realized for every complete bit-quantum map if the required transformations of classical statistical probability distributions can be implemented. Specializing to static memory materials the possibilities will be restricted by the properties of classical statistical step evolution operators. The answer will now depend on the generalized Ising model, and on the selection of the quantum subsystem. For a given number of qubits a large variety of quantum gates can be realized by a finite number of Ising spins.

Quantum subsystems are realized only for a subset of classical density matrices $\rho^{\prime}(t)$. The local probabilistic state at $t$ has to obey certain ``quantum conditions``. The quantum conditions guarantee the positivity of the density matrix $\rho(t)$. The classical statistical evolution with $S(t)$ has to be compatible with the quantum conditions. In particular, for pure quantum states $\rho^{\prime}$ has to obey pure state quantum conditions. If $S$ is compatible with these conditions, this will imply a unitary evolution of the quantum subsystem. In this paper we only discuss briefly the preparation of quantum states by an initial sequence of step evolution operators. We focus on the realization of unitary quantum gates.

An important issue for quantum computing are the scaling properties with the number of qubits $Q$. The density matrix for $Q$ qubits has $2^{2Q}-1$ independent elements. If each element is represented by the expectation value of an independent macroscopic two-level observable, the number of needed observables would increase very rapidly. We argue that this increase can be reduced dramatically if correlations between classical Ising spins are used for the determination of the elements of $\rho$. We propose a ``correlation map'' for which a modest number of $3Q$ classical bits is sufficient for the description of $Q$ qubits. In this case suitable $n$-point correlation functions with $n\leq Q$ have to be employed for $Q$ qubits. Limiting the degree $n$ of the correlation functions makes the reduction in the needed macroscopic observables less dramatic, but still very substantial.  We present arguments that the correlation map could be complete, showing how contradictions to generalized Bell's inequalities are avoided. A proof of completeness is not yet available, however. 

In sect.~\ref{sec:Quantum_jump_in_classical_statistical_systems} we present a simple generalized Ising model with only three bits or Ising spins, $s_{k}(t)=\pm 1$, $k=1,2,3$, at every layer $t$. This realizes discrete quantum gates for a single qubit. Many characteristic features of quantum mechanics are already visible for this simple classical statistical model: unitary evolution, complex structure, representation of observables as hermitian non-commuting operators, quantum rule for the computation of expectation values, relation between eigenvalues of operators and outcomes of possible measurements. We proceed in sect.~\ref{sec:Two_entangled_quantum_spins} to the realization of two entangled qubits by fifteen Ising spins. 

Sect.~\ref{sec:Bit_quantum_maps} discusses more generally the possible bit-quantum maps for an arbitrary number of quantum spins. We establish the correlation map that reduces greatly the number of needed classical bits by employing correlations of the Ising spins. Properties of this map are further discussed in appendix \ref{app:Correlation_map_for_two_qubits}. In sect.~\ref{sec:V} we discuss the realization of unitary transformations for quantum subsystems by maps between classical statistical probability distributions. We establish that for every complete bit-quantum map every unitary transformation of the quantum subsystem can be realized by a suitable, in general non-linear, transformation of the classical probability distribution for the Ising spins. If these non-linear transformations can be realized, universal quantum computing can be performed with a finite number of classical bits. We also discuss restricted transformations of the classical probability distribution, as computing with random operations and cellular automata. For these linear transformations universal quantum computing needs an infinite number of classical bits.

In sects~\ref{sec:Quantum_jump_in_classical_statistical_systems}-\ref{sec:V} we only employ the local probabilities $p_{\tau}(t)$ to find a given spin configuration $\tau$ at $t$. They correspond to the diagonal elements of the classical density matrix. This setting will be extended to the classical density matrix in the subsequent sections. We also concentrate in sects~\ref{sec:Quantum_jump_in_classical_statistical_systems}-\ref{sec:V} on deterministic bit-operations that map each given spin configuration at $t$ to precisely another spin configuration at $\te$. The corresponding particular ``unique jump step evolution operators'' are a type of cellular automata. They are a special case of more general probabilistic step evolution operators that will be discussed in the following sections.

In sect.~\ref{sec:Static_Memory_Materials} we turn to possible realizations of quantum gates by static memory materials. We introduce the classical density matrix $\rho^{\prime}(t)$ and discuss the partial loss of information as one proceeds from the boundary at $t_{in}$ to a location in the bulk at $t$. We construct linear bit-quantum maps from classical density matrices to quantum density matrices. In general, the information in the classical probability distribution is not sufficient for the computation of the quantum density matrix for the subsystem, such that probabilistic information in the classical density matrix beyond its diagonal elements is needed. Sect.~\ref{sec:Probabilistic_Computing} addresses probabilistic computing by static memory materials or other stochastic systems. Manipulations of classical bits are no longer deterministic, but follow probabilistic laws. We propose a general formalism that goes far beyond the limitations of standard Markov chains. It includes static memory materials, stochastic chains, quantum computing or deterministic computing as special cases.

Sect.~\ref{sec:Quantum_Computing} discusses possible realizations of quantum computers by classical Ising spins. A moderate number of classical bits can implement a rich structure of unitary quantum gates already by simple spin transformations. Not all arbitrary unitary transformations can be implemented, however, by a finite number of classical bits in this way. Arbitrary unitary transformations need an infinite number of classical bits. This ``infinity'' may correspond to continuous variables. In sect.~\ref{sec:Real_classical_variables} we consider classical statistical variables that are real numbers. Real variables admit indeed an infinite number of yes/no observables or macroscopic Ising spins. We construct explicit probability distributions over real numbers in $\mathbb{R}^3$ that account for all aspects of a single quantum spin. In sect.~\ref{sec:Quantum_Mechanics} we discuss more systematically an infinite (continuous) number of Ising spins, analogous to the infinite number of bits needed for the precise location of a point on a circle. In this limit arbitrary unitary operations for an arbitrary number of quantum spins can be realized by generalized Ising models. This demonstrates the embedding of quantum mechanics in classical statistics. The discussion of our results in sect.~\ref{sec:Discussion} points to interesting directions for further developments in the realization of novel computing structures, models for neural networks or the brain, and foundations of quantum mechanics.

\section{Quantum jumps in classical statistical systems}
\label{sec:Quantum_jump_in_classical_statistical_systems}

In this section we discuss the three-spin chain for Ising spins \cite{CWQF}. The three classical bits $s_{k}(t)=\pm 1$ at every layer $t$ realize a quantum density matrix $\rho(t)$ for a single quantum spin. Suitable unique jump step evolution operators induce a discrete unitary evolution of the density matrix. The operations between layers discussed in this section are deterministic. The probabilistic aspects appear here only through the probability distributions at a given layer on which the operations act. Despite its simplicity basic properties of quantum mechanics as non-commutativity, particle-wave duality and quantum correlations can already be found in this system. It also demonstrates the relation between the quantum subsystem and its environment, both embedded into a common classical statistical system.

\subsection{Quantum density matrix for classical statistical ensemble of Ising spins}

The key idea for realizing quantum operations by classical statistical systems consists in the identification of a suitable quantum subsystem \cite{CWA,CWB}. At every $t$ the classical probabilistic information is mapped to a quantum density matrix. In turn, the statistical information contained in the density matrix defines the quantum subsystem. In general, the map is not invertible, such that part of the classical probabilistic information is lost for the subsystem. The probabilistic information of the classical statistical ensemble that goes beyond the one in the subsystem concerns the environment of the subsystem and will play no role. In consequence, a probabilistic description of the subsystem deals with ``incomplete statistics'' \cite{CWICS,CWB}.

As a simple example we consider three Ising spins $s_{k}=\pm 1$, $k=1,\ldots ,3$. They stand for arbitrary macroscopic observables that can take two different values. At every $t$ the system is described by a classical statistical ensemble with eight classical states labeled by $\tau=1,\ldots ,8$, corresponding to the $2^{3}=8$ possibilities to have three spins up or down. The probability distribution associates to each state a probability $p_{\tau}$,
\begin{equation}\label{eq:1} 
p_{\tau}\geq 0\, ,\quad \sum_{\tau}\, p_{\tau}=1.
\end{equation}
Classical observables $A$ have a fixed value $A_{\tau}$ in each state $\tau$, with expectation value
\begin{equation}\label{eq:2} 
\exval{A}=\sum_{\tau}\, A_{\tau}p_{\tau}.
\end{equation}
We may number the states by $(0,0,0)$, $(0,0,1)$, $(0,1,0)$, $(0,1,1)$, $(1,0,0)$, $(1,0,1)$, $(1,1,0)$, $(1,1,1)$, where $1$ denotes spin up and $0$ stands for spin down. The expectation value of $s_{3}$ is then given by $\exval{s_{3}}=-p_{1}+p_{2}-p_{3}+p_{4}-p_{5}+p_{6}-p_{7}+p_{8}$, or for $A=s_{1}s_{2}$ one has $\exval{s_{1}s_{2}}=p_{1}+p_{2}-p_{3}-p_{4}-p_{5}-p_{6}+p_{7}+p_{8}$. We denote the expectation values of the three spins by $\rho_{z}$, $z=1,\ldots ,3$,
\begin{equation}\label{eq:3} 
\rho_{z}=\exval{s_{z}}\, ,\quad -1\leq \rho_{z}\leq 1.
\end{equation}
(At this stage there is no difference between the indices $z$ and $k$. We employ here $z$ for the purpose of compatibility with a later more general labeling.)

The density matrix is a complex $2\times 2$ matrix, constructed as
\begin{equation}\label{eq:4} 
\rho=\dfrac{1}{2}\left (1+\rho_{z}\tau_{z}\right )\, ,\quad \rho^{\dagger}=\rho\, ,\quad \tr(\rho)=1\, ,
\end{equation}
with $\tau_{z}$ the three Pauli-matrices. (Summation over repeated indices is implied unless stated otherwise.) It can be used to compute the expectation values of the three spins according to the usual quantum rule. For this purpose, we associate to the three spins $s_{z}$ the operators $L_{z}=\tau_{z}$,
\begin{equation}\label{eq:5} 
\exval{s_{z}}=\tr\left (L_{z}\rho\right )=\rho_{z}.
\end{equation}
The subsystem describes a single quantum spin or qubit, with spin operators $\hat{S}_k$ in the three directions given by $\tau_k$.

\subsection{Quantum subsystem and environment}

\begin{figure}[t!]
\includegraphics[scale=0.35]{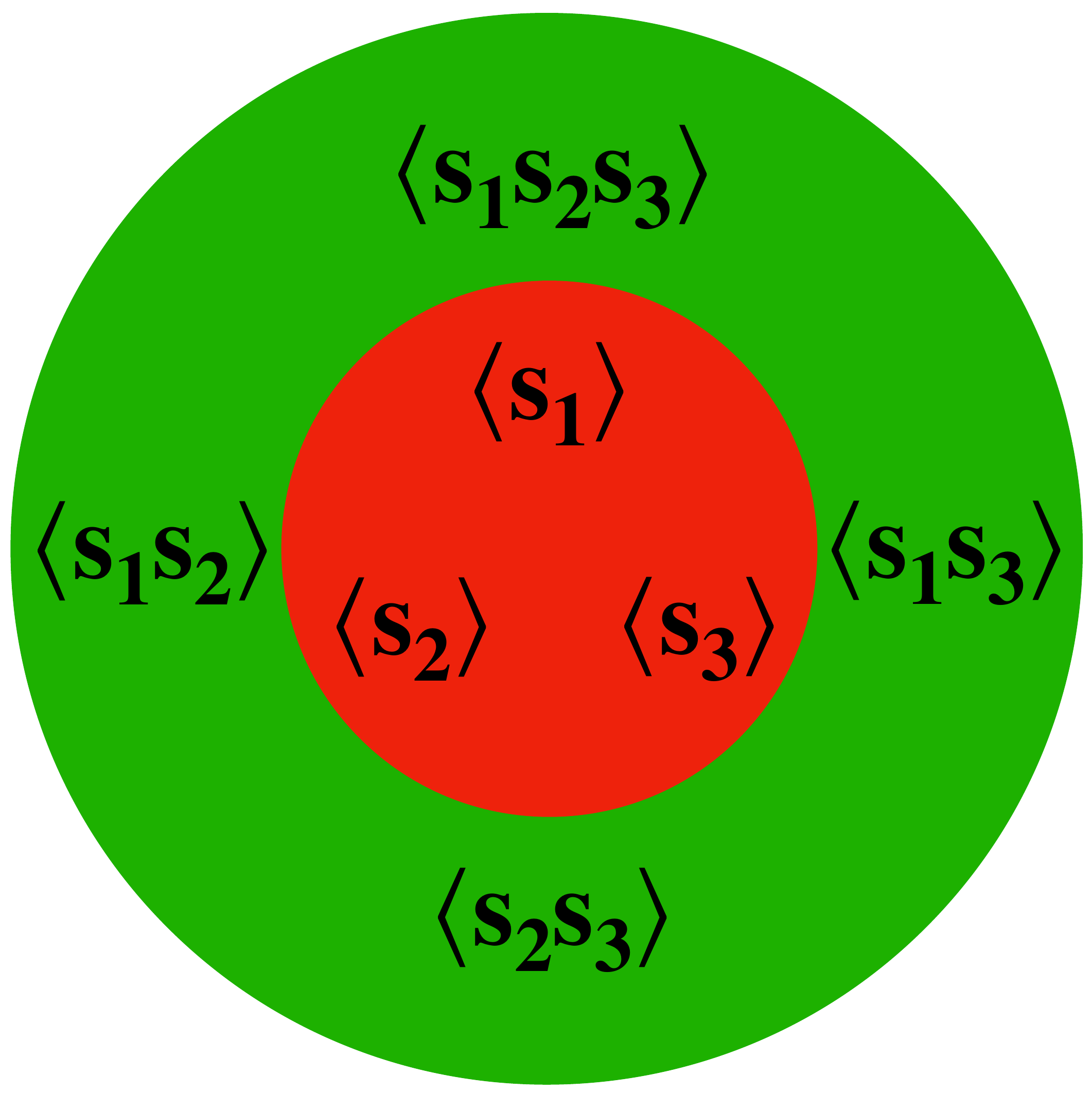}
\caption{Schematic embedding of the quantum subsystem within the classical statistical system in the space of correlation functions. The inner region (red) comprises the quantum subsystem, and the outer region (green) constitutes the environment. In contrast, a ``classical subsystem'' would eliminate $s_{1}$, consisting of the correlations $\exval{s_{2}}$, $\exval{s_{3}}$ and $\exval{s_{2}s_{3}}$. The quantum subsystem is clearly not of this type.}\label{fig:2} 
\end{figure}

The statistical information contained in the density matrix involves only three linear combinations of classical probabilities,
\begin{align}\label{eq:5A}
&\rho_{1}=\prob{-}{-}{-}{-}{+}{+}{+}{+}\, ,\nn\\
&\rho_{2}=\prob{-}{-}{+}{+}{-}{-}{+}{+}\, ,\nn\\
 &\rho_{3}=\prob{-}{+}{-}{+}{-}{+}{-}{+}.
\end{align}
They constitute the only statistical information available for the subsystem. Out of the seven independent probabilities the subsystem employs only the information contained in three independent numbers. Since the subsystem contains only reduced statistical information, it is not surprising that this information is insufficient for the computation of all expectation values of classical observables. For example, the classical correlations as $\exval{s_{1}s_{2}}$ cannot be computed from the subsystem. They need information about the environment. The subsystem realized by the three numbers $\rho_{z}$ is a simple example for ``incomplete statistics'' \cite{CWA,CWB}. We note that it cannot be obtained by simply omitting one or two of the three spins.

One may visualize the quantum subsystem as a submanifold in the space of correlation functions. For the three classical Ising spins one can define seven independent correlation functions $\exval{s_{k}}$, $\exval{s_{k}s_{l}}$ for $l\neq k$, and $\exval{s_{1}s_{2}s_{3}}$. They correspond to the seven independent probabilities. The quantum subsystem consists of $\exval{s_{k}}$, while $\exval{s_{k}s_{l}}$ and $\exval{s_{1}s_{2}s_{3}}$ constitute the environment. We have depicted this setting in Fig.~\ref{fig:2}.

\subsection{Quantum condition}

\begin{figure}[t!]
\includegraphics[scale=0.25]{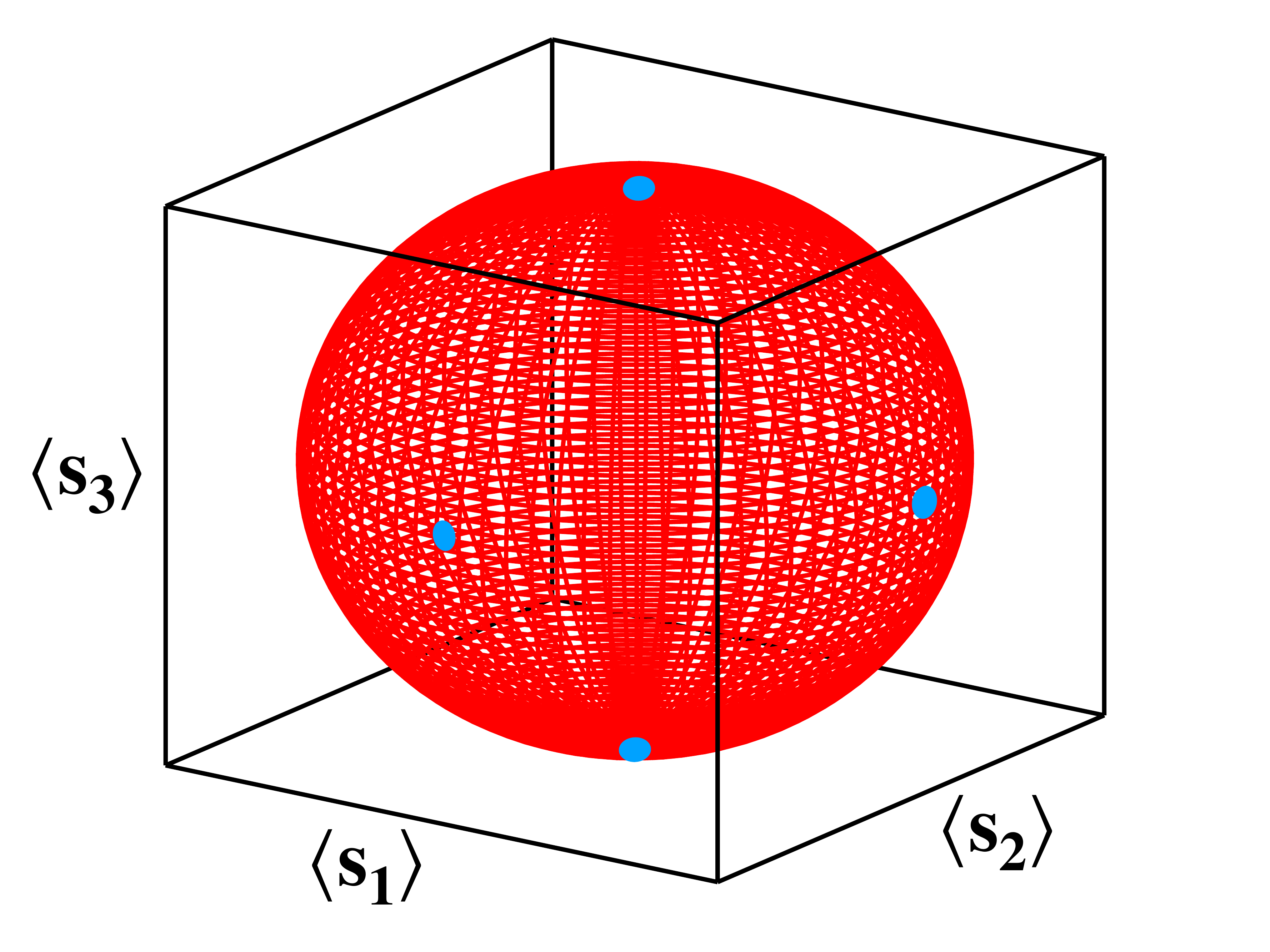}
\caption{Quantum condition. For a quantum subsystem the expectation values $\exval{s_{z}}$ must be inside or on the Bloch sphere. Points on the Bloch sphere are pure quantum states. Points outside the Bloch sphere correspond to classical statistical probability distributions that do not realize a quantum subsystem. Corners of the cube have $|\exval{s_{k}}|=1$ for all $k$ and are not compatible with the quantum subsystem. The Bloch sphere touches the cube at the points indicated at the center of its surfaces.}\label{fig:3} 
\end{figure}

A quantum mechanical density matrix has to obey the positivity condition that all eigenvalues $\lambda_{\alpha}$ must be positive definite, $\lambda_{\alpha}\geq 0$. In particular, all diagonal elements have to obey $\rho_{\alpha\alpha}\geq 0$. This imposes the ``quantum condition``
\begin{equation}\label{eq:7} 
\rho_{z}\rho_{z}\leq 1.
\end{equation}
Indeed, with $\tr(\rho)=1$ the positivity condition for a $2\times 2$ matrix is met if $\det(\rho)>0$. With $\det(\rho)=(1/4)(1-\rho_{z}\rho_{z})$ this coincides with the condition \eqref{eq:7}. We conclude that the quantum subsystem cannot be realized for arbitrary classical probability distributions, but only for a subset obeying the quantum condition. For example ,the classical probability distribution $p_{\tau}=\delta_{\tau 1}$ leads to $\rho_{1}=-1$, $\rho_{2}=-1$, $\rho_{3}=-1$ and therefore violates the condition \eqref{eq:7}. The upper boundary of eq.~\eqref{eq:7}, $\rho_{z}\rho_{z}=1$, corresponds to a pure quantum state with $\rho^{2}=\rho$. For classical probability distributions obeying the quantum condition the expectation values $\exval{s_{z}}$ are on or inside the Bloch sphere, as depicted in Fig.~\ref{fig:3}. Points outside the Bloch sphere correspond still to valid classical statistical probability distributions, but do not realize a quantum subsystem.

\subsection{Quantum operations}

Quantum operations are realized by unitary transformations of the density matrix, 
\begin{equation}\label{eq:8} 
\rho\raw U\rho\, U^{\dagger}\com U^{\dagger}U=1.
\end{equation}
For pure quantum states they act in the space of expectation values $\exval{s_{k}}$ as rotations on the Bloch sphere, see Fig.~\ref{fig:3}. This extends to mixed quantum states, with length of the rotated vector now smaller than one.

For example, the Hadamard gate is given by
\begin{equation}\label{eq:9} 
U_{H}=\dfrac{1}{\sqrt{2}}\left (\begin{array}{cc}
1 & 1 \\ 
1 & -1
\end{array} \right ).
\end{equation}
The corresponding transformation for the coefficients $\rho_{z}$ reads
\begin{equation}\label{eq:10}
\rho_{1}\raw\rho_{3}\com \rho_{2}\raw-\rho_{2}\com\rho_{3}\raw\rho_{1}.
\end{equation}
On the level of the classical spins it can be realized by the exchange $s_{1} \leftrightarrow s_{3}$, $s_{2} \rightarrow -s_{2}$. This translates on the level of the classical probabilities $p_{\tau}$ to
\begin{equation}\label{eq:11}
p_{1}\leftrightarrow p_{3}\com p_{2}\leftrightarrow p_{7}\com p_{4}\leftrightarrow p_{5}\com p_{6}\leftrightarrow p_{8}\, ,
\end{equation}
realizing eq.~\eqref{eq:10}. The transformation \eqref{eq:11} is a unique jump operation where each state $\tau$ jumps precisely to another state $\tau^{\prime}$. It can obviously be performed by a deterministic manipulation of the Ising spins.

There are other unique jump operations among the $p_{\tau}$ that realize suitable unitary transformations of the density matrix. We list a few simple ones that we specify by their (deterministic) action on the classical spins
\begin{widetext}
\begin{align}\label{eq:11A}
& s_{1}\raw s_{2}\, ,\, s_{2}\raw -s_{1}\, :\, \rho_{1}\raw \rho_{2}\, ,\,\rho_{2}\raw -\rho_{1}\, :\, U_{12}=\left (\begin{array}{cc}
1 & 0 \\ 
0 & -i
\end{array} \right )\, ,\nonumber\\
& s_{3}\raw s_{1}\, ,\, s_{1}\raw -s_{3}\, :\, \rho_{3}\raw \rho_{1}\, ,\,\rho_{1}\raw -\rho_{3}\, :\, U_{31}=\dfrac{1}{\sqrt{2}}\left (\begin{array}{cc}
1 & 1 \\ 
-1 & 1
\end{array} \right )\, ,\nonumber\\
& s_{1}\raw- s_{1}\, ,\, s_{2}\raw -s_{2}\, :\,\rho_{1}\raw -\rho_{1}\, ,\,\rho_{2}\raw -\rho_{2}\, :\, U_{Z}=\left (\begin{array}{cc}
1 & 0 \\ 
0 & -1
\end{array} \right )\, ,\nonumber\\
& s_{1}\raw- s_{1}\, ,\, s_{3}\raw -s_{3}\, :\, \rho_{1}\raw -\rho_{1}\, ,\,\rho_{3}\raw -\rho_{3}\, :\, U_{Y}=\left (\begin{array}{cc}
0 & 1 \\ 
-1 & 0
\end{array} \right )\, ,\nonumber\\
& s_{2}\raw- s_{2}\, ,\, s_{3}\raw -s_{3}\, :\, \rho_{2}\raw -\rho_{2}\, ,\,\rho_{3}\raw -\rho_{3}\, :\, U_{X}=\left (\begin{array}{cc}
0 & 1 \\ 
1 & 0
\end{array} \right )\, .
\end{align}
\end{widetext}
The group structure of the unit jump operations is preserved. For example, performing first the transformation $U_{X}$ yields $\rho_{2}^{\prime}=-\rho_{2}$, $\rho_{3}^{\prime}=-\rho_{3}$. Applying subsequently the transformation $U_{31}$, $\rho_{1}^{\prime\prime}=-\rho_{3}^{\prime}=\rho_{3}$, $\rho_{3}^{\prime\prime}=\rho_{1}^{\prime}=\rho_{1}$, $\rho_{2}^{\prime\prime}=\rho_{2}^{\prime}=-\rho_{2}$, results in the transformation \eqref{eq:10}, as reflected by the corresponding matrix multiplication yielding eq.~\eqref{eq:9}, $U_{H}=U_{31}U_{X}$. One realizes $U_{12}^{2}=U_{Z}$, $U_{31}^{2}=U_{Y}$, $U_{X}=U_{Z}U_{Y}$. With $(U_{31}U_{12})^{3}=(1+i)/\sqrt{2}$ one finds that the transformation $U_{31}U_{12}$ has period three, noting that the overall phase does not matter. The transformations $U_{12}$ or $U_{31}$ have period four.

A particular unique jump operation
\begin{equation}\label{eq:11B}
s_{2}\raw -s_{2}\com \rho_{2}\raw -\rho_{2}\com \rho\raw\rho^{*}
\end{equation}
transforms the density matrix into its complex conjugate. This operation cannot be expressed as a unitary transformation \eqref{eq:8}. We may combine it with transformations of the type \eqref{eq:11A}. We observe that not every unit jump operation is reflected by a unitary transformation \eqref{eq:8}. For example, the exchange $p_{1}\leftrightarrow p_{3}$, $p_{2}\leftrightarrow p_{4}$, corresponds to a conditional change of $s_{2}$. It is flipped if $s_{1}=-1$, and left invariant for $s_{1}=1$. This transformation leaves $\rho_{1}$ and $\rho_{3}$ invariant, while $\rho_{2}$ changes to $\rho_{2}^{\prime}=\prob{+}{+}{-}{-}{-}{-}{+}{+}$. The latter cannot be expressed in terms of $\rho_{1}$, $\rho_{2}$ and $\rho_{3}$, resulting therefore in an environment dependent change of the density matrix rather than a quantum operation.

Every unitary transformation \eqref{eq:8} can be expressed (in a non-unique way) by some transformation among the classical probabilities $p_{\tau}$. This follows from the simple observation that for every set of $\exval{s_{k}}$ which admits a quantum subsystem at $t$ one can find a classical statistical probability distribution $\lbrace p_{\tau}(t)\rbrace$ realizing it. The same holds for the set of $\exval{s_{k}}$ at $\te$ with associated probability transformation $\lbrace p_{\tau}(\te)\rbrace$. Thus arbitrary $\rho(t)$ and arbitrary $\rho(\te)$ can be realized by suitable $\lbrace p_{\tau}(t)\rbrace$ and $\lbrace p_{\tau}(\te)\rbrace$, respectively. Any arbitrary unitary transformation $\rho(t)\raw\rho(\te)$ can be implemented by a corresponding map $\lbrace p_{\tau}(t)\rbrace\raw \lbrace p_{\tau}(\te)\rbrace$. Neither the probability distributions $\lbrace p_{\tau}(t)\rbrace$, $\lbrace p_{\tau}(\te)\rbrace$, nor the map are unique. The map needs not to be linear.

Only a subset of transformations of probability distributions corresponds, however, to the unique jump operations which form the discrete eight-dimensional permutation group. Maps among probability distributions respecting the quantum condition correspond to maps of points inside or on the Bloch sphere in Fig.~\ref{fig:3}. They are not necessarily unitary operations, since the length of the rotated vector can change. If, however, a map between probability distributions maps every pure quantum state to a pure quantum state, the associated map $\rho(t)\raw\rho(\te)$ is unitary. In this case the vector $(\exval{s_{1}},\exval{s_{2}},\exval{s_{3}})$ is rotated, with length preserved.

\subsection{Non-commuting observables in the quantum \\
system}

The three operators $\hat{S}_{k}$ for a single quantum spin in the $x$, $y$ and $z$ direction are given by $\tau_{1}$, $\tau_{2}$ and $\tau_{3}$. With the quantum rule for expectation values,
\begin{equation}\label{eq:12A}
\exval{S_{k}}=\tr(\rho\hat{S}_{k})=\rho_{k}=\exval{s_{k}}\, ,
\end{equation}
the expectation values of the quantum spin in each of the three directions coincides with the expectation value of a particular classical Ising spin, that we may associate to the same direction.

The three operators $\hat{S}_{k}=\tau_{k}$ do not commute. This may be surprising at first sight, since classical statistics is often believed to be commutative, and no sign of non-commutativity is visible directly for the three Ising spins $s_{k}$. The non-commutative operator representation of the observables is a property of the subsystem. Incomplete statistics \cite{CWICS,CWIS} for the subsystem does not permit the computation of classical correlations as $\exval{s_{1}s_{2}}$. In other words, the probabilistic information in the density matrix characterizing the subsystem is not sufficient for the determination of \emph{simultaneous} probabilities to find a given value $\pm 1$ for $s_{1}$ \emph{and} to find a value $\pm 1$ for $s_{2}$. This lack of information enforces non-commutative operators. If $\hat{S}_{1}$ and $\hat{S}_{2}$ would commute, the simultaneous probabilities would be computable from $\rho$.

\subsection{Particle-wave duality}

The discreteness of quantum mechanics, e.g. the association of the possible outcomes of measurements with the discrete eigenvalues of the quantum operators $\hat{S}_{k}$, follows directly from the discreteness of the classical Ising spins. The possible measurement values correspond indeed to the spectrum of the associated operators. Also the quantum rule \eqref{eq:12A} for the computation of expectation values follows directly from the standard classical statistical setting \eqref{eq:2}.

On the other hand, expectation values are continuous, and the density matrix involves continuous variables as well. For pure quantum states the density matrix can be expressed as a bilinear in the quantum wave function $\psi$, $\rho_{ij}=\psi_{i}\psi_{j}^{*}$. The wave function is continuous. Unitary transformations $\psi(t)\raw\psi(\te)$ are described by a linear (discrete) Schr{\"o}dinger equation. In this sense particle-wave duality is nothing else than the compatibility of discrete measurement values (particle aspect) with a continuous probabilistic description by $\psi$ or $\rho$ (wave aspect).

\subsection{Measurements and conditional probabilities}

Even though the issue of measurements is somewhat besides the main topic of this paper and not needed for our purposes, a brief discussion may be useful for an understanding why no conflict with Bell's inequalities arises in our approach. A valid measurement at $t$ of a given Ising spin, say $s_{1}$, with measurement value $s_{1}=1$, will set after the measurement the probability of finding $s=-1$ to zero. This does not yet tell what happens during the measurement to the two other spins $s_{2}$ and $s_{3}$. Different measurement procedures for a measurement of $s_{1}$ will, in general, result in different outcomes for the probability distribution concerning $s_{2}$ and $s_{3}$. We define an ideal quantum measurement as a measurement that is compatible with the quantum subsystem. If the probability distribution obeys the quantum condition before the measurement, it should still obey the quantum condition after an ideal quantum measurement. The combination of $\exval{s_{1}}=1$ and the quantum condition fixes the state after the ideal quantum measurement as a pure quantum state with $\exval{s_{1}}=1$, $\exval{s_{2}}=\exval{s_{3}}=0$.

For a sequence of two measurements one is interested in conditional probabilities, say the probability to find at $\te$ $s_{2}=+1$ or $-1$ after $s_{1}$ has been measured at $t$ to be $+1$. We denote the conditional probabilities to find $s_{2}(\te)=1$ or $-1$ by $(w_{+}^{2})^{1}_{+}$ and $(w_{-}^{2})^{1}_{+}$ respectively, and similarly by $(w_{+}^{2})^{1}_{-}$ and $(w_{-}^{2})_{-}^{1}$ if $s_{1}(t)$ was measured to be $-1$. The conditional probabilities for an ideal quantum measurement obey
\begin{align}\label{eq:16A} 
(w_{+}^{2})^{1}_{+}+(w_{-}^{2})^{1}_{+}&=1\, ,\nn\\
(w_{+}^{2})^{1}_{+}-(w_{-}^{2})^{1}_{+}&=\tr \left[ \tau_{2}\, U(t)\, \rho_{(1+)}U^{-1}(t)\right] \, ,
\end{align}
with defining $U(t)$ the unitary evolution of the density matrix from $t$ to $\te$ and $\rho_{(1+)}=(1+\tau_{1})/2$ the density matrix for a quantum state with $\exval{s_{1}(t)}=1$. Similar relations hold for $(w_{+}^{2})^{1}_{-}$ and $(w_{-}^{2})^{1}_{-}$, with $\rho_{(1+)}$ replaced by $\rho_{(1-)}=(1-\tau_{1})/2$. The expectation value of $\exval{s_{2}(\te)}$ under the condition that $s_{1}(t)$ was measured to be $+1$ is given by the second equation \eqref{eq:16A}. It can be obtained by the ``reduction of the density matrix`` to $\rho(t)=\rho_{1+}$ after $s_{1}(t)=1$ has been measured, a subsequent time evolution from $\rho(t)$ to $\rho(\te)$ and finally evaluating the expectation value $\exval{s_{2}(\te)}$ according to eq.~\eqref{eq:12A}. If we define ideal quantum measurements by the conditional probabilities \eqref{eq:16A}, the ``reduction of the density matrix`` can be seen as a pure mathematical prescription for an efficient computation of the conditional probabilities. It does not need to be implemented by an actual physical change of the quantum subsystem by the measurement.

The measurement correlation weighs the expectation values $\exval{s_{2}(\te)}_{\pm}$ after $s_{1}(t)$ has been found to be $+1$ or $-1$ with the probabilities that $s_{1}(t)$ is found to be $+1$ or $-1$ in the state characterized by $\rho(t)$, and with the corresponding value of $s_{1}(t)$,
\begin{align}\label{eq:16B}
\exval{s_{2}(\te)s_{1}(t)}_{m}&=\left [(w_{+}^{2})^{1}_{+}-(w_{-}^{2})^{1}_{+}\right ]\dfrac{1+\exval{s_{1}(t)}}{2}\nn\\
&\quad -\left [(w_{+}^{2})^{1}_{-}-(w_{-}^{2})^{1}_{-}\right ]\dfrac{1-\exval{s_{1}(t)}}{2}\nn\\
&=\dfrac{1}{2}\tr\left (\left\lbrace \hat{S}_{2H}(\te)\, \hat{S}_{1H}(t) \right\rbrace\, \rho\right )\, .
\end{align}
Here $\hat{S}_{1H}(t)=\tau_{1}$ and $\hat{S}_{2H}(\te)=U^{\dagger}(t)\, \tau_{2}\, U(t)$ are the operators for $s_{1}(t)$ and $s_{2}(\te)$ in the Heisenberg picture (with reference point $t$) and we employ the identities
\begin{equation}\label{eq:16C}
2(1\pm\exval{s_{1}})\, \rho_{1\pm}=(1\pm\tau_{1})\, \rho\, (1\pm\tau_{1})\, .
\end{equation}
The relation \eqref{eq:16B} is precisely the quantum correlation involving the operator product. (For more details on conditional probabilities and measurement correlations see \cite{CWB}.)

\section{Two entangled quantum spins}\label{sec:Two_entangled_quantum_spins}

As a next step we want to realize quantum operations on a system of two entangled quantum spins. Two qubits are described by a hermitian $4\times 4$ density matrix with elements $\rho_{\alpha\beta}$, $\alpha,\beta=1,\ldots ,4$. Hermiticity and the normalization $\tr(\rho)=1$ imply that $\rho$ is characterized by $15$ independent elements. In the present section we realize the two-qubit quantum subsystem by $15$ independent classical bits. In the next section we discuss a realization employing only six classical bits. Some elements of the density matrix will then be defined by classical correlation functions.

A typical quantum operation is the CNOT-gate, corresponding to
\begin{equation}\label{eq:12}
U_{C}=\left (\begin{array}{cc}
1 & 0 \\ 
0 & \tau_{1}
\end{array} \right )=U_{C}^{\dagger}\, .
\end{equation}
It has the property that it maps a direct product state to an entangled state. The CNOT-gate, together with the Hadamard gate $U_{H}$ and a $\pi/8$-rotation in Hilbert space (to be discussed in sect.~\ref{sec:Quantum_Computing}), form a basic set of quantum gates from which all unitary transformations for two qubits can be obtained by suitable sequential applications. We will discuss which classical statistical systems can realize this quantum gate, as well as other quantum gates.

\subsection{Quantum density matrix}

For the density matrix,
\begin{equation}\label{eq:13}
\rho=\dfrac{1}{4}\left (1+\rho_{z}L_{z}\right )=\dfrac{1}{4}\left (1+\rho_{\mu\nu} L_{\mu\nu}\right )\, ,
\end{equation}
we need $15$ numbers $\rho_{z}$ in a suitable basis for the generators $L_{z}$ of $SU(4)$. We first realize this system by fifteen classical Ising spins $s_z = s_{\mu\nu}$, $\mu,\nu=0,1,2,3$, with $s_{00}$ omitted. For this realization we define $\sigma_{\mu\nu}=s_{\mu\nu}$ for $(\mu,\nu)\neq (0,0)$, and $\sigma_{00}=0$. Later, we will discuss other expressions of $\sigma_{\mu\nu}$ in terms of Ising spins. Similar to eq.~\eqref{eq:3} we define
\begin{equation}\label{eq:14}
\rho_{z}=\rho_{\mu\nu}=\exval{\sigma_{\mu\nu}}\, .
\end{equation}
(Recall that $\rho_{00}$ does not figure among the fifteen $\rho_{z}$.) We label the $SU(4)$ generators $L_{z}$ by
\begin{equation}\label{eq:15}
L_{z}=L_{\mu\nu}=\tau_{\mu}\otimes \tau_{\nu}\, ,
\end{equation}
with Pauli matrices $\tau_{k}$ and $\tau_{0}=1$. (Again $L_{00}=1$ is not part of the $L_{z}$.) We observe the relations 
\begin{align}\label{eq:16}
\tr(L_{\mu\nu})&=0\com L_{\mu\nu}^{\dagger}=L_{\mn}\, ,\nn \\
\tr( L_{\mn}L_{\rho\lambda})&=4\delta_{\mu\rho}\delta_{\nu\lambda}\, ,
\end{align}
implying
\begin{equation}\label{eq:17}
\tr(L_{\mn}\rho)=\rho_{\mn}=\exval{\sigma_{\mn}}.
\end{equation}
We can therefore represent the observables $\sigma_{\mn}$ by the operators $L_{\mn}$. The eigenvalues of $L_{\mn}$ are $\pm 1$, corresponding to the possible measurement values $\sigma_{\mn}=\pm 1$ in the $2^{15}$ states $\tau$ of the classical statistical ensemble. 

The quantum condition for the positivity of the density matrix requires now
\begin{equation}\label{eq:18}
\rho_{z}\rho_{z}\leq 3 \, .
\end{equation}
This ensures $\tr(\rho^{2})\leq 1$, according to
\begin{align}\label{eq:19}
\tr(\rho^{2})&=\dfrac{1}{16}\, \tr\left( 1+2\rho_{z}L_{z}+\rho_{z}\rho_{y}L_{z}L_{y}\right) \nonumber \\
&=\dfrac{1}{4}\left( 1+\rho_{z}\rho_{z}\right )\, .
\end{align}
Pure quantum states with $\rho^{2}=\rho$ have to obey $\rho_{z}\rho_{z}=3$. Indeed, for pure states one needs $\tr(\rho^{2})=1$, such that $\rho_{z}\rho_{z}=3$ is a necessary requirement. This condition is, however, not sufficient \cite{CWB}. For mixed quantum states one has $\rho_{z}\rho_{z}<3$.

A pure state density matrix can be written in terms of a complex wave function $\psi$,
\begin{equation}\label{eq:20}
\rho_{\alpha\beta}=\psi_{\alpha}\psi^{*}_{\beta}\com\psi_{\alpha}\psi_{\alpha}^{*}=1\, .
\end{equation}
The particular pure state
\begin{equation}\label{eq:21}
\rho_{\alpha\beta}^{(0)}=\delta_{\alpha 1}\delta_{\beta 1}\com \psi_{\alpha}^{(0)}=\delta_{\alpha 1}
\end{equation}
corresponds to $\rho_{30}=\rho_{03}=\rho_{33}=1$, with all other $\rho_{\mn}$ vanishing. It obeys $\rho_{z}\rho_{z}=3$. An arbitrary pure state can be obtained from $\psi_{\alpha}^{(0)}$ by a unitary transformation, $V^{\dagger}V=1$,
\begin{equation}\label{eq:22}
\psi=V\psi^{(0)}\com \rho=V\rho^{(0)}V^{\dagger}\, .
\end{equation}
Unitary transformations do not change $\tr(\rho^{2})$, leaving therefore $\rho_{z}\rho_{z}=3$ unchanged. The manifold of $\rho_{z}$ corresponding to pure states is obtained from eq.~\eqref{eq:22} and corresponds to the complex projective space $\bC\bP^{4}$. This is a submanifold of the space $S^{15}$ defined by $\rho_{z}\rho_{z}=3$. The latter condition is therefore not sufficient for the realization of a positive quantum density matrix. The ``pure state condition'' involves additional restrictions on the $\rho_{\mn}$, which can be formulated as the condition $\rho^{2}=\rho$. Pure state density matrices have one eigenvalue equal to one and all others zero, and are therefore positive. The positivity extends to mixed quantum states for which the density matrix is a weighted sum of pure states density matrices $\rho^{(i)}$, $\rho=\sum_{i}\, w_{i}\rho^{(i)}$, $0\leq w_{i}\leq 1$, $\sum_{i}\, w_{i}=1$.

\subsection{Quantum operations on two qubits}

We may take the three components of the first quantum spin operator to correspond to $L_{k0}=\tau_{k}\otimes 1$, while the operators associated to the second quantum spin are given by $L_{0k}=1\otimes \tau_{k}$. The classical Ising spins are labeled as $\sigma_{k0}=s_{k}^{(1)}$, $\sigma_{0k}=s^{(2)}_{k}$, such that the expectation values of the two quantum spins in the three cartesian directions coincide with the ones for the classical Ising spins $s_{k}^{(i)}$. Unitary transformations acting only on the spin one are represented by $U=U^{(1)}\otimes 1$, while those acting on spin two read $U=1\otimes U^{(2)}$. Here $U^{(1)}$, $U^{(2)}$ are unitary $2\times 2$ matrices, given by eq.~\eqref{eq:11A}. On the level of the classical spins they are realized by the same transformations of the $s_{k}^{(i)}$ as in sect.~\ref{sec:Quantum_jump_in_classical_statistical_systems}, while $\sigma_{kl}$ transforms the same way as $s_{k}^{(1)}s_{l}^{(2)}$. For example, the Hadamard gate for spin one corresponds to $\sigma_{10}\leftrightarrow \sigma_{30}$, $\sigma_{20}\leftrightarrow -\sigma_{20}$, $\sigma_{0k}$ invariant, $\sigma_{1k}\leftrightarrow \sigma_{3k}$, $\sigma_{2k}\leftrightarrow -\sigma_{2k}$. Similarly, the Hadamard gate for spin two is realized by $\sigma_{01}\leftrightarrow \sigma_{03}$, $\sigma_{02}\leftrightarrow -\sigma_{02}$, $\sigma_{k0}$ invariant, $\sigma_{k1}\leftrightarrow \sigma_{k3}$, $\sigma_{k2}\leftrightarrow -\sigma_{k2}$. These transformations constitute unique mappings between the states $\tau$, and correspondingly between the $2^{15}$ probabilities $p_{\tau}$. The transformation 
\begin{align}\label{eq:30A}
L_{\mn}=(\tau_{\mu}\otimes\tau_{\nu})\raw \left (1\otimes U^{(2)}\right )(\tau_{\mu}\otimes\tau_{\nu})\left (1\otimes \bigl (U^{(2)}\bigl )^{\dagger}\right )
\end{align}
 implies that the $L_{k 0}$ are invariant, while $L_{\mu k}\raw \tau_{\mu}~\otimes~U^{(2)}\,\tau_{k}\, U^{(2)\dagger}$. Thus the above transformations of $\sigma_{\mn}$, and correspondingly of $\rho_{\mn}$, indeed realize the unitary transformations for spin two. 

A transformation acting on both spin one and spin two is the exchange $s^{(1)}_{k}\leftrightarrow s^{(2)}_{k}$, accompanied by $s_{kl}\leftrightarrow s_{lk}$. It results in $\rho_{\mn}\leftrightarrow \rho_{\nu\mu}$. The corresponding unitary transformation has therefore to obey
\begin{equation}\label{eq:27}
U_{S}L_{\mn}U_{S}^{\dagger}=L_{\nu\mu},
\end{equation}
as realized by the swap gate
\begin{equation}\label{eq:28}
U_{S}=\left (\begin{array}{cccc}
1 & 0 & 0 & 0 \\ 
0 & 0 & 1 & 0 \\ 
0 & 1 & 0 & 0 \\ 
0 & 0 & 0 & 1
\end{array} \right )=U_{S}^{\dagger}\, .
\end{equation}
The swap gate maps unitary operations on the first spin to unitary operations acting on the second spin, e.g.
\begin{equation}\label{eq:29A} 
U_{S}\, U_{12}^{(1)}\, U_{S} = U_{12}^{(2)}\, .
\end{equation}
Combination with the operations \eqref{eq:11A} permits unitary transformations with period eight, e.g.
\begin{equation}\label{eq:29B}
\left( U_{12}^{(1)}\, U_{S} \right)^{2} = \diag(1,-i,-i,-1)\com 
\left( U_{12}^{(1)}\, U_{S} \right)^{8} = 1\, .
\end{equation}

We next turn to the CNOT gate \eqref{eq:12}. For the wave function of a pure state we may choose four basis functions
\begin{align}\label{eq:23}
&\psi_{1}=(1,0,0,0)=\ket{\ua\ua}\com &\psi_{2}=(0,1,0,0)=\ket{\ua\da}\, ,\nonumber\\
&\psi_{3}=(0,0,1,0)=\ket{\da\ua}\com &\psi_{4}=(0,0,0,1)=\ket{\da\da}\, ,
\end{align}
where the arrows denote, as usual, eigenvalues of $L_{30}$ and $L_{03}$. The CNOT-operation \eqref{eq:12} exchanges $\psi_{3}\leftrightarrow\psi_{4}$. It switches the second spin if the first one is down, and leaves the second spin invariant if the first one is up. It can be seen as a conditional quantum spin flip. Starting from a product state
\begin{align}
\label{eq:24}\psi_{in}&=\dfrac{1}{\sqrt{2}}\left (\ket{\ua}-\ket{\da}\right )\ket{\da}=\dfrac{1}{\sqrt{2}}\left (\ket{\ua\da}-\ket{\da\da}\right )\\
&=\dfrac{1}{\sqrt{2}}\left (0,1,0,-1\right )\, ,\nonumber
\end{align}
the CNOT-gate produces an entangled state
\begin{equation}\label{eq:25}
\psi_{f}=U_{C}\psi_{in}=\dfrac{1}{\sqrt{2}}\left (\ket{\ua\da}-\ket{\da\ua}\right )=\dfrac{1}{\sqrt{2}}\left (0,1,-1,0\right ).
\end{equation}
While $\psi_{in}$ corresponds to fixed values $s_{1}^{(1)}=-1$, $s_{3}^{(2)}=-1$, $\sigma_{13}=1$, with all other expectation values vanishing, the non-vanishing expectation values for $\psi_{f}$ are $\sigma_{11}=\sigma_{22}=\sigma_{33}=-1$. In the entangled state $\psi_{f}$ the correlations rather than the individual spins take sharp values.

The unitary transformation \eqref{eq:12} of the CNOT-gate, $\rho\raw U\, \rho\, U^{\dagger}$, is realized by the map
\begin{align}
 &\rho_{10}\leftrightarrow \rho_{11}\com \rho_{20}\leftrightarrow \rho_{21}\com \rho_{13}\leftrightarrow -\rho_{22}\, ,\nonumber\\
& \rho_{02}\leftrightarrow \rho_{32}\com \rho_{03}\leftrightarrow \rho_{33}\com \rho_{23}\leftrightarrow \rho_{12}\, ,\nonumber\\
\label{eq:26}&\rho_{30},\,\rho_{01},\,\rho_{31}\,  \text{  invariant}.
\end{align}
On the level of the classical Ising spins this is simply realized by the corresponding map between the $s_{\mn}$. This constitutes a direct proof that entangled quantum spin states, and operations changing product quantum states to entangled states, can be represented by classical Ising spins and operations among them \cite{CWA}.

More generally, exchanges and sign changes of spins, $\rho_{\mn}\raw \rho^{\prime}_{\mn}=\pm \rho_{\mu^{\prime}\nu^{\prime}}$, can be realized by unitary transformations provided that $\rho^{\prime}=(1+\rho^{\prime}_{\mn}L_{\mn})/4=(1+\rho_{\mn}L^{\prime}_{\mn})/4$, $L^{\prime}_{\mn}=UL_{\mn}U^{\dagger}$. (We occasionally use primes for maps, e.g. $\rho\raw\rho^{\prime}$ stands for $\rho(t)\raw\rho(\te)$. In this case $\rho^{\prime}$ should not be confounded with the classical density matrix.) Generators $L^{\prime}$ and $L$ related by a unitary transformation have to obey the consistency condition for operator products
\begin{align}\label{eq:29}
L_{\mn}L_{\tau\rho}=C_{\mn\tau\rho\lambda\sigma}L_{\lambda\sigma}\;\Rightarrow\; L^{\prime}_{\mn}L^{\prime}_{\tau\rho}=C_{\mn\tau\rho\lambda\sigma}L^{\prime}_{\lambda\sigma}.
\end{align}
For example, this property is responsible for the important minus sign in $\rho_{13}\leftrightarrow -\rho_{22}$. Indeed, from $L_{13}=L_{10}L_{03}$ one infers $L^{\prime}_{13}=L_{10}^{\prime}L_{03}^{\prime}=L_{11}L_{33}=(\tau_{1}\otimes \tau_{1})(\tau_{3}\otimes \tau_{3})=\tau_{1}\tau_{3}\otimes \tau_{1}\tau_{3}=-\tau_{2}\otimes \tau_{2}=-L_{22}$.

\subsection{Bell's inequalities}

The classical Ising spins $s_{z}$ may be considered as hidden variables. The classical correlation functions $\exval{s_{z}s_{y}}$ have therefore to obey Bell's inequalities \cite{BELL} and generalizations thereof. For example, the identity
\begin{equation}\label{eq:BI1}
s_u\,(s_v - s_w) = s_u s_v\, (1-s_v s_w) = \pm (1 - s_v s_w)
\end{equation}  
implies an inequality for classical correlation functions
\begin{equation}\label{eq:BI2}
|\langle s_u s_v \rangle - \langle s_u s_w \rangle| \leq 1 - \langle s_v s_w \rangle\, .
\end{equation}
These classical correlation functions are, however, part of the environment, as visible by generalizing the setting of Fig.~\ref{fig:2} to fifteen Ising spins. The values of the classical correlations can be computed in the classical statistical system only if the full probabilistic information is available. They are not properties of the quantum subsystem, and their values cannot be computed from the incomplete statistical information of the quantum subsystem. Classical probability distributions with different values of the classical correlation $\exval{s_{z}s_{y}}$ are mapped to the same quantum subsystem.

In contrast, the measurement correlation for ideal quantum measurements are properties of the quantum subsystem. They can be computed from the information contained in $\rho(t)$ and its evolution with $t$ \cite{CWPO}, \cite{CWA}, \cite{CWB}. The measurement correlations do not have to obey Bell's inequalities. The simple reason why our setting is not in conflict with Bell's inequalities is that these equalities apply to quantities that are of no interest in our context. In the following we will not consider intermediate measurements or sequences of measurements. We concentrate on sequences of quantum gates, assuming that necessary measurements of the outcome are performed at the end of the sequence.

\section{Bit-quantum maps}\label{sec:Bit_quantum_maps} 

The bit-quantum maps map the probabilistic information for classical Ising spins to a quantum density matrix for quantum spins. In our context these maps are performed for every $t$-layer separately. In the present section the probabilistic information about the Ising spins concerns their expectation values and correlation functions. In turn, they are given by the probabilities to find different spin combinations. In sect.~\ref{sec:Static_Memory_Materials} we will extend this to a more general formulation in terms of the classical density matrix. The bit-quantum maps are not unique. We first extend the construction of sects~\ref{sec:Quantum_jump_in_classical_statistical_systems}, \ref{sec:Two_entangled_quantum_spins} to an arbitrary number of quantum spins and subsequently address possible other realizations that employ suitable correlation functions of the Ising spins in addition to their expectation values. Further bit-quantum maps will be discussed in sects~\ref{sec:Quantum_Computing}-- \ref{sec:Quantum_Mechanics}. The question which quantum gates can be realized by classical Ising spins depends on the choice of the bit-quantum map.

For a given generalized Ising model the quantum density matrix $\rho$ defines the quantum-subsystem. Different bit-quantum maps correspond to different choices of subsystems. On the other hand, for a given $\rho$ the different bit-quantum maps for different generalized Ising models account for the different possibilities to realize a given quantum subsystem by generalized Ising models.

A particularly interesting bit-quantum map is the ``correlation map``. It realizes a quantum subsystem for $Q$ qubits by $3Q$ classical Ising spins. If this map can be exploited for practical applications, it leads to highly attractive scaling properties of classical statistical realizations of quantum operations. The key element of the correlation map is the association of some of the elements of $\rho$ to correlations of classical Ising spins $s_{k}^{(i)}$. For $Q$ qubits one exploits $n$-point functions up to the degree $n=Q$. The rapid increase of the number of $n$-point functions with $n$ matches the rapid increase of the number of independent elements of $\rho$ with $Q$. The establishment of this map is the key topic of this section.

\subsection{Arbitrary number of qubits}

Quantum operations realized by classical Ising spins can be extended to an arbitrary number of quantum spins. As a first possibility we may consider a realization of $Q$ quantum spins by $M=2^{2Q}-1$ independent classical Ising spins $\sigma_{\mu_{1}\mu_{2}\ldots\mu_{Q}}=s_{\mu_{1}\mu_{2}\ldots\mu_{Q}}$, with $\sigma_{00\ldots 0}=0$. The density matrix is given by
\begin{equation}\label{eq:30}
\rho=2^{-Q}\left (1+\rho_{\mu_{1}\ldots\mu_{Q}} L_{\mu_{1}\ldots\mu_{Q}}\right )\com \rho_{\mu_{1}\ldots\mu_{Q}}=\exval{\sigma_{\mu_{1}\ldots\mu_{Q}}}\, ,
\end{equation}
with $SU(2^{Q})$-generators
\begin{equation}\label{eq:31}
L_{\mu_{1}\mu_{2}\ldots\mu_{Q}}=\tau_{\mu_{1}}\otimes\tau_{\mu_{2}}\otimes\ldots\otimes \tau_{\mu_{Q}}\, .
\end{equation}
(The unit matrix $L_{00\ldots 0}$ is not a generator of $SU(2^{Q})$.) Single spin quantum operations are realized by transformations of the Ising spins where $\sigma_{\mu_{1}\mu_{2}\ldots\mu_{Q}}$ transforms in the same way as the product $\sigma_{\mu_{1}0\ldots 0}\sigma_{0\mu_{2}\ldots 0}\cdots\sigma_{00\ldots \mu_{Q}}$. Similarly, operators acting on a pair of quantum spins $i$ and $j$, as the CNOT-gate or the permutation $i\leftrightarrow j$ (SWAP gate), are realized if $\sigma_{\mu_{1}\ldots\mu_{i}\mu_{j}\ldots\mu_{Q}}$ transforms as $\sigma_{\mu_{1}0\ldots 0}\cdots\sigma_{0\ldots\mu_{i}\mu_{j}\ldots 0}\cdots\sigma_{00\ldots \mu_{Q}}$. (This extends to orders of spins where $i$ and $j$ are not neighboring.)

In particular, one can use the permutation of two spins, $P_{ij}:\, i\leftrightarrow j$, for a cyclic transport of the spin states to spins at neighboring ``sites'' on a spin chain $S^{(1)}S^{(2)}\ldots S^{(Q)}$. The ``transport'' 
\begin{equation}\label{eq:32}
P_{+}=P_{Q-1,Q}\, P_{Q-2,Q-1}\cdots P_{3,4}\, P_{2,3}\, P_{1,2}
\end{equation}
effectively transforms $S^{(1)}\ldots\,  S^{(Q)}$ to $S^{(2)}S^{(3)}\ldots\, S^{(Q)}S^{(1)}$. The corresponding unitary transformation acting on the density matrix (or quantum wave function for pure states) obtains by matrix multiplication of matrices $U_{S}$, cf. eq.~\eqref{eq:28}, for the pairs corresponding to $P_{ij}$. It is clear that a great number of quantum operations can be realized in this way by operations on classical Ising spins. The prize to pay is the very rapid increase of the number of classical Ising spins with the number $Q$ of quantum spins. Already for $Q=20$ quantum spins $2^{40}$ Ising spins are needed.

\subsection{Correlation map}

For a practical use of the realization of quantum operations by Ising spins it would be a great advantage if the fast increase of $M$ with $Q$ could be avoided. We next explore the possibility to use only $M=3Q$ classical Ising spins $s^{(m)}_{k}$, $m=1,\ldots , Q$. For two quantum spins ($Q=2$) one has to realize the density matrix by probabilistic information about six Ising spins $s_{k}^{(1)}$, $s_{k}^{(2)}$. For this purpose we use in eq.~\eqref{eq:13} the classical correlation function $\exval{s_{k}^{(1)}s_{l}^{(2)}}$, namely
\begin{equation}\label{eq:33}
\rho_{k0}=\exval{s_{k}^{(1)}}\com \rho_{0k}=\exval{s_{k}^{(2)}}\com \rho_{kl}=\exval{s_{k}^{(1)}s_{l}^{(2)}}\, .
\end{equation}
The elements $\sigma_{kl}$ in eq.~\eqref{eq:14} are now products of spins, $\sigma_{kl}=s_{k}^{(1)}s_{l}^{(2)}$, while $\sigma_{k0}=s_{k}^{(1)}$, $\sigma_{0k}=s_{k}^{(2)}$. We observe that only part of the classical correlation functions are used for the quantum subsystem. Correlation functions as $\exval{s_{k}^{(i)}s_{l}^{(i)}}$ or three point correlations belong to the environment in a suitable generalization of Fig.~\ref{fig:2}.

The generalization to $Q$ quantum spins is straightforward. Elements $\rho_{\muQ}$ with $s$ indices $k_{i}=1,\ldots ,3$ and $t=Q-s$ indices $0$ are given by $s$-point correlations $\exval{s^{(m_{1})}_{k_{1}}s^{(m_{2})}_{k_{2}}\ldots s^{(m_{s})}_{k_{s}}}$, where $m_{i}$ denotes the position of index $k_{i}$. (An example is $\rho_{0kl0m}=\exval{s_{k}^{(2)}s_{l}^{(3)}s_{m}^{(5)}}$.) Employing $s_{0}^{(m)}=1$ we can write
\begin{equation}\label{eq:33A}
\rho=\dfrac{1}{2^{Q}} \left( 1 + \exval{s_{\mu_{1}}^{(1)}s_{\mu_{2}}^{(2)}\ldots s_{\mu_{Q}}^{(Q)}}\, L_{\muQ} \right).
\end{equation}
We note that the classical spins appearing in the correlations used for the quantum subsystem are all different. Eq.~\eqref{eq:33A} defines the correlation map for $Q$ qubits, involving $3Q$ classical bits.

The quantum condition on $\rho$ is essential for the possibility to realize arbitrary density matrices by classical correlation functions. The main point may be demonstrated for $Q=2$ by a matrix with $\rho_{30}=1$, $\rho_{03}=1$, $\rho_{33}=-1$, with all other $\rho_{\mn}$ vanishing. Obviously, it cannot be realized by classical correlation functions since $\exval{s_{3}^{(1)}s_{3}^{(2)}}=-1$ is incompatible with $\exval{s_{3}^{(1)}}=1$, $\exval{s_{3}^{(2)}}=1$. While $\rho_{z}\rho_{z}=3$ is obeyed, the matrix $\rho=\diag(1,1,1,-1)/2$ has a negative eigenvalue and therefore does not obey the quantum condition. 

It is important that the operator $L_{kl}=L_{k0}\cdot L_{0l}$ can be associated with a matrix product of commuting quantum operators. The expectation value $\tr(L_{kl}\rho)=\tr(L_{k0}L_{0l}\rho)$ can be seen as the quantum correlation of the quantum observables $S_{k}^{(1)}$ and $S_{l}^{(2)}$. (We use $s_{k}$ for classical spins and $S_{k}$ for quantum spins.) Quantum correlations for non-commuting observables may violate inequalities for classical correlation functions. In this case they cannot be represented by classical correlation functions. Since $\rho_{kl}$ involves different quantum spins the associated quantum spin operators $L_{k0}$ and $L_{0l}$ indeed commute, such that simultaneous probabilities to find eigenvalues of $L_{k0}$ and $L_{0l}$ are part of the information in the quantum subsystem for this pair. In contrast, quantum correlations for different components of the same spin, as $\exval{S_{k}^{(1)}S_{l}^{(1)}}$, $k\neq l$, involve a noncommuting operator product $L_{k0}L_{l0}$. These quantum correlations are not represented by classical correlation functions. Inversely, the information on the classical correlation function $\exval{s_{k}^{(1)}s_{l}^{(1)}}$, $k\neq l$, cannot be computed from $\rho$. The necessary statistical information involves the environment and is not part of the quantum subsystem. Similar to Fig.~\ref{fig:2} this classical correlation lies outside the quantum subsystem. Combinations of such classical correlation functions obey Bell's inequalities, but they are of no relevance for the quantum subsystem.

The realization of the correlation map \eqref{eq:33A} as a valid bit-quantum map for arbitrary  quantum systems for $Q$ qubits needs completeness, as discussed below. While the reduction to $3Q$ classical bits for $Q$ qubits is striking, the reader should keep in mind that the cost is not trivial. For the exact representation of a density matrix one needs $2^{2Q}-1$ correlation functions. If all of them have to be stored in memory during a calculation, the needed memory space increases dramatically for large $Q$. One may hope, however, that not all details of the density matrix are relevant for practical computations, making some approximation to $\rho$ sufficient. If the correlation map is complete, and if efficient algorithms for the handling of correlations for $3Q$ classical bits can be found, it is not excluded that the BPP computational complexity class contains the BQP class, which would be rather surprising. This issue is beyond the scope of this paper.

\subsection{Restrictions on classical correlations and \\
quantum condition}

A key ingredient for the realization of a bit-quantum map is completeness. For a complete bit-quantum map every possible density matrix $\rho$ needs to have at least one realization in terms of the probabilistic information for the classical statistical system. Completeness ensures that all possible states of the quantum system can be reached in principle. If, in addition, all unitary operations can be realized for a given number $Q$ of qubits, the classical statistical system can perform universal quantum computations for this number of qubits. 

If a bit-quantum map is not complete, not all quantum density matrices $\rho$ can be realized by the classical system. There may still be a rather dense subset of quantum density matrices that can be realized by an incomplete bit-quantum map. In this case the possible unitary transformations that can be performed by classical bits are not arbitrary, but can be for every realized $\rho$ a very dense subgroup of the unitary group $SU(2^Q)$. The classical system can perform arbitrary unitary transformations with an error that may be rather small.

For the correlation map \eqref{eq:33A} completeness requires that all values of elements
\begin{equation}\label{eq:41A} 
\rho_{\mu_{1}\mu_{2}\ldots \mu_{Q}}=\exval{s_{\mu_{1}}^{(1)}s_{\mu_{2}}^{(2)}\ldots s_{\mu_{Q}}^{(Q)}}
\end{equation}
that are consistent with the quantum condition can indeed be realized by the corresponding classical correlation functions. Since classical correlation functions obey restrictions, this property is not automatic. We will present arguments that the correlation map \eqref{eq:33A} could indeed be complete, while a proof is not yet achieved.

We start with $Q=2$ and generalize subsequently to arbitrary $Q$. Classical correlation functions of two Ising spins $s_{1}$ and $s_{2}$ have to be realized by a positive probability distribution. They obey the necessary and sufficient conditions
\begin{align}\label{eq:34}
&|\exval{s_{1}}|\leq 1\com |\exval{s_{2}}|\leq 1\com |\exval{s_{1}s_{2}}|\leq 1\com\nn\\
&-1+|\exval{s_{1}}+\exval{s_{2}}|\leq \exval{s_{1}s_{2}}\leq 1-|\exval{s_{1}}-\exval{s_{2}}|\, .
\end{align}
In our setting this translates to bounds for the coefficients $\rho_{\mn}$. For all pairs $(k,l)$ they have to obey
\begin{align}\label{eq:35}
&|\rho_{k0}|\leq 1\com |\rho_{0l}|\leq 1\com |\rho_{kl}|\leq 1\com\nn\\
&-1+|\rho_{k0}+\rho_{0l}|\leq \rho_{kl}\leq 1-|\rho_{k0}-\rho_{0l}|\, .
\end{align}
We want to show that every density matrix \eqref{eq:13} which obeys the quantum condition automatically obeys the inequalities \eqref{eq:35} for every pair $(k,l)$. There is therefore no obstruction from relations of the type \eqref{eq:34} to realize an arbitrary density matrix by a suitable probability distribution for the Ising spins.

The proof uses the possibility to make basis changes for quantum systems. In every basis all diagonal elements $\rho_{\alpha\alpha}$ of a positive density matrix obey $\rho_{\alpha\alpha}\geq 0$. This can be exploited to establish the inequalities \eqref{eq:35}. Let us consider for an arbitrary pair $(k,l)$ the three generators $L_{k0}$, $L_{0l}$ and $L_{kl}$. They commute and can therefore be diagonalized simultaneously. The entries of these diagonal matrices are the eigenvalues $\pm 1$. In the basis where both $L_{k0}$ and $L_{0l}$ (and therefore also $L_{kl}$) are diagonal, the diagonal elements of $\rho$ can be associated with the probabilities $p_{++}$, $p_{+-}$, $p_{-+}$, $p_{--}$ to find for $(L_{k0},L_{0l})$ the eigenvalues $(++)$, $(+-)$, $(-+)$ and $(--)$. Indeed, $p_{++}$ is defined as the sum over all diagonal elements $\rho_{\alpha\alpha}$ for which $L_{k0}$ and $L_{0l}$ both have the eigenvalue one, and similar for the other combinations. The positivity of the density matrix implies $\rho_{\alpha\alpha}\geq 0$ for every $\alpha$ separately - it is at this place where the quantum condition enters crucially. Thus $p_{++}$, $p_{+-}$, $p_{-+}$ and $p_{--}$ are all positive semidefinite. The normalization $\tr(\rho)=1$ implies $p_{++}+p_{+-}+p_{-+}+p_{--}=1$, such that the set $(p_{++},\, p_{+-},\, p_{-+},\, p_{--})$ has all properties of a probability distribution. In this diagonal basis eq.~\eqref{eq:17} yields
\begin{align}\label{eq:36}
\rho_{k0}&=\tr(L_{k0}\, \rho) = p_{++}+p_{+-}-p_{-+}-p_{--}\com\nn\\
\rho_{0l}&=\tr(L_{0l}\, \rho) = p_{++}-p_{+-}+p_{-+}-p_{--}\com\nn\\
\rho_{kl}&=\tr(L_{kl}\, \rho) = p_{++}-p_{+-}-p_{-+}+p_{--}\, .
\end{align}
The relations \eqref{eq:35} follow from eq.~\eqref{eq:36} for every probability distribution $(p_{++},\, p_{+-},\, p_{-+},\, p_{--})$. 

For six Ising spins the constraints \eqref{eq:34} on the two-point correlation functions and expectation values have to hold for any arbitrary pair of spins. This is indeed guaranteed by the positivity of the density matrix. The question arises if this is sufficient to realize arbitrary sets of quantum correlations $S_k^{(1)} S_l^{(2)}$ by classical correlation functions. So far we did not find any counterexample. In particular, it is straightforward to construct explicitly classical probability distributions realizing particular entangled quantum states, as for example, the entangled pure state $\psi = (0,\, \cos(\vartheta),\,\sin(\vartheta),\, 0)$. This includes for $\vartheta = -\pi/4$ the maximally entangled state \eqref{eq:25}. The construction can be found in Appendix \ref{app:Correlation_map_for_two_qubits}, where we explicitly establish the classical probability distributions for selected families of entangled quantum states, and discuss further a few general aspects of the correlation map. Even without a proof of completeness it is already clear by explicit construction that a rather dense set of quantum density matrices can be obtained from the correlation map.

\subsection{Restrictions on correlation functions and generalized Bell inequalities}

For more than two classical spins there exist inequalities for linear combinations of classical correlation functions of the type \eqref{eq:BI2}. In contrast to the quantum-bit map with fifteen classical spins for $Q=2$, the correlation map identifies some of the quantum correlations with classical correlations. It is a necessary condition for the completeness of the correlation map that those quantum correlations that are identified with classical correlations obey the generalized Bell inequalities. The quantum correlations not identified with classical correlations need not obey such restrictions.

For the correlation map the particular quantum correlations between the three orthogonal spins $S_k^{(1)}$ and $S_l^{(2)}$ equal the classical correlation functions
\begin{equation}\label{eq:49A}
\langle S_k^{(1)} S_l^{(2)} \rangle = \tr \big( S_k^{(1)} S_l^{(2)}\, \rho \big) = \rho_{kl}
    = \langle s_k^{(1)} s_l^{(2)} \rangle\, .
\end{equation}
The classical correlation functions obey generalized Bell's inequalities as the CHSH inequality \cite{CHSH}. If the quantum correlations on the r.h.s. of eq.~\eqref{eq:49A} violated these inequalities, this would show that the correlation map cannot be complete. For the CHSH inequality we define
\begin{align}\label{eq:49B}
A = \pm \,s_k^{(1)}\, , \quad A' = \pm\, s_l^{(1)}\, , \notag \\
B = \pm \,s_m^{(2)}\, , \quad B' = \pm\, s_n^{(2)}\, ,
\end{align}
and consider
\begin{equation}\label{eq:49C}
C = AB + AB' + A'B - A'B' = A\,(B + B') + A'\,(B - B')\, .
\end{equation}
With $A$, $A'$, $B$, $B'$ taking values $\pm 1$ one has either $B' = B$ or $B' = -B$. For $B' = B$ one infers $C = 2AB$ and therefore $|C| \leq 2$. The same holds for $B' = - B$, with $C = 2A'B$. Since $|C| \leq 2$ holds for every classical state, one infers for the corresponding combination of classical correlation functions the CHSH inequality
\begin{equation}\label{eq:49D}
|\langle C \rangle| = |\langle AB \rangle + \langle AB'\rangle + \langle A' B\rangle - \langle A' B' \rangle| \leq 2\, .
\end{equation}
This inequality includes the original Bell's inequality as a special case.

One may consider the maximally entangled pure quantum state \eqref{eq:25} with
\begin{equation}\label{eq:49E}
\rho = \frac{1}{2} \begin{pmatrix}
0 & 0 & 0 & 0 \\
0 & 1 & -1 & 0 \\
0 & -1 & 1 & 0 \\
0 & 0 & 0 & 0
\end{pmatrix}\, ,
\end{equation}
and evaluate the quantum correlations with classical spins $s_k^{(i)}$ replaced by quantum spin operators $S_k^{(i)}$. For the maximally entangled state one finds the quantum correlations (no summation over indices)
\begin{equation}\label{eq:49F}
\langle S_k^{(1)} S_k^{(2)} \rangle = -1\, , \quad \langle S_k^{(1)} S_l^{(2)} \rangle = 0 
    \quad \text{for } k \neq l\, .
\end{equation}
For arbitrary assignments in eq.~\eqref{eq:49B} the inequality \eqref{eq:49D} is obeyed for the quantum correlations, and no contradiction to the completeness of the correlation map emerges. More in detail, we may take for $A' \neq A$ without loss of generality $A = S_1^{(1)}$, $A' = S_3^{(1)}$. For $B = - S_3^{(2)}$, $B' = S_3^{(2)}$ one has $\langle A' B \rangle = 1$, $\langle A' B'\rangle = -1$, $\langle AB \rangle = \langle A B' \rangle = 0$, such that the bound \eqref{eq:49D} is saturated, $\langle C \rangle = 2$. Switching the sign either of $B$ or of $B'$ yields $\langle C \rangle = 0$. Choosing instead $B' = \pm \,S_2^{(2)}$ results in $\langle C \rangle = 1$, while for $B' = S_1^{(2)}$ one has $\langle C \rangle = 0$, and $B' = -S_1^{(2)}$ implies $\langle C \rangle = 2$. Other similar cases obtain by rotations of the axes or exchange of $(A,\, A') \leftrightarrow (B,\, B')$. Finally, one has for $A' = A$ the relation $\langle C \rangle = 2\,\langle AB \rangle$ which obeys the inequality $|\langle C \rangle| \leq 2$ for all choices of $A$ and $B$, and similarly for $A' = -A$. 

It is important in this respect that only for orthogonal spins the quantum correlations equal the classical correlations. If we considered for $B$, $B'$ spin vectors $S^{(2)}$ in arbitrary directions, say with an angle of $\pi/4$ to the $3$-axis, the quantum correlations would violate the CHSH inequality. This observation has an important consequence for possible extensions of the correlation map: only for sets of three orthogonal spins $S_k^{(j)}$ the quantum correlations can be equal to the classical correlations. The choice of the three axes is arbitrary, but once the correlation map identifies matrix elements of $\rho$ with correlation functions of classical spins with a given choice of axes, the quantum correlations for spins oriented along other directions cannot be equal to any classical correlation function. In this sense the correlation map uses a maximum number of classical correlations that remains consistent with generalized Bell inequalities.

For a given density matrix the set of conditions \eqref{eq:35} holds for all pairs $(k,\,l)$ simultaneously. The same is true for the classical correlation functions \eqref{eq:34}. We observe that the classical inequalities for correlation functions can all be saturated simultaneously. This means that for each pair $(k,\, l)$ one can have either $\langle s_k s_l\rangle = 1$ or $\langle s_k s_l \rangle = -1$. A simple way to saturate all inequalities are fixed values for all classical spins. Which side of the inequality for $\langle s_k s_l \rangle$ can be satisfied depends, however, on the other correlation functions. For example, one has the relation
\begin{equation}\label{eq:49G}
\langle s_k s_l \rangle = \langle s_k s_{l'} \rangle = \langle s_{k'} s_l \rangle = 1 \; 
    \Rightarrow \; \langle s_{k'} s_{l'} \rangle = 1\, ,
\end{equation}
which follows from the observation that equal values of $s_k$ and $s_l$ and equal values of $s_k$ and $s_{l'}$ imply equal values of $s_l$ and $s_{l'}$, and in turn equal values of $s_{k'}$ and $s_l$ require equal values of $s_k$ and $s_{k'}$. The three relations on the l.h.s. of eq.~\eqref{eq:49G} can only be realized by configurations for which all four spins $s_k$, $s_{k'}$, $s_l$, $s_{l'}$ are equal. This implies $\langle s_{k'} s_{l'}\rangle = 1$ and saturates the corresponding CHSH inequality. Quantum correlations can saturate less of the inequalities simultaneously. The condition \eqref{eq:18} implies that at most three inequalities between spins $S_k^{(1)}$ and $S_l^{(2)}$ can be saturated simultaneously. This suggests that inequalities involving more than one correlation function could be more restricting for quantum systems than for classical systems, and obstructions to completeness may therefore be avoided.

The problem of completeness of the correlation map may be posed as follows. A positive density matrix $\rho$ defines probability distributions for all subsystems of pairs $(S_k^{(1)},\, S_l^{(2)})$. Correspondingly, $\rho$ ``constructs'' probabilistic subsystems for corresponding pairs of classical spins $(s_k^{(1)},\, s_l^{(2)})$. The probability distribution for each pair is associated to a ``reduced probability distribution'' obtained from ``integrating out'' the four complementary spins in some overall probability distribution $p_6$ for six classical spins. Completeness is realized if an overall probability distribution exists such that all nine subsystems for pairs $(s_k^{(1)},\, s_l^{(2)})$ have the reduced probability distributions dictated by the quantum density matrix. 

Consider four spins $S_1^{(1)}$, $S_2^{(1)}$, $S_1^{(2)}$ and $S_2^{(2)}$. Let us take the three pairs $(S_1^{(1)},\, S_1^{(2)} )$, $(S_1^{(1)},\, S_2^{(2)} )$ and $(S_2^{(1)},\, S_1^{(2)} )$, and fix the reduced classical probability distributions for the corresponding subsystems according to the values dictated by a given density matrix $\rho$. The probability distributions for these three subsystems determine all four expectation values $\langle S_1^{(1)}\rangle$, $\langle S_2^{(1)}\rangle$, $\langle S_1^{(2)}\rangle$, $\langle S_2^{(2)}\rangle$, as well as three correlations $\langle S_1^{(1)} S_1^{(2)} \rangle$, $\langle S_1^{(1)} S_2^{(1)} \rangle$ and $\langle S_2^{(1)} S_1^{(2)} \rangle$. Let us assume that a family of classical probability distributions $p_4$ for the four spins $(s_1^{(1)},\, s_2^{(1)},\, s_1^{(2)},\, s_2^{(2)})$ realizes these seven expectation values. The density matrix also fixes the subsystem of the pair $(S_2^{(1)} S_2^{(2)})$ and therefore in addition the correlation function $\langle S_2^{(1)} S_2^{(2)}\rangle$. Given the other seven expectation values, the possible values of $\langle S_2^{(1)} S_2^{(2)}\rangle$ are limited by the quantum constraint of positivity of the density matrix. On the other hand, the family of classical probability distributions $p_4$ enforces ``classical constraints'' on the possible values of $\langle s_2^{(1)} s_2^{(2)} \rangle$ in function of the other seven expectation values. The question arises if for every $\rho$ one can find $p_4$, such that the classical constraints are not in contradiction to the values of $\langle s_2^{(1)} s_2^{(2)}\rangle$ allowed by the quantum constraint. This is precisely the issue of the CHSH inequality which specifies classical constraints for this setting. If the CHSH inequalities are the strongest  classical constraints for sets of four correlation functions $\langle s_k^{(1)} s_l^{(2)} \rangle$, and if the quantum correlations $\langle S_k^{(1)} S_l^{(2)} \rangle$ obey the CHSH inequalities for all $\rho$, one can infer that suitable subsystems $p_4$ exist for all subsets of two pairs of spins.

There may still be additional constraints for the correlation functions involving three pairs of spins, as needed for the correlation map. They concern the existence of the overall probability distribution $p_6$ which should be consistent with all reduced probability distributions $p_4$. The corresponding inequalities typically will involve combinations of six or more correlation functions. Such inequalities could act as obstructions to the completeness of the correlation map. We know, however, that no obstructions can exist for all quantum states for which we have constructed associated classical overall probability distributions explicitly, as the ones discussed in appendix \ref{app:Correlation_map_for_two_qubits}. Since those contain the maximally entangles state, we could have ``predicted'' that for this state the quantum correlations $\langle S_k^{(1)} S_l^{(2)}\rangle$ are all compatible with the CHSH inequalities, even without explicit computation.

\subsection{Correlation map for $Q\geq 3$}

The setting generalizes in a straightforward way to an arbitrary number $Q$ of quantum spins. The classical $n$-point functions employed in eq.~\eqref{eq:41A} obey constraints similar to eq.~\eqref{eq:34}, translating to constraints on the elements of the density matrix analogous to eq.~\eqref{eq:35}. One has to establish that the positivity of the density matrix ensures these constraints. All constraints on the classical correlation functions arise from the requirement of an underlying positive probability distribution. On the other hand, every set of classical correlation functions obeying the constraints can be realized by a suitable probability distribution. For finite $M$ there is an invertible map between the space of probability distributions and the space of classical correlation functions.

For establishing completeness of the correlation map one would like to show that the positivity of the density matrix restricts the possible values of quantum correlations $\rho_{\mu_1\cdots \mu_Q}$ sufficiently, such that all values compatible with these restrictions can be obtained as classical correlations from suitable classical probability distributions. Then the constraints on classical correlation functions are obeyed by $\rho_{\mu_{1}\ldots\mu_{Q}}$, and no contradiction of the particular quantum correlations $\rho_{\mu_1\cdots\mu_Q}$ with Bell's inequalities for classical correlations can occur. The relation between the two sets of correlation functions
\begin{align}\label{eq:49H}
\rho_{\mu_1\cdots \mu_Q} &= \langle \sigma_{\mu_1}^{(1)} \sigma_{\mu_2}^{(2)} \cdots
    \sigma_{\mu_Q}^{(Q)} \rangle_{\text{cl}} \notag \\
&= \sum_{\{s\}} p[s]\, \sigma_{\mu_1}^{(1)} 
    \sigma_{\mu_2}^{(2)} \cdots \sigma^{(Q)}_{\mu_Q} = \langle S_{\mu_1}^{(1)}
    S_{\mu_2}^{(2)} \cdots S_{\mu_Q}^{(Q)} \rangle_{\text{q}} \notag \\
&= \tr\big\{ \rho\, S_{\mu_1}^{(1)} S_{\mu_2}^{(2)} \cdots S_{\mu_Q}^{(Q)} \big\}
\end{align}  
is then consistent. Here we distinguish for clarity the classical and quantum correlations by subscripts cl and q, and the quantum operators $S_\mu^{(i)}$ are
\begin{equation}\label{eq:49I}
S_k^{(i)} \quad \text{for} \; k=1,2,3, \quad S_0^{(i)} = 1\, .
\end{equation}

We concentrate first on the correlation function $\exval{s_{k_{1}}^{(1)}s_{k_{2}}^{(2)}\ldots s_{k_{Q}}^{(Q)}}$ for a given set $(k_{1},k_{2},\ldots ,k_{Q})$, together with the sets of correlation functions for which some of the $s_{k_{i}}^{(i)}$ are replaced by one. All constraints for this particular set arise from the necessary positive probability distributions determining these correlations. It is sufficient that a positive probability distribution exists for the classical subsystem for which all Ising spins different from the selected $s_{k_{i}}^{(i)}$ are integrated out.

In analogy to $Q=2$ we show that all classical constraints involving only spins in a given subset are matched by equivalent constraints for a positive density matrix. For any arbitrary given set $(k_{1},k_{2},\ldots ,k_{Q})$ one chooses a basis where all $Q$ generators $L_{k_{1}00\ldots 0}$, $L_{0k_{2}0\ldots 0}$, ..., $L_{0\ldots 0k_{Q}}$ (one index $k_{i}=1,2,3$, all other indices $0$) are diagonal. This is possible since these generators commute. In the same basis all products of these generators are diagonal as well. The diagonal elements $\rho_{\alpha\alpha}$ in this basis permit the definition of probabilities to find for these $Q$ selected operators the values $+1$ or $-1$. The positivity of $\rho_{\alpha\alpha}$ ensures again the positivity of these probabilities. The $2^Q$ positive elements $\rho_{\alpha\alpha}$ define a probability distribution for the $Q$ different spins in the selected set. From this distribution all quantum correlation functions can be computed according to the rule of classical statistics. They correspond to classical correlation functions, evaluated with the probabilities $p_\alpha = \rho_{\alpha\alpha}$. It is crucial in this respect that all quantum operators in the selected set commute, such that simultaneous probabilities to find a given sequence of values, say $(+1,\, -1,\, -1,\, \dots,\, +1)$ can be defined as $p_{+--\dots +}$, similar to $p_{++}$, $p_{+-}$, $p_{-+}$ and $p_{--}$ for $Q = 2$. In the diagonal basis every product $A$ of spins $S_k$ in the set has a value $A_\alpha = \pm 1$ for each $\alpha$, and the expectation value obeys
\begin{equation}\label{eq:49J}
\langle A \rangle = \sum_{\alpha} \rho_{\alpha\alpha} A_\alpha = \sum_\alpha p_\alpha
    A_\alpha\, .
\end{equation}
Being derived from a probability distribution, the elements $\rho_{\mu_{1}\mu_{2}\ldots\mu_{Q}}$ with $\mu_{i}$ either zero or $k_{i}$ obey all constraints for the classical correlation functions of the selected set of $Q$ two-level observables.

These constraints are necessary constraints for the classical correlation functions $\exval{s_{\mu_{1}}^{(1)}s_{\mu_{2}}^{(2)}\ldots s_{\mu_{Q}}^{(Q)}}$ with $\mu_{i}=0$ or $k_{i}$. This follows by ``integrating out`` the spins $s_{l}^{(i)}$ for $l\neq k_{i}$, thereby establishing positive probability distributions for classical subsystems consisting of $Q$ classical spins. Since the choice of $Q$ generators by the set $(k_{1}\ldots k_{Q})$ is arbitrary, one establishes a set of constraints for $n$-point functions for $Q$ spins with $n\leq Q$, and arbitrary selections of the spins $s_{k_{1}}^{(1)}$, $s_{k_{2}}^{(2)}$, ..., $s_{k_{Q}}^{(Q)}$. These constraints hold both for the set of correlation functions for the sets of $Q$ classical spins and for the associated elements of the density matrix.

Similar to $Q=2$, a given quantum density matrix $\rho$ defines classical probability distributions for all subsystems with $Q$ classical spins, with the other $2Q$ classical spins being integrated out. In order to show completeness of the correlation map one finally needs to establish that the sets of constraints on $n$-point function for arbitrary subsets $(k_{1}\ldots k_{Q})$ of selected spins are sufficient for the existence of an ``overall probability distribution`` for $3Q$ classical spins that realizes all these correlations. One has to establish that no obstructions arise from inequalities for linear combinations of classical correlations, which generalize the CHSH inequalities. For the time being a proof of the sufficiency of the discussed constraints on the classical level is not available. Establishing the completeness of the correlation map by finding classical probability distributions for $3Q$ classical spins that can describe an arbitrary density matrix for $Q$ qubits could have far-reaching consequences for scaling properties of computations with classical bits. Typically, such overall classical probability distributions will not be unique, since correlation functions for $m$ spins with $m>Q$ are not specified. It is sufficient to establish the existence of at least one possible overall probability distribution.

\subsection{Properties of bit-quantum maps}

Our construction defines a ``bit-quantum-map'' as a map from a subset of probability distributions for Ising spins to a positive quantum density matrix. The subset is defined by suitable expectation values and classical correlations obeying the quantum condition. If arbitrary quantum states are to be covered, the bit-quantum map should be complete. It is, however, not invertible - different probability distributions for Ising spins realize the same density matrix. On the other hand, for a complete map every quantum density matrix can be realized by a suitable probability distribution for Ising spins. The bit-quantum map is a map from the probabilistic information about classical bits or Ising spins to the probabilistic information about qubits or quantum spins, as encoded in the quantum density matrix. It is possible whenever the most general quantum density matrix can be constructed from classical expectation values and correlations which obey the restrictions of the type \eqref{eq:34}, as well as all other inequalities for linear combinations of classical correlations.

More precisely, a general map from the probability distribution for Ising spins to the density matrix $\rho$ is defined by eq.~\eqref{eq:30}, with $\sigma_{\muQ}$ observables that can be expressed in terms of classical Ising spins. This map is a complete bit-quantum map if arbitrary $\rho_{\muQ}$ consistent with the quantum condition can be realized by expectation values of the observables $\sigma_{\muQ}$ for suitable classical probability distributions. Different bit-quantum maps are distinguished by different associations of observables $\sigma_{\mu_{1}\ldots\mu_{Q}}$ to Ising spins or products thereof. Examples are the correlation map \eqref{eq:33A} and the map \eqref{eq:30} for independent spins for each $\sigma_{\mu_{1}\ldots\mu_{Q}}$.

There are many intermediate possibilities of bit-quantum maps between the correlation map and the map associating each $\sigma_{\mu_{1}\ldots \mu_{Q}}$ to an independent Ising spin. For example, we may replace in the correlation map for $Q=2$ for some of the $\sigma_{kl}$ the correlation function $\exval{s_{k}^{(1)}s_{l}^{(2)}}$ by an independent Ising spin $s_{kl}$. If the correlation map is complete, this defines a new complete bit-quantum map, since the condition $|\exval{s_{kl}}|\leq 1$ is weaker than the restriction on the classical correlation function. Not every map that defines $\rho$ by expectation values of $\sigma_{\muQ}$ is a bit-quantum map. For example, we can define for two quantum spins a different map by replacing in the map $\rho_{\mn}=\exval{s_{\mu}^{(1)}s_{\nu}^{(2)}}$ the element $\rho_{22}$ by $-\exval{s_{2}^{(1)}s_{2}^{(2)}}$, while leaving all other elements as before. This is not a bit-quantum map since the matrix $\rho$ defined in this way is not positive. Positivity requires that for $\rho_{20}=1$, $\rho_{02}=1$ only $\rho_{22}=1$ is possible. For classical Ising spins the expectation values $\exval{s_{2}^{(1)}}=1$, $\exval{s_{2}^{(2)}}=1$ require for the classical correlation $\exval{s_{2}^{(1)}s_{2}^{(2)}}=1$. This contradicts the association $\rho_{22}=-\exval{s_{2}^{(1)}s_{2}^{(2)}}$.

\section{Unitary transformations by maps of probability distributions}\label{sec:V} 

This section discusses maps of classical statistical probability distributions for Ising spins that realize unitary transformations for the density matrix of appropriate quantum subsystems. If the quantum conditions for the subsystem are obeyed and the bit-quantum map is complete, arbitrary unitary transformations can be induced by non-linear transformations of the classical probability distributions. Linear transformations of the probability distributions only realize a subset of unitary transformations. As particular examples for linear transformations of probability distributions we discuss the standard computing with random operations, as well as unique jump operations or cellular automata. We emphasize that the transformation of the probability distribution between layers of a neural network needs not to be linear, such that restrictions arising from the linearity of the transformation do not apply in general.

\subsection{Unitary quantum transformations by maps of probability distributions for classical bits}

Unitary transformations are compatible with the quantum condition. For $\rho(t)$ obeying the quantum condition, also $\rho(\te)=U\rho(t)\, U^{\dagger}$ obeys the quantum condition. For a complete bit-quantum map every quantum density matrix $\rho(t)$ can be obtained from suitable probability distributions $\lbrace p_{\tau}(t)\rbrace$ for Ising spins. As a consequence, every $U$ can be realized by a map from a probability distribution $\lbrace p_{\tau}(t)\rbrace$ realizing $\rho(t)$, to a different probability distribution $\lbrace p_{\tau}(\te)\rbrace$ which realizes $\rho(\te)$. The existence of the map from $\lbrace p_{\tau}(t)\rbrace$ to $\lbrace p_{\tau}(\te)\rbrace$ can be visualized from Fig.~\ref{fig:1} if one replaces the classical density matrices $\rho^{\prime}$ by probability distributions (that consist of the diagonal elements of $\rho^{\prime}$, see sect.~\ref{sec:Static_Memory_Materials}), and interprets $S(t)$ as this map. The map between probability distributions is not unique, since the probability distributions $\lbrace p_{\tau}(\te)\rbrace$ realizing a given $\rho(\te)$ are not unique. 

The possible realization of an arbitrary unitary transformation by a suitable map of classical probability distributions is a central finding of the present work. For all bit-quantum maps which allow to obtain arbitrary $\rho$ obeying the quantum condition from suitable classical statistical probability distributions, all unitary transformations of $\rho$ can be realized by suitable maps of classical probability distributions. If the required maps of the classical probability distributions can be realized in practice, universal quantum computing can be performed.

In particular, we can construct the space of $\rho_{\mn}$ which realize pure quantum states by starting from a given pure state density matrix $\rho^{(0)}$ and applying unitary transformations. For $\rho^{(0)}$ we may take the pure state $\rho_{30}=\exval{s_{3}^{(1)}}=1$, $\rho_{03}=\exval{s_{3}^{(2)}}=1$, $\rho_{33}=\exval{s_{3}^{(1)}s_{3}^{(2)}}=1$, with all other $\rho_{\mn}$ vanishing. This correlation function is realized if $p_{\tau}=0$ for all states with $s_{3}^{(1)}=-1$ or $s_{3}^{(2)}=-1$, and equal probabilities for all states with $s_{3}^{(1)}=s_{3}^{(2)}=1$, see appendix \ref{app:Correlation_map_for_two_qubits}. (This construction of $\rho^{(0)}$ is easily generalized to an arbitrary number of qubits $Q$.) All pure state wave functions can be obtained from the one associated to $\rho^{(0)}$ by a suitable $U$. This holds up to the overall phase of the wave function $\psi$ which drops out in $\rho$, and up to a discrete transformation $\psi\raw\psi^{*}$, $\rho\raw\rho^{*}$, which in turn can be represented by a map among the classical probabilities. Thus, up to complex conjugation, one can obtain all pure state density matrices by unitary transformations from $\rho^{(0)}$. Since every pure state density matrix can be realized by a suitable classical probability distribution for Ising spins, there exist maps between probability distributions that realize an arbitrary unitary transformation of a pure quantum state.

At this stage we have established the existence of a map $p_{\tau}(t)\raw p_{\tau}(\te)$ for any arbitrary unitary transformation of the quantum subsystem. We have not specified its properties. In general, it may be a complicated non-linear map. In some cases, as for artificial neural networks learning the map, knowledge of its precise form is not needed and may not be of much interest. On the other hand, one may want to derive particular maps between probability distributions that realize tasks of quantum computing. In the following we will concentrate on maps between classical statistical probability distributions that have a clear prescription for their realization. We first discuss the standard approach to probabilistic computing in terms of transition probabilities. In the next section we discuss static memory materials which open a wider approach for the transmission of probabilistic information by computational steps.

\subsection{Computing with random operations}

One way of probabilistic computing transforms a given spin configuration $\tau$ to another spin configuration $\sigma$ with a probability $W_{\sigma\tau}$, 
\begin{equation}\label{eq:PCA} 
W_{\sigma\tau}\geq 0\com \sum_{\sigma}\, W_{\sigma\tau}=1\, .
\end{equation}
In practice this may be done by using some type of random number, whose value decides to which state $\sigma$ a given state $\tau$ is transformed. If at $t$ the probability to find a state $\tau$ is given by $p_{\tau}(t)$, the probability distribution at $\te$ obeys the linear relation
\begin{equation}\label{eq:PCB} 
p_{\rho}(\te)=W_{\rho\tau}(t)\, p_{\tau}(t)\, .
\end{equation}
For positive transition probabilities $W_{\rho\tau}$ the positivity of $p_{\rho}(\te)$ is guaranteed for arbitrary positive $p_{\tau}(t)$. For restricted $p_{\tau}(t)$ positive $p_{\rho}(\te)$ may be compatible with negative elements $W_{\tau\rho}$, but the interpretation as transition probabilities is lost and a practical realization is not obvious.

We are interested in the question which unitary transformations for a quantum subsystem can be realized by a random operation of the type \eqref{eq:PCB}. A unitary transformation of the density matrix amounts to a linear map for the coefficients
\begin{equation}\label{eq:PCC} 
\rho^{\prime}_{z}=b_{zy}\rho_{y}\, ,
\end{equation}
where $\rho_{z}^{\prime}$ stands for $\rho_{z}(\te)$ and $\rho_{y}$ for $\rho_{y}(t)$ (and similar for $p^{\prime}_{\rho}$ and $p_{\tau}$). A matrix $b$ representing a unitary transformation obeys constraints - if the $\rho_{y}$ obey the conditions for a pure quantum state, this has to hold for $\rho^{\prime}_{z}$ as well. Our question asks which $b_{zy}$ obeying these constraints can be realized by the linear map
\begin{equation}\label{eq:40A}
p^{\prime}_{\rho}=W_{\rho\tau}p_{\tau}\, .
\end{equation}
With
\begin{equation}\label{eq:40B}
\rho_{z}=d_{z\tau}p_{\tau}\com \rho_{z}^{\prime}=d^{\prime}_{z\tau}p_{\tau}=d_{z\rho}p_{\rho}^{\prime}
\end{equation}
the transition probabilities $W_{\rho\tau}$ and the matrix elements of $b$ have to obey
\begin{equation}\label{eq:40C}
d^{\prime}_{z\tau}=d_{z\rho}W_{\rho\tau}=b_{zy}d_{y\tau}\, .
\end{equation}

For those unitary transformations that can be realized by eq.~\eqref{eq:40A} the relation \eqref{eq:40C} does not fix $W_{\rho\tau}$ uniquely. A suitable $W_{\rho\tau}$ can be inferred by extending the sets of correlation functions $\rho_{z}$ and $\rho_{z}^{\prime}$ to complete full sets of correlation functions by adding, rather arbitrarily, a sufficient number of additional correlations. The map between a complete set of correlation functions and the probabilities $p_{\tau}$ is invertible, such that the coefficients $W_{\rho\tau}$ are uniquely fixed in this case. The large freedom in the choice of $W_{\rho\tau}$ arises from the large variety of possible extensions to complete sets of correlation functions. 

The class of unitary transformations that can be realized by the ``random computation steps'' \eqref{eq:40A} is restricted by the properties of the matrix $W$. The transition probabilities $W_{\tau\rho}$ are elements of a nonnegative matrix $W$, and for a finite number of Ising spins $M$ the matrix $W$ has finite dimension $2^{M}$. Finite-dimensional nonnegative matrices have a largest set of eigenvalues $\lambda_{i}$. The second equation \eqref{eq:PCA} implies $\sum_{\tau}\, p_{\tau}^{\prime}=\sum_{\tau}\, p_{\tau}$, such that the largest eigenvalues obey $|\lambda_{i}|=1$. Indeed, if the largest eigenvalues would be $|\lambda_{i}|<1$, the transformation $p_{\tau}^{(m)}=(W^{m})_{\tau\rho}\, p_{\rho}$ would lead to vanishing $p_{\tau}^{(m)}=0$ for $m\raw\infty$. Similarly, for $|\lambda_{i}|>1$ some $p_{\tau}^{(m)}$ would exceed one or become negative for large enough $m$. This contradicts the fact that the $p_{\tau}^{(m)}$ form a valid probability distribution for all $m$.

If all eigenvalues of $W$ are maximal eigenvalues with $|\lambda_{i}|=1$, one concludes that $W$ is an orthogonal matrix and therefore a unique jump operation. The operation is then deterministic rather than probabilistic - for every $\rho$ one element $W_{\tau\rho}=W_{\tau(\rho),\rho}$ equals one, whereas all other elements $W_{\tau\rho}$ for $\tau\neq \tau(\rho)$ are zero. The configuration $\rho$ is mapped uniquely to the configuration $\tau(\rho)$. This is the type of operations that we have discussed in sects~\ref{sec:Quantum_jump_in_classical_statistical_systems} and \ref{sec:Two_entangled_quantum_spins}.

For genuine probabilistic operations some of the eigenvalues of $W$ have absolute magnitude smaller than one. By suitable regular transformations one may render $W$ block diagonal, one block with eigenvalues $|\lambda_{i}|=1$, and the other block with eigenvalues $|\lambda_{k}|<1$. By arguments similar to the ones above only the first block is involved for unitary transformations of the quantum subsystems. The maximal eigenvalues obey $\lambda=\exp(2\pi i k_{i}/h_{i})$, $0\leq k_{i}\leq h_{i}$, with periods $h_{i}$. The maximal period $\bar{h}=N=2^{M}$ can only be realized by unique jump operators. Only unitary transformations with at most the maximal period can be achieved. We conclude that linear random maps between probability distributions that realize unitary transformations with period larger than $2^{M}$ can only be implemented by linear transformations \eqref{eq:40A} with some of the coefficients $W_{\tau\rho}$ being negative, or by non-linear transformations. Linear transformations \eqref{eq:38} involving some negative elements $W_{\tau\rho}$ do not yield a valid positive probability distribution $\lbrace p_{\rho}^{\prime}\rbrace$ for arbitrary probability distributions $\lbrace p_{\tau}\rbrace$. They could, however, map an arbitrary probability distributions $\lbrace p_{\tau}\rbrace$ obeying the quantum conditions to other probability distributions $\lbrace p^{\prime}_{\rho}\rbrace$ obeying the quantum conditions. 

For positive $W$ the linear transformations \eqref{eq:PCB} are standard Markovian chains. The unitary transformations that can be realized by such chains can be classified to be a subset of those that can be realized by cellular automata. These are transformations that have a maximal period $2^{M}$, as described in some detail in refs.~\cite{CWIT,CWQF}. This is in contrast to more general transformations for which arbitrary unitary transformations can be realized. We emphasize that the more general transformations do not involve memory effects. Knowledge of the probabilities $p_{\tau}(t)$ at $t$ is sufficient for determining at $\te$ the probabilities according to any one of the (non-unique) maps that realize a given quantum operation.

\subsection{Unitary transformations by unique jump \\
operations}

While for every bit-quantum map arbitrary transformations of $\rho$ can be realized by maps between classical probability distributions, which are a general form of probabilistic computing, this does not hold for deterministic manipulations of the Ising spins. In particular, the realization of matrix elements $\rho_{z}$ by correlation functions imposes new restrictions which unitary quantum operators can be realized by deterministic Ising spin operations. 
  
Let us first consider simple permutations and sign changes of the Ising spins. Maps on Ising spins induce corresponding maps on correlation functions. For the example of two quantum spins the map $s_{1}^{(1)}\leftrightarrow s_{3}^{(1)}$ induces the map $\rho_{10}\leftrightarrow \rho_{30}$ as well as $\exval{s_{1}^{(1)}s_{k}^{(2)}}\leftrightarrow \exval{s_{3}^{(1)}s_{k}^{(2)}}$ or $\rho_{1k}\leftrightarrow \rho_{3k}$. For maps consistent with this restriction we can take over the previous discussion in sect.~\ref{sec:Two_entangled_quantum_spins} by replacing the spins $s_{kl}$ by products $s_{k}^{(1)}s_{l}^{(2)}$. Both $s_{kl}$ and $s_{k}^{(1)}s_{l}^{(2)}$ can only take values $\pm 1$, such that $\sigma_{kl}=s_{k}^{(1)}s_{l}^{(2)}$ can be considered as a composite Ising spin.
 
The single-spin quantum operations \eqref{eq:11A} are consistent with this ``{compositeness} restriction``. Indeed, for $Q=2$ the transformation of $\sigma_{kl}$ is the same as $s_{k}^{(1)}s_{l}^{(2)}$. The same holds for the permutation $s_{k}^{(1)}\leftrightarrow s_{k}^{(2)}$. This extends to $Q$ spins. These quantum operations can be realized in a simple way by transformations on the $3Q$ classical Ising spins $s_{k}^{(m)}$.

The issue of a possible realization of the CNOT-gate \eqref{eq:12} by operations on the statistical information for six classical Ising spins is more complicated. As for any unitary transformation we have shown above that maps between probability distributions for Ising spins exist that realize eq.~\eqref{eq:26} and therefore the CNOT gate. For our economical realization of the $Q$-qubit density matrix by probabilities for $3Q$ Ising spins and appropriate correlation functions we can answer positively the conceptual question if the CNOT-gate or arbitrary unitary transformations of the quantum system can be realized by a change of the probability distribution for Ising spins. The crucial practical question remains if such a change of probability distributions can be realized in a practical way.

We investigate in appendix \ref{app:Unique_jump_realizations_of_the_CNOT-gate} the question if the CNOT-gate can be realized by a unique jump operation which amounts to a deterministic manipulation of Ising spins. A direct investigation finds that no unique jump operation exists that realizes the map \eqref{eq:25} for arbitrary probability distributions. The question if such an operation is possible on the restricted set of probability distributions obeying the quantum constraint has not yet found a full answer. We discuss this issue in appendix \ref{app:Unique_jump_realizations_of_the_CNOT-gate} as well.

\subsection{Infinitesimal unitary transformations and Hamilton operator}

It is instructive to consider for $Q=2$ infinitesimal unitary transformations ($\tilde{\varepsilon}_{\mn}$ real)
\begin{equation}\label{eq:37}
U=1+i\tilde{\varepsilon}_{\mn}L_{\mn}\, ,
\end{equation}
inducing an infinitesimal change of the density matrix
\begin{align}\label{eq:38}
U\rho\, U^{\dagger}&=\rho+\dfrac{i}{4}\tilde{\varepsilon}_{\mn}\rho_{\tau\rho}\, [L_{\mn},L_{\tau\rho}]\nn\\
&=\rho-\dfrac{1}{2}\tilde{\varepsilon}_{\mn}\rho_{\tau\rho}\, \tilde{f}_{\mn \tau\rho\lambda\sigma}\, L_{\lambda\sigma}\, .
\end{align}
Here the real coefficients $\tilde{f}$ are related to the structure constants of $SU(4)$ by
\begin{equation}\label{eq:39}
[L_{\mn},L_{\tau\rho}]=2i\tilde{f}_{\mn \tau\rho\lambda\sigma}\, L_{\lambda\sigma}.
\end{equation}
Eq.~\eqref{eq:38} determines the infinitesimal change of the coefficients $\delta\rho_{\lambda\sigma}$.

If we express $\rho_{\lambda\sigma}$ by expectation values of correlation functions this amounts to a map of correlation functions. In turn, for arbitrary real infinitesimal $\tilde{\varepsilon}_{\mn}$ the infinitesimal change of the expectation values and correlations,
\begin{align}\label{eq:40}
\delta\rho_{\lambda\sigma}&=\delta\exval{s_{\lambda}^{(1)}s_{\sigma}^{(2)}}=-2\tilde{\varepsilon}_{\mn}\tilde{f}_{\mn \tau\rho\lambda\sigma}\, \rho_{\tau\rho}\nn\\
&=-2\tilde{\varepsilon}_{\mn}\tilde{f}_{\mn \tau\rho\lambda\sigma}\,\exval{s_{\tau}^{(1)}s_{\rho}^{(2)}}\, ,
\end{align}
needs to be expressed by a suitable change of the probability distribution. The non-vanishing components of $\tilde{f}$ read
\begin{align}\label{eq:41}
\tilde{f}_{k\mu l0 m\nu}&=\tilde{f}_{\mu k 0l \nu m}=\tilde{f}_{k 0 l\mu m\nu}=\tilde{f}_{ 0k\mu l \nu m}\nn\\
&=\tilde{f}_{k\mu l \nu m0}=\tilde{f}_{\mu k \nu l0m}=\varepsilon_{klm}\delta_{\mn}\, .
\end{align}
The changes required for a unitary transformation of $\rho$ are therefore
\begin{align}\label{eq:42}
\delta\exval{ s_{j}^{(1)}}&=-2\varepsilon_{klj}\left (\tilde{\varepsilon}_{k0}\exval{ s_{l}^{(1)}} +\tilde{\varepsilon}_{km}\exval{ s_{l}^{(1)} s_{m}^{(2)}}\right )\, ,\nn\\
\delta\exval{ s_{j}^{(2)}}&=-2\varepsilon_{klj}\left (\tilde{\varepsilon}_{0k}\exval{ s_{l}^{(2)}} +\tilde{\varepsilon}_{mk}\exval{ s_{m}^{(1)} s_{l}^{(2)}}\right )\, ,\nn\\
\delta\exval{ s_{i}^{(1)} s_{j}^{(2)}}&=-2\varepsilon_{klj}\left (\tilde{\varepsilon}_{0k}\exval{ s_{i}^{(1)} s_{l}^{(2)}} +\tilde{\varepsilon}_{ik}\exval{s_{l}^{(2)}}\right )\nn\\
&\quad -2\varepsilon_{kli}\left (\tilde{\varepsilon}_{k0}\exval{ s_{l}^{(1)} s_{j}^{(2)}} +\tilde{\varepsilon}_{kj}\exval{s_{l}^{(1)}}\right ) .
\end{align}
 The quantum condition for the expectation values $\exval{s_{k}^{(1)}}$, $\exval{s_{l}^{(2)}}$ and $\exval{s_{k}^{(1)}s_{l}^{(2)}}$ guarantees that $\exval{s_{k}^{(1)}+\delta s_{k}^{(1)}}$, $\exval{s_{l}^{(2)}+\delta s_{l}^{(2)}}$ and $\exval{(s_{k}^{(1)}+\delta s_{k}^{(1)})(s_{l}^{(2)}+\delta s_{l}^{(2)})}$ also obey the quantum condition. The latter ones therefore obey the constraints \eqref{eq:34} or \eqref{eq:35}. This is not easily visible directly from eq.~\eqref{eq:42}.

If the ratio $\tilde{\varepsilon}_{\mn}/\epsilon$ remains finite as the distance between steps $\epsilon=t^{\prime}-t$ goes to zero, the continuum limit yields the von Neumann equation
\begin{equation}\label{eq:CL1} 
\partial_{t}\rho=\dfrac{1}{2\epsilon} \left( \rho(\te)-\rho(t-\epsilon) \right) =
- i \left[ H(t),\, \rho(t) \right]\, ,
\end{equation}
with Hamiltonian operator
\begin{equation}\label{eq:CL2} 
H=-\dfrac{\tilde{\varepsilon}_{\mn}}{\epsilon}L_{\mn}\, .
\end{equation}
For $\tilde{\varepsilon}_{\mn}\raw 0$ the period of $U$ diverges to infinity. Infinitesimal transformations can therefore be realized by random operations of the type \eqref{eq:PCB}, or by the subclass of unique jump operations, only for an infinite number of Ising spins $M\raw\infty$. This may not be surprising, since non-trivial continuous functions require an infinite number of bits for their precise characterization. On the other hand, for any complete bit-quantum map infinitesimal unitary transformations can be realized for finite $M$ if arbitrary changes of the classical probability distributions are admitted. In this case the continuous change of functions exploits the continuous probability distributions.

\section{Static memory materials}\label{sec:Static_Memory_Materials}

Concerning practical realizations of quantum jumps by classical Ising spins we have so far mainly discussed deterministic operations on the Ising spins that are associated to unique jump operations. (An exception are the stochastic chains \eqref{eq:PCB}.) We have not yet discussed the general realization of the probabilistic aspects, as the probabilistic quantum constraints or probabilistic operations for computational steps. For a quantum computer the dynamical behavior of quantum spins in time or space realizes automatically these probabilistic aspects. In the same way, classical statistical systems can also implement the probabilistic aspects of a computation by the dynamical behavior of Ising spins. This may happen for the time evolution of Ising spins, or for generalized equilibrium states with $t$ a position variable. In this way no explicit random number generator is needed for probabilistic computation. Nature does this job by being probabilistic itself. 

In this section we begin the discussion of probabilistic computing operations by investigating static memory materials. These are generalized Ising models for which part of the boundary information is transported into the bulk. The variable $t$ corresponds then to the position of hypersurfaces. The transport of probabilistic information from one hypersurface to the next one is governed by the transfer matrix or step evolution operator \cite{CWIT,CWQF}.

The generalized Ising models in a static (generalized equilibrium) state are by far not the only possible implementations of probabilistic computing. They offer, however, a definite framework since all aspects can be treated by standard methods of equilibrium classical statistics, or be numerical simulations \cite{SEW}. More general settings for probabilistic computing could be realized, for example, by artificial neural networks. They will be discussed in sect.~\ref{sec:Probabilistic_Computing}.

\subsection{Generalized Ising models}

A possible way to realize a change of classical probability distributions employs the change of local probabilities in generalized Ising models. We take discrete $t$-points on a one-dimensional lattice with lattice distance $\epsilon$. On each point $t$ we place $M$ classical Ising spins $s_{\gamma}(t)$, $\gamma=1,\ldots ,M$. The probability for an arbitrary configuration of all classical spins $\lbrace s_{\gamma}(t)\rbrace$ is given by
\begin{equation}\label{eq:54}
w[s]=\exp \left(-S_{cl}[s] \right)\, b(s_{in},\, s_{f})\, ,
\end{equation}
with normalized partition function
\begin{equation}\label{eq:55}
Z=\int\cD s\, w[s]=1\, .
\end{equation}
The classical action
\begin{equation}\label{eq:56}
S_{cl}[s]=\sum_{t=t_{in}}^{t_{f}-\epsilon} \, \cL \left(s_{\gamma}(t+\epsilon), \,
 s_{\gamma}(t);\, t \right)
\end{equation}
describes interactions between spins on neighboring $t$-layers. The boundary term $b$ involves only the spins at the initial boundary $s_{\gamma,in}=s_{\gamma}(t_{in})$ and final boundary $s_{\gamma,f}=s_{\gamma}(t_{f})$, where $t_{in}\leq t\leq t_{f}$. If $b$ is positive for all configurations of $s_{in}$ and $s_{f}$, the probabilities $w[s]$ are all positive semidefinite for real $S_{cl}$. The ``functional integral'' $\int\cD s$ denotes the sum over all spin configurations, such that eq.~\eqref{eq:55} ensures the proper normalization of the probability distribution $w$.

A given $\cL(t)$ will be associated with a quantum gate at $t$. A sequence of different quantum gates will employ different $\cL(t)$ on different $t$-layers. The factor $\exp(-S_{cl})$ can be written as a product of factors $\exp(-\cL(t))$. This will be reflected in the multiplication of transformations performed by different gates. As an example, for a given $t$ the factor $\exp(-\cL(t))$ can represent a restricted Boltzmann machine with $M$ visible and $M$ hidden nodes, provided that $\cL(t)$ is of the form
\begin{equation}\label{eq:BM1} 
\cL(t)=-a_{\gamma}(t)s_{\gamma}(\te)-b_{\gamma}s_{\gamma}(t)-W_{\gamma\delta}(t)s_{\gamma}(\te)s_{\delta}(t)\, .
\end{equation}
The generalized Ising model represents then a chain of such Boltzmann machines coupled sequentially, with visible nodes of one layer playing the role of the hidden nodes at the next layer. We do not discuss here a possible learning of the coefficients $a$, $b$, $W$ for the optimization of given tasks, but rather consider fixed values (that may be the result of a training process). Another example are cellular automata for which $\cL(t)$ diverges for all neighboring spin configurations that do not correspond to the particular map of configurations of the automaton at $t$. We will not impose here any restrictions on the form of $\cL(t)$.

A ``local probability distribution'' at $t$ obtains by ``integrating out'' all spins at sites $t^{\prime}\neq t$
\begin{equation}\label{eq:57}
p(t)=p[s_{\gamma}(t)]=\prod_{t^{\prime}\neq t}\,\prod_{\gamma}\,\sum_{s_{\gamma}(t^{\prime})=\pm 1}\, w[s]\, .
\end{equation}
It associates to each configuration of local spins $[s_{\gamma}(t)]$ a probability, with
\begin{equation}\label{eq:58}
Z=\prod_{\gamma}\,\sum_{s_{\gamma}(t)=\pm 1}\, p[s_{\gamma}(t)]=1\, .
\end{equation}
If the local probability distribution $p(t)$ obeys the quantum constraint, we can associate to it a quantum density matrix $\rho(t)$ according to eq.~\eqref{eq:33}.

Consider now the local probability distribution $p(t+\epsilon)$ on a neighboring $t$-layer. If it obeys again the quantum condition we define the quantum density matrix $\rho(t+\epsilon)$. The map from $p(t)$ to $p(t+\epsilon)$ is a map between probability distributions for Ising spin configurations, as discussed in the preceding sections. It induces a map between density matrices, $\rho(t)\raw \rho(t+\epsilon)$, provided the quantum condition is preserved. This map will depend on the choice of $\cL(t)$ in eq.~\eqref{eq:56}. We are interested in models where the transformation is unitary, $\rho(t+\epsilon)=U(t)\rho(t)U^{\dagger}(t)$. Such models can realize ``static memory materials'' \cite{CWIT,CWQF}. If the density matrix $\rho(t_{in})$ at the initial boundary is non-trivial, the information in $\rho(t_{in})$ cannot be lost by a unitary evolution. It is still present in $\rho(t)$ for arbitrary positions $t$. The ``bulk'' keeps memory of the boundary conditions. 

\subsection{Unitary quantum operations by static memory materials}

For simplicity, we restrict the discussion here to two quantum spins, $Q=2$, and to the bit-quantum map involving classical correlations \eqref{eq:33}. Extensions to arbitrary $Q$ are conceptually straightforward.

We consider a chain of Ising spins $s_{k}^{(m)}(t)=s_{\gamma}(t)$, with six Ising spins $s_{\gamma}(t)= ( s_{k}^{(1)}(t),s_{k}^{(2)}(t) )$ at every point $t$ on the chain. Simple deterministic spin operations can be realized by $\cL(t)$ inducing unique jumps. As an example we consider
\begin{align}\label{eq:59}
\cL(t)&=-\beta\, \biggl \{s_{1}^{(1)\prime}s_{3}^{(1)}+s_{3}^{(1)\prime}s_{1}^{(1)}-s_{2}^{(1)\prime}s_{2}^{(1)}\nn\\
&\quad +\sum_{k}\, s_{k}^{(2)\prime}s_{k}^{(2)}-6\biggl \}\, ,
\end{align}
where $s_{\gamma}^{\prime}$ stands for $s_{\gamma}(t+\epsilon)$ and $s_{\gamma}$ for $s_{\gamma}(t)$. For 
\begin{equation}\label{eq:60}
s_{1}^{(1)\prime}=s_{3}^{(1)}\com s_{3}^{(1)\prime}=s_{1}^{(1)}\com s_{2}^{(1)\prime}=-s_{2}^{(1)}\com s_{k}^{(2)\prime}=s_{k}^{(2)}
\end{equation}
one has $\cL(t)=0$. If $s_{\gamma}^{\prime}$ deviates from the values \eqref{eq:60} one has
\begin{equation}\label{eq:61}
\cL(t)=2m\beta\, ,
\end{equation}
with $m$ a positive (non-zero) integer. Consider now the limit $\beta\raw\infty$. For all configurations of $(s_{\gamma}^{\prime}, s_{\gamma})$ for which $m>0$ the factor $\exp(-\cL(t))$ in eqs~\eqref{eq:54}, \eqref{eq:56} vanishes. The probability $w[s]$ for such a configuration is zero, such that all configurations with non-zero probability must indeed obey eq.~\eqref{eq:60}. This realizes the single-spin unitary Hadamard gate \eqref{eq:9}, \eqref{eq:10} for $S^{(1)}$. 

For $\beta\raw\infty$ the generalized Ising model therefore realizes a ``unique jump operation'' which associates to every configuration of spins at $t$ a unique configuration of spins at $t+\epsilon$, similar to cellular automata \cite{CA, TH, TH2, EL}. If $p(t)$ obeys the quantum condition such that $\rho(t)$ is positive, $\rho(t+\epsilon)=U(t)\rho(t)U^{\dagger}(t)$ obtains by the map of probabilities $p(t)\raw p(t+\epsilon)$ induced by the spin transformations \eqref{eq:60}. Other simple spin transformations can be constructed in a similar way. For example, the spin exchange $s_{k}^{(1)}=s_{k}^{(2)}$, $s_{k}^{(2)}=s_{k}^{(1)}$ corresponding to the unitary transformation \eqref{eq:28} is realized, for $\beta\raw \infty$, by
\begin{equation}\label{eq:62}
\cL(t)=-\beta\left (s_{k}^{(1)\prime}s_{k}^{(2)}+s_{k}^{(2)\prime}s_{k}^{(1)}-6\right ).
\end{equation}
A sequence of unitary transformations $U_{3}U_{2}U_{1}$ can be realized by a sequence $\cL_{1}(t)$, $\cL_{2}(t+\epsilon)$, $\cL_{3}(t+2\epsilon)$ realizing $U_{1}$, $U_{2}$ and $U_{3}$, respectively.

The limit $\beta\raw\infty$ may be associated to the zero temperature limit (limit of vanishing ratio temperature/excitation gap). This limit does not lead to unique jump operations for arbitrary forms of $\cL(t)$. As an example we may consider
\begin{align}\label{eq:63}
\cL(t)&=-\beta\, \biggl \{s_{3}^{(1)\prime}s_{3}^{(2)\prime}s_{3}^{(2)}+s_{3}^{(2)\prime}s_{3}^{(1)}s_{3}^{(2)}\nn\\
&\;\vphantom{\biggl (}\hphantom{=-\beta\biggl \lbrace } +s_{3}^{(1)\prime}s_{2}^{(2)\prime}s_{2}^{(2)}+s_{2}^{(2)\prime}s_{3}^{(1)}s_{2}^{(2)}\nn\\
&\;\vphantom{\biggl (}\hphantom{=-\beta\biggl \lbrace } +s_{1}^{(1)\prime}s_{1}^{(2)\prime}s_{1}^{(1)}+s_{1}^{(1)\prime}s_{1}^{(1)}s_{1}^{(2)}\nn\\
&\;\vphantom{\biggl (}\hphantom{=-\beta\biggl \lbrace } +s_{2}^{(1)\prime}s_{1}^{(2)\prime}s_{2}^{(1)}+s_{2}^{(1)\prime}s_{2}^{(1)}s_{1}^{(2)}\\
&\;\vphantom{\biggl (}\hphantom{=-\beta\biggl \lbrace } +\gamma\biggl (s_{1}^{(1)\prime}s_{2}^{(2)\prime}s_{2}^{(1)}s_{3}^{(2)}+s_{2}^{(1)\prime}s_{3}^{(2)\prime}s_{1}^{(1)}s_{2}^{(2)}\nn\\
&\;\vphantom{\biggl (}\hphantom{=-\beta\biggl \lbrace } -s_{2}^{(1)\prime}s_{2}^{(2)\prime}s_{1}^{(1)}s_{3}^{(2)}-s_{1}^{(1)\prime}s_{3}^{(2)\prime}s_{2}^{(1)}s_{2}^{(2)}\biggl )\nn\\
&\;\vphantom{\biggl (}\hphantom{=-\beta\biggl \lbrace } + s_{3}^{(1)\prime}s_{3}^{(1)}+ s_{1}^{(2)\prime}s_{1}^{(2)}+\delta s_{3}^{(1)\prime}s_{1}^{(2)\prime}s_{3}^{(1)}s_{1}^{(2)}-\Delta \biggl \}\nn.
\end{align}
It is chosen to resemble in some aspects the CNOT gate \eqref{eq:12}, \eqref{eq:26}. We observe that not all fifteen terms in the sum can equal one. This is due to the minus sign for the eleventh and twelfth term. Having the ninth and eleventh term both equal requires $s_{1}^{(1)\prime}s_{2}^{(1)}=-s_{2}^{(1)\prime}s_{1}^{(1)}$. On the other hand, the fourteenth term can only equal one if $s_{1}^{(2)\prime}=s_{1}^{(2)}$, such that in turn setting the fifth and seventh term equal one requires $s_{1}^{(1)\prime}s_{1}^{(1)}=s_{2}^{(1)\prime}s_{2}^{(1)}=s_{1}^{(2)}$. Multiplying our first relation with $s_{1}^{(1)}$ yields $s_{1}^{(1)\prime}s_{1}^{(1)}s_{2}^{(1)}=-s_{2}^{(1)\prime}$, and further multiplication with $s_{2}^{(1)}$ results in $s_{1}^{(1)\prime}s_{1}^{(1)}=-s_{2}^{(1)\prime}s_{2}^{(1)}$, contradicting the requirement from the fifth and seventh term. Interactions of the type \eqref{eq:63} may be called frustrated since not all terms can be minimized simultaneously. The additive constant $\Delta$ reflects this fact. It is chosen such that $\cL(t)=0$ for the spin values for which $\cL(t)$ is minimal.

For $\gamma=0$ and $\delta=1$ eq.~\eqref{eq:60} represents a unique jump operation that realizes the conditional $C$-transformation discussed in appendix \ref{app:Unique_jump_realizations_of_the_CNOT-gate}. If $s_{3}^{(1)}=-1$ the signs of $s_{2}^{(2)}$ and $s_{3}^{(2)}$ are flipped, and if $s_{1}^{(2)}=-1$ one changes the sign of $s_{1}^{(1)}$ and $s_{2}^{(1)}$. One has $\Delta=11$ in this case. The same operation is realized for $\gamma=1$, $\delta=1$, again with $\Delta=11$. For $\gamma=0,1$ the value of $\delta$ does actually not matter if $\Delta=10+\delta$. For $\gamma=2$, $\delta=0$, $\Delta=10$ eq.~\eqref{eq:63} no longer represents a unique jump operation. For a given configuration $s_{\gamma}(t)$ there are now three different configurations $s_{\gamma}(t+\epsilon)$ that all lead to a minimal value $\cL(t)=0$. All three possibilities are attained with equal probability, such that the interaction \eqref{eq:63} realizes genuine probabilistic computing. We need a formalism for describing this type of situation. This also holds for finite $\beta$. Even for a fixed configuration of $s_{\gamma}(t)$ the probabilities for many configurations of $s_{\gamma}(\te)$ do not vanish.

\subsection{Classical density matrix}

For unique jump operations the computation of $p(t+\epsilon)$ from $p(t)$ is easy - every state $\tau$ at $t$ corresponds precisely to one state $\tau^{\prime}$ at $t+\epsilon$, inducing a map $p_{\tau^{\prime}}(t+\epsilon)=p_{\tau}(t)$. The generalization to arbitrary $\cL(t)$ needs an ``evolution law'' how $p(t+\epsilon)$ is related to $p(t)$. In general, no simple form of such an evolution law can be formulated in terms of the local probabilities alone. The appropriate formulation is the quantum formalism for classical statistics \cite{CWIT,CWQF}. It extends the local probabilistic information to a classical density matrix $\rho^{\prime}(t)$ with elements $\rho^{\prime}_{\tau\rho}(t)$. Similar to quantum mechanics the local probabilities are the diagonal elements of $\rho^{\prime}$, $p_{\tau}(t)=\rho^{\prime}_{\tau\tau}(t)$. We consider here an arbitrary number $M$ of Ising spins $s_{\gamma}(t)$, $\gamma=1,\ldots ,M$, such that the number of ``classical states'' $\tau$ is given by $2^{M}$, and $\rho^{\prime}$ is a real $2^{M}\times 2^{M}$-matrix. The classical density matrix obeys a linear evolution law
\begin{equation}\label{eq:Z1} 
\rho^{\prime}(t+\epsilon)=S(t)\rho^{\prime}(t)S^{-1}(t)\, ,
\end{equation}
with the step evolution operator $S(t)$ corresponding to the transfer matrix in a particular normalization. The step evolution operator can be computed from $\cL(t)$. 

In order to briefly recapitulate this quantum formalism for classical statistics we consider in eq.~\eqref{eq:54} a boundary term of the product form
\begin{equation}\label{eq:Z2}
b(s_{in},\, s_{f})=f_{in}(s_{in})\,\bar{f}_{f}(s_{f})\, .
\end{equation}
The classical wave function $f(t)$ obtains by integrating out the Ising spins at $t^{\prime}<t$, 
\begin{equation}\label{eq:Z3} 
f(t)=\int\,\cD s(t_{in}\leq t^{\prime}<t)\, \exp\left (-\sum_{t^{\prime}<t}\, \cL(t^{\prime})\right )\, f_{in}(s_{in})\, .
\end{equation}
This wave function depends only on the local spins $s_{\gamma}(t)$. It obeys a linear evolution law
\begin{equation}\label{eq:Z4}
f(t+\epsilon)=\int\,\cD s(t)\,\exp\left (-\cL(t)\right )\, f(t)\, .
\end{equation}
Similarly, for the conjugate classical wave function $\bar{f}(t)$ one integrates out the spins at $t^{\prime}>t$
\begin{equation}\label{eq:Z5} 
\bar{f}(t)=\int\,\cD s(t < t^{\prime}\leq t_{f})\, \exp\left (-\sum_{t^{\prime}\geq t}\, \cL(t^{\prime})\right )\, \bar{f}_{f}(s_{f})\, .
\end{equation}
Again, it depends on $s_{\gamma}(t)$ and obeys a linear evolution law
\begin{equation}\label{eq:Z7}
\bar{f}(t-\epsilon)=\int\,\cD s(t)\,\exp\left (-\cL(t-\epsilon)\right )\, \bar{f}(t)\, .
\end{equation}

 The local probability distribution is a bilinear of the classical wave functions
\begin{equation}\label{eq:Z6}
p(t)=f(t)\bar{f}(t)\, .
\end{equation}
We may distinguish the argument of $\bar{f}(t)$ from the one of $f(t)$, choosing $\bar{s}(t)$ for $\bar{f}(t)$. The classical density matrix
\begin{equation}\label{eq:Z8} 
\tilde{\rho}(t)=f(s(t))\,\bar{f}(\bar{s}(t))
\end{equation}
depends then on the two sets of local spins $s$ and $\bar{s}$. With this differentiation the density matrix is associated to an integration of all spins at $t^{\prime}\leq t$, and we can reinstore general boundary conditions $b(s_{in},s_{f})$. The local probability distribution $p(t)$ obtains from the ``diagonal elements'' of $\tilde{\rho}(t)$ for which $\bar{s}(t)$ is identified with $s(t)$.

In the occupation number basis \cite{CWIT} we can define basis functions $h_{\tau}(s)$ that equal one precisely for the spin configuration that corresponds to $\tau$, and are zero otherwise. The classical wave functions are expanded as
\begin{align}\label{eq:Z6A}
f(s(t))&=\tilde{q}_{\tau}(t)\, h_{\tau}(s(t))\, ,\nn\\
\bar{f}(\bar{s}(t))&=\bar{q}(t)\, h_{\tau}(\bar{s}(t))\, .
\end{align}
Bilinears can be expanded as
\begin{align}\label{eq:Z9} 
&\tilde{\rho}(t)=h_{\tau}(s)\, \rho^{\prime}_{\tau\rho}(t)\, h_{\rho}(\bar{s})\, ,\nn\\
&\cL(t)=h_{\tau}(s(t+\epsilon))\, M_{\tau\rho}(t)\, h_{\rho}(s(t))\, .
\end{align}
This yields the transfer matrix $\bar{S}$
\begin{equation}\label{eq:Z10} 
\exp\left (-\cL(t)\right )=h_{\tau}(s(t+\epsilon))\, \bar{S}_{\tau\rho}(t)\, 
h_{\rho}(s(t))\, ,
\end{equation}
with positive semidefinite elements
\begin{equation}\label{eq:Z11} 
\bar{S}_{\tau\rho}(t)=\exp\left (- M_{\tau\rho}(t)\right ).
\end{equation}
We choose the additive normalization in $\cL(t)$ such that the largest absolute value of eigenvalues of $\bar{S}(t)$ equals one. With this normalization $S(t)$ becomes the step evolution operator. From eq.~\eqref{eq:Z11} we infer the important restriction that $S(t)$ is a nonnegative matrix. All matrix elements $S_{\tau\rho}$ are positive real numbers or vanish.

In the occupation number basis the evolution laws for the wave vectors $\tilde{q}$ and $\bar{q}$ are linear matrix equations
\begin{align}\label{eq:Z11A} 
& \tilde{q}(t+\epsilon) = S(t)\tilde{q}(t)\, ,\nn\\
& \bar{q}(t) = \bar{q}(t+\epsilon)S(t)\, , \notag \\
& \bar{q}(t+\epsilon) = \bar{q}(t)\, S^{-1}(t)\, .
\end{align}
Solutions obey the superposition principle familiar from quantum mechanics. The local probability distribution at $t$ is given in the occupation number basis by the probabilities
\begin{equation}\label{eq:A19A}
p_\tau(t) = q_{\tau}(t)\,\bar{q}_\tau(t)\, .
\end{equation} 
While the classical wave functions obey a simple linear evolution law, no such simple law can be formulated for the local probabilities in the general case. An exception are Markov processes or deterministic unique jump chains.

The close analogy of the linear evolution law to quantum mechanics allows us to take over insights from quantum mechanics to classical statistics. An example is the no-cloning theorem. The theorem states that no invertible step evolution operator exists that can clone arbitrary unknown wave functions, and therefore arbitrary unknown probability distributions. One may clone specific probability distributions, as a fixed sequence of bits with probability one, by a simple copy-and-paste operation. No step evolution operator can clone all arbitrary probability distributions, however. By cloning one understands that a product state
\begin{align}\label{eq:A19B}
q_\tau &= q_{\alpha\beta} = \varphi_\alpha e_\beta\, , \quad q = \ket{\varphi}
    \ket{e} \, , \notag \\ 
\bar{q}_\tau &= \bar{q}_{\alpha\beta} = \bar{\varphi}_\alpha \bar{e}_\beta\, , \quad 
    \bar{q} =  \bra{\bar{\varphi}} \bra{\bar{e}}\, ,
\end{align}
with arbitrary factor (ancilla state) $(e,\,\bar{e})$, normalized with $\bar{e}_\beta e_\beta = 1$, is transformed by the step evolution operator to a product of identical states
\begin{align}\label{eq:A19C}
& S_{\alpha\beta\gamma\delta} \, \varphi_\gamma\, e_\delta = \varphi_\alpha
    \varphi_\beta\, , \quad S \ket{\varphi}\ket{e} = \ket{\varphi}\ket{\varphi}\, ,
    \notag \\
& \bar{\varphi}_\gamma\, \bar{e}_\delta\, S^{-1}_{\gamma\delta\alpha\beta} = 
    \bar{\varphi}_\alpha \bar{\varphi}_\beta\, , \quad \ket{\bar{\varphi}} 
    \ket{\bar{e}} S^{-1} = \bra{\bar{\varphi}} \bra{\bar{\varphi}} \, .
\end{align}
This operation clones the probability distribution $p_\tau = p_\alpha \bar{p}_\beta \rightarrow p_\alpha p_\beta$. 

Consider now two different probabilistic states characterized by different pairs of wave functions $(\varphi,\,\bar{\varphi})$ and $(\psi,\,\bar{\psi})$, taken as direct product with the same state $(e,\, \bar{e})$. We want to know if both can be cloned with the same invertible step evolution operator. If one assumes that this is the case one finds for the product
\begin{align}\label{eq:A19D}
\bra{\bar{\psi}} \bra{e} \ket{\varphi} \ket{e} &= \braket{\bar{\psi}| \varphi} = 
    \bra{\bar{\psi}} \bra{\bar{e}} S^{-1} S\, \ket{\varphi} \ket{e} \notag \\
& = \bra{\bar{\psi}} \bra{\bar{\psi}} \ket{\varphi} \ket{\varphi} = 
    \big( \braket{\bar{\psi} | \varphi} \big)^2\, ,
\end{align}
implying that $\braket{\bar{\psi} | \varphi}$ equals either one or zero. This constitutes a restriction on two states that can be cloned simultaneously, demonstrating that two arbitrary states cannot be cloned by the same evolution operator. Sets of different states $(\bar{\varphi}_i,\, \varphi_i)$ that can be cloned by the same $S$ typically need to be orthogonal, $\braket{\bar{\psi}_i | \varphi_j} = \delta_{ij}$. This is the case for fixed bit sequences with probabilities one. They are all orthogonal, and arbitrary such sequences can be copied.

Initial conditions of the type \eqref{eq:Z2} realize ``pure classical states'' with a product form of the classical density matrix
\begin{equation}\label{eq:XXAA} 
\rho^{\prime}_{\tau\rho}=\tilde{q}_{\tau}(t)\, \bar{q}_{\rho}(t)\, .
\end{equation}
Translating eqs~\eqref{eq:Z4}, \eqref{eq:Z7}, \eqref{eq:Z8} to the occupation number basis yields the matrix equation \eqref{eq:Z1} which generalizes to arbitrary initial conditions. This linear evolution equation will be the basic equation for our approach to probabilistic computation. It requires local probabilistic information beyond the local probabilities $p_{\tau}(t)$. In general, the computation of $p_{\tau}(\te)$ requires knowledge of diagonal as well as off-diagonal elements of $\rho^{\prime}(t)$.

A unique jump step evolution operator $S(t)$ has precisely one element equal to one in each row and column, while all other elements vanish. This is realized, for $\beta\raw\infty$, by our examples \eqref{eq:59} and \eqref{eq:62}, or by eq.~\eqref{eq:60} for $\gamma=0,1$. Unique jump evolution operators are orthogonal matrices, $S^{T}S=1$. They do not change the length of the classical wave vector,
\begin{equation}\label{eq:Z12}
\tilde{q}_{\tau}(t+\epsilon)=S_{\tau\rho}(t)\tilde{q}_{\rho}(t)\, ,
\end{equation}
by mapping each element $\tilde{q}_{\tau}(t)$ to precisely one element $\tilde{q}_{\tau^{\prime}}(t+\epsilon)$, and similar for the conjugate wave function. Correspondingly, each local probability $p_{\tau}(t)=\rho^{\prime}_{\tau\tau}(t)=\tilde{q}_{\tau}(t)\bar{q}_{\tau}(t)$ (no sum over $\tau$ here) is mapped to precisely one $p_{\tau^{\prime}}(t+\epsilon)$. 

\subsection{Loss of memory}

For a discussion of step evolution operators that are not unique jump operations we first investigate a system with four states $\tau$, $M=2$, and consider the particular step evolution operator
\begin{equation}\label{eq:Z13}
S_{P}=\dfrac{1}{2}\left (\begin{array}{cccc}
0 & 1 & 1 & 0 \\ 
1 & 0 & 0 & 1 \\ 
1 & 0 & 0 & 1 \\ 
0 & 1 & 1 & 0
\end{array} \right ).
\end{equation}
The eigenvalues of $S_{P}$ are
\begin{equation}\label{eq:Z14}
\lambda_{1}=1\com\lambda_{2}=-1\com\lambda_{3}=\lambda_{4}=0\, ,
\end{equation}
with (unnormalized) eigenfunctions
\begin{align}\label{eq:Z15}
& q^{(1)}=\left (\begin{array}{c}
1 \\ 
1 \\ 
1 \\ 
1
\end{array} \right )\com q^{(2)}=\left (\begin{array}{c}
1 \\ 
-1 \\ 
-1 \\ 
1
\end{array} \right )\, ,\nn\\
& q^{(3)}=\left (\begin{array}{c}
1 \\ 
0 \\ 
0 \\ 
-1
\end{array} \right )\com q^{(4)}=\left (\begin{array}{c}
0 \\ 
1 \\ 
-1 \\ 
0
\end{array} \right ).
\end{align}
For an arbitrary wave function $\tilde{q}(t)$ the wave function $\tilde{q}(t~+~\epsilon)=S(t)\tilde{q}(t)$ is a linear combination of only $q^{(a)}=\frac{1}{2}(q^{(1)}+q^{(2)})$ and $q^{(b)}=\frac{1}{2}(q^{(1)}-q^{(2)})$. The memory of information contained in the subspace spanned by the two combinations $q^{(3)}(t)$ and $q^{(4)}(t)$ is erased. Applying once more $S(t+\epsilon)=S_{P}$ one obtains a jump $q^{(a)}(t+2\epsilon)=q^{(b)}(t+\epsilon)$, $q^{(b)}(t+2\epsilon)=q^{(a)}(t+\epsilon)$. This amounts to a unique jump operation for a subspace.

The erase of memory is a limiting case of a more smooth loss of memory of boundary conditions. For 
\begin{equation}\label{eq:Z16}
S=\left (\begin{array}{cccc}
0 & w & u & 0 \\ 
u & 0 & 0 & w \\ 
w & 0 & 0 & u \\ 
0 & u & w & 0
\end{array} \right )\com w=1-u\com 0\leq u\leq 1\, , 
\end{equation}
the eigenvalues are $\lambda_{1,2}=\pm 1$, $\lambda_{3,4}=\pm i(1-2u)$. After $P$-steps with equal $S$, $\tilde{q}(t+P\epsilon)=S^{P}\tilde{q}(t)$, the eigenvectors to $\lambda_{3}$ and $\lambda_{4}$ are suppressed by a factor $(1-2u)^{P}$. For $u$ not too close to the boundaries at $u=0,1$, and sufficiently large $P$, the part of the memory of boundary conditions stored in the corresponding eigenvectors $q^{(3)}$ and $q^{(4)}$ is effectively lost, and only the memory in the space of eigenvectors $q^{(1)}$ and $q^{(2)}$ survives. For $u\neq 1/2$ the step evolution operator $S$ is invertible. We may consider eq.~\eqref{eq:Z13} as the limit of eq.~\eqref{eq:Z16} for $u$ approaching $1/2$. In this limit $S$ acts as a projector on a subspace, combined with an exchange of $\tilde{q}^{(a)}$ and $\tilde{q}^{(b)}$.

In the other limit of $u$ close to zero or one the loss of information is only slow. Many steps with equal $S$ would be needed before a substantial part of the boundary information is lost. The small quantities $u$ or $w$ may be considered as ``errors'' in a deterministic computation. This extends to more general ``approximate unique jump operators''. Our approach may be used as a formalism to deal with a probabilistic error in deterministic computations.

The matrix $S_{P}$ can be used for a simple demonstration that knowledge of the local probabilities $p_{\tau}(t)$ is insufficient for the computation of $p_{\tau}(\te)$. Let us assume that $S(t)$ is given by $S_{P}$, while $S(t^{\prime}<t)$ and $S(t^{\prime}>t)$ are unknown step evolution operators. The loss of memory also occurs for the conjugate wave function $\bar{q}$, such that $\bar{q}(t)=\bar{q}_{a}(t)q^{(a)}+\bar{q}_{b}(t)q^{(b)}$ or $\bar{q}_{4}(t)=\bar{q}_{1}(t)$, $\bar{q}_{3}(t)=\bar{q}_{2}(t)$. At $t$ the four probabilities are 
\begin{align}
p_{1}(t) &= \bar{q}_{1}(t)\tilde{q}_{1}(t)\, , \notag \\
p_{2}(t) &= \bar{q}_{2}(t)\tilde{q}_{2}(t)\, , \notag \\
p_{3}(t) &= \bar{q}_{2}(t)\tilde{q}_{3}(t)\, , \notag \\
p_{4}(t) &= \bar{q}_{1}(t)\tilde{q}_{4}(t)\, , 
\end{align}
while at $t+\epsilon$ one has 
\begin{align}
p_{1}(t+\epsilon) &= \bar{q}_{1}(t+\epsilon)(\tilde{q}_{2}(t)+\tilde{q}_{3}(t))/2\, , \notag \\ 
p_{2}(t+\epsilon) &= \bar{q}_{2}(t+\epsilon)(\tilde{q}_{1}(t)+\tilde{q}_{4}(t))/2\, , \notag \\ 
p_{3}(t+\epsilon) &= \bar{q}_{3}(t+\epsilon)(\tilde{q}_{1}(t)+\tilde{q}_{4}(t))/2\, , \notag \\
p_{4}(t+\epsilon) &= \bar{q}_{4}(t+\epsilon)(\tilde{q}_{2}(t)+\tilde{q}_{3}(t))/2 \, .
\end{align}
In general, $p_{\tau}(t+\epsilon)$ cannot be computed from $p_{\tau}(t)$ alone -- the four components $\bar{q}_{\tau}(t+\epsilon)$ are needed. Only if we assume $\bar{q}_{1}(t+\epsilon)=\bar{q}_{4}(t+\epsilon)$ and $\bar{q}_{2}(t+\epsilon)=\bar{q}_{3}(t+\epsilon)$, induced, for example, by a factor $S(t+\epsilon)$ given by eq.~\eqref{eq:Z13}, the situation changes. Now $\bar{q}(t+\epsilon)$ is computable from $\bar{q}(t)$, $\bar{q}_{1}(t+\epsilon)=\bar{q}_{4}(t+\epsilon)=\bar{q}_{2}(t)$, $\bar{q}_{2}(t+\epsilon)=\bar{q}_{3}(t+\epsilon)=\bar{q}_{1}(t)$. In this case the step evolution operator \eqref{eq:Z13} generates a map of probabilities
\begin{align}\label{eq:Z17}
p_{1}(t+\epsilon) &= p_{4}(t+\epsilon)=\dfrac{1}{2}\left (p_{2}(t)+p_{3}(t)\right )\nn\\
p_{2}(t+\epsilon) &= p_{3}(t+\epsilon)=\dfrac{1}{2}\left (p_{1}(t)+p_{4}(t)\right ).
\end{align}
More generally, for invertible $S$, as for eq.~\eqref{eq:Z16}, the computation of $p_{\tau}(t+\epsilon)$ from the local statistical information at $t$ needs knowledge of all elements of the classical density matrix according to eq.~\eqref{eq:Z1}. Information about the local probabilities $p_{\tau}(t)$ is not sufficient.

Classical statistical systems with a complete or partial transport of boundary information into the bulk may appear somewhat unfamiliar. They correspond to step evolution operators for which a certain number of elements vanishes. We have concentrated on those $S$ that actually can be used for purposes of computing. There are, of course, also many statistical systems for which the boundary information is rapidly lost in the bulk. An example is given by the step evolution operator
\begin{equation}\label{eq:Z22}
S=\dfrac{1}{3}\left (\begin{array}{cccc}
1 & 1 & 1 & 0 \\ 
1 & 1 & 0 & 1 \\ 
1 & 0 & 1 & 1 \\ 
0 & 1 & 1 & 1
\end{array} \right ).
\end{equation}
The eigenvalues are $\lambda_{1}=1$, $\lambda_{2}=\lambda_{3}=1/3$, $\lambda_{4}=-1/3$. A sequence $S^{P}$ projects for large $P$ onto the unique equipartition state $\tilde{q}\sim (1,1,1,1)$, amounting to a complete loss of boundary information. The interaction \eqref{eq:63} with $\gamma=2$, $\delta=0$, $\Delta=10$ realizes a block diagonal step evolution operator, with blocks given by eq.~\eqref{eq:Z22}.

\subsection{Preparation of quantum conditions}

For a quantum subsystem the associated classical statistical probability distribution $\lbrace p_{\tau}\rbrace$ has to obey the quantum conditions. The general solution of the quantum constraints are genuinely probabilistic, i.e. some probabilities $p_{\tau}$ differ from one or zero. The question arises how to realize in practice a genuinely probabilistic distribution $p_{\tau}(t)$. One possibility involves the use of many different initial conditions with occurrence determined by random numbers, such that appropriate weights determine the wanted expectation values for the initial quantum density matrix $\rho(t_{in})$. Another, perhaps more interesting, possibility enforces a single distribution $\tau_{in}$ of Ising spins at $t_{in}$. At $t_{in}$ the probability distribution obeys then $p_{\tau_{in}}(t_{in})=1$, $p_{\tau\neq \tau_{in}}(t_{in})=0$. For $t>t_{in}$ further inside the bulk a genuinely probabilistic $\lbrace p_{\tau}(t)\rbrace$ may then be achieved by the probabilistic nature of the information transport encoded in the step evolution operator $S$. This requires $S$ to be different from a unique jump operator.

For a given computational purpose one may ``initialize'' the system by preparing at $\bar{t}_{in}>t_{in}$ a quantum density matrix $\rho(\bar{t}_{in})$ by a sequence of step evolution operators $S_{in}=S(\bar{t}_{in}-\epsilon)\ldots S(t_{in}+\epsilon)S(t_{in})$, starting from a fixed probability distribution $\lbrace p_{\tau}(t_{in})\rbrace$. The latter may reflect a single spin configuration at $t_{in}$. The role of $S_{in}$ can be twofold. First, it may enforce that $\lbrace p_{\tau}(\bar{t}_{in})\rbrace$ obeys the quantum conditions even if this does not hold for $\lbrace p_{\tau}(t_{in})\rbrace$. Second, it may realize a genuinely probabilistic distribution $\lbrace p_{\tau}(\bar{t}_{in})\rbrace$ even if $\lbrace p_{\tau}({t}_{in})\rbrace$ corresponds to a single spin configuration $\tau_{in}$. The second part is automatic whenever $S_{in}$ is not a chain of unique jump operators. For the first role we present next a simple example.

One may employ the projector $S_{P}$ in eq.~\eqref{eq:Z13} for ``preparing'' a quantum state at $t+\epsilon$ even if $\lbrace p_{\tau}({t})\rbrace$ does not obey the quantum conditions. Consider three Ising spins and the $8\times 8$ matrices
\begin{equation}\label{eq:Z18}
S(t+\epsilon)=S(t)=\hat{S}_{P}\com \hat{S}_{P}=\left (\begin{array}{cc}
S_{P} & 0 \\ 
0 & S_{P}
\end{array} \right ).
\end{equation}
One obtains $\tilde{q}_{3}(t+\epsilon)=\tilde{q}_{2}(t+\epsilon)$, $\tilde{q}_{4}(t+\epsilon)=\tilde{q}_{1}(t+\epsilon)$, $\tilde{q}_{7}(t+\epsilon)=\tilde{q}_{6}(t+\epsilon)$, $\tilde{q}_{8}(t+\epsilon)=\tilde{q}_{5}(t+\epsilon)$, and similar relations for $\bar{q}_{\tau}(t+\epsilon)$. This results in relations for the local probabilities
\begin{align}\label{eq:Z19}
&p_{3}(t+\epsilon)=p_{2}(\te)\com p_{4}(t+\epsilon)=p_{1}(\te)\, ,\nn\\
& p_{7}(t+\epsilon)=p_{6}(\te)\com p_{8}(t+\epsilon)=p_{5}(\te)\, .
\end{align}
Using the numbering of states after eq.~\eqref{eq:2}, and the quantum density matrix defined by eqs~\eqref{eq:4}, \eqref{eq:3}, this yields
\begin{align}\label{eq:Z20}
&\rho_{1}(\te)=p_{+}(\te)-p_{-}(\te)\, ,\nn\\
& p_{+}=p_{5}+p_{6}+p_{7}+p_{8}\com p_{-}=p_{1}+p_{2}+p_{3}+p_{4}\, ,
\end{align}
and
\begin{equation}
\rho_{2}(\te)=\rho_{3}(\te)=0.
\end{equation}
The quantum condition $\rho_{z}\rho_{z}\leq 1$ is obeyed since $0\leq p_{+}\leq 1$, $0\leq p_{-}\leq 1$ implies
\begin{equation}\label{eq:Z21}
|\rho_{1}(\te)|\leq 1 \, .
\end{equation}
This holds for arbitrary boundary conditions. A pure state is realized for $\tilde{q}_{1}(t)=\tilde{q}_{2}(t)=\tilde{q}_{3}(t)=\tilde{q}_{4}(t)=0$ or $\tilde{q}_{5}(t)=\tilde{q}_{6}(t)=\tilde{q}_{7}(t)=\tilde{q}_{8}(t)=0$, such that either $p_{+}(\te)$ or $p_{-}(\te)$ equals one. Other quantum states, e.g. with $|\rho_{3}|\leq 1$ or $|\rho_{2}|\leq 1$, can be prepared by similar projection operators.

As an example, we may start at $t_{in}$ with a fixed spin configuration $s_{1}(t_{in})=1$, $s_{2}(t_{in})=s_{3}(t_{in})=-1$, or $p_{5}(t_{in})=1$, $p_{\tau\neq 5}(t_{in})=0$, e.g. $\tau_{in}=5$. According to eq.~\eqref{eq:5A} one has $\rho_{1}(t_{in})=1$, $\rho_{2}(t_{in})=\rho_{3}(t_{in})=-1$. At $t_{in}$ one has $\rho_{z}\rho_{z}=3$, and the quantum condition \eqref{eq:7} for a single quantum spin is violated. In the next step, at $\bar{t}_{in}=t_{in}+\epsilon$, the probabilities $p_{\tau}^{\prime}=p_{\tau}(t_{in}+\epsilon)$ read
\begin{equation}\label{eq:93AA} 
p^{\prime}_{6}=p^{\prime}_{7}=\dfrac{1}{2}\com p^{\prime}_{1}=p^{\prime}_{2}=p^{\prime}_{3}=p^{\prime}_{4}=p^{\prime}_{5}=p^{\prime}_{8}=0\, .
\end{equation}
This implies $\rho_{2}(\bar{t}_{in})=\rho_{3}(\bar{t}_{in})=0$, $\rho_{1}(\bar{t}_{in})=1$. The quantum condition \eqref{eq:7} is obeyed at $\bar{t}_{in}$. Thus, $S_{in}=S(t_{in})=\bar{S}_{P}$ ``prepares a pure quantum state, starting at $t_{in}$ from a classical probability distribution that does not describe a quantum subsystem. The role of $\bar{S}_{P}$ is easy to understand. It does not change the value of $s_{1}$. If $s_{2}$ and $s_{3}$ have the same sign, it produces equal probabilities for the configuration $\ket{\uparrow\uparrow}$ and $\ket{\downarrow\downarrow}$ for $s_{2}^{\prime}$ and $s_{3}^{\prime}$ at $\te$, independently of the relative weight for $\ket{\uparrow\uparrow}$ and $\ket{\downarrow\downarrow}$ for $s_{2}$ and $s_{3}$ at $t$. Similarly, if the sign of $s_{2}$ and $s_{3}$ is opposite at $t$, it produces at $\te$ equal probabilities for $\ket{\uparrow\downarrow}$ and $\ket{\downarrow\uparrow}$. In short, $\bar{S}_{P}$ erases the memory of the difference between the states $\ket{\uparrow\uparrow}$ and $\ket{\downarrow\downarrow}$, as well as $\ket{\uparrow\downarrow}$ and $\ket{\downarrow\uparrow}$, for $s_{2}$ and $s_{3}$. This loss of memory prepares the quantum state, with the associated uncertainty relations appropriate for a single quantum spin. We observe that $\bar{S}_{P}$ maps an arbitrary fixed spin configuration at $t$ to a pure quantum state for the quantum subsystem at $\te$.

Replacing $S_{P}$ by $S(u)$ in eq.~\eqref{eq:Z16}, with $u$ not too far away from $1/2$, results after a few steps in a similar loss of memory and (approximate) preparation of a quantum subsystem. Such an approximate preparation of a quantum subsystem may be used by nature for quantum computations by neurons, or by artificial neuronal networks. ``Learning'' may enhance the capabilities of producing quantum subsystems and performing unitary operations if parameters in the step evolution operators can be adapted according to success or failure of trials.

\subsection{Bit-quantum map for density matrices}

For general step evolution operators the information in the classical density matrix $\rho^{\prime}(t)$ is needed for a computation of the quantum density matrix at $\te$. If the off-diagonal elements of $\rho^{\prime}(t)$ play a role for the computation of the probabilities $p_{\tau}(\te)$, the formulation of the bit-quantum map should be based on the classical density matrix $\rho^{\prime}$, rather than invoking only the local classical probabilities $p_{\tau}$.

The parameters $\rho_{z}$, which parametrize the quantum density matrix, can be computed from the classical density matrix as
\begin{equation}\label{eq:WA}
\rho_{z}(t)=\tr\left (A^{\prime}_{(z)}\rho^{\prime}(t)\right )\, ,
\end{equation}
with $A^{\prime}_{(z)}$ appropriate ``classical'' operators. For the example of a single qubit realized by three classical Ising spins, the three operators $A^{\prime}_{(z)}$ are diagonal, 
\begin{align}\label{eq:WA1}
A^{\prime}_{(1)} &= \diag(-1,-1,-1,-1,\,1\, ,\,1\, ,\,1\, , \,1\,)\, , \notag \\
A^{\prime}_{(2)} &= \diag(-1,-1,\,1\, ,\, 1\, ,-1,-1,\,1\, ,\, 1\, )\, , \notag \\
 A^{\prime}_{(3)} &= \diag(-1,\, 1\, ,-1, \,1\, ,-1, \,1\, ,-1, \,1\,)\, ,
\end{align}
 where we use the numbering of sect.~\ref{sec:Quantum_jump_in_classical_statistical_systems} for the eight classical states $\tau$. The trace sums over $\tau$.

We define a linear bit-quantum map from the classical density matrix $\rho^{\prime}$ to the quantum density matrix $\rho$ by
\begin{equation}\label{eq:WC1}
\rho=2^{-Q}\left [1+L_{z}\, \tr\left (A^{\prime}_{(z)}\rho^{\prime}\right )\right ]=\rho(\rho^{\prime}) \, .
\end{equation}
Eq.~\eqref{eq:WC1} defines the map $\rho^{\prime}=f(\rho)$ shown in Fig.~\ref{fig:1}. If this map is complete, arbitrary unitary transformations of $\rho$ can be realized if one admits arbitrary transformations of $\rho^{\prime}$, as easily visible from Fig.~\ref{fig:1}. One the other hand, if the transformations of $\rho^{\prime}$ are restricted by eq.~\eqref{eq:Z1}, the possible realizations of unitary quantum transformations are limited by the properties of $S$.

For $M$ classical bits and $Q$ qubits this is a linear map from real $2^{M}\times 2^{M}$-matrices to hermitian $2^{Q}\times 2^{Q}$-matrices,
\begin{equation}\label{eq:WCA}
\rho(\alpha\rho_{1}^{\prime}+\beta\rho^{\prime}_{2})=\alpha\, \rho(\rho^{\prime}_{1})+\beta\, \rho(\rho^{\prime}_{2})\, .
\end{equation}
If $\rho_{1}^{\prime}$ is some ``classical realization'' of the quantum density matrix $\rho_{1}$, $\rho(\rho^{\prime}_{1})=\rho_{1}$, and similarly $\rho(\rho^{\prime}_{2})=\rho_{2}$, a classical realization for the linear combination $\alpha\rho_{1}+\beta\rho_{2}$ is provided by $\alpha\rho_{1}^{\prime}+\beta\rho_{2}^{\prime}$. We recall that the map $\rho(\rho^{\prime})$ is not invertible - many classical realizations of a given quantum density matrix exist. In appendix \ref{app:CNOT-gate_in_probabilistic_computing} we briefly discuss some consequences of the linearity of the bit-quantum map for the realization of unitary quantum operations by suitable step evolution operators.

While the map \eqref{eq:WC1} can be defined for arbitrary $\rho^{\prime}$, a positive quantum density matrix is obtained only for those $\rho^{\prime}$ for which the quantum condition holds. If $\rho_{1}$ and $\rho_{2}$ obey the quantum constraint, the linear combination $\alpha\rho_{1}+\beta\rho_{2}$ obeys the quantum constraint only under restrictions for the coefficients, e.g. $0\leq \alpha\leq 1$, $\beta=1-\alpha$. 

Incidentally, the quantum subsystem is an example for a subsystem in classical statistics that is not obtained by simply integrating out some of the states $\tau$. The construction \eqref{eq:WC1} can be extended to subsystems of quantum systems. Replacing $\rho^{\prime}$ by a density matrix in quantum mechanics constitutes an example for a subsystem in quantum mechanics that is not obtained by a partial trace.

The definition \eqref{eq:WA} yields for the quantum density matrix at $\te$
\begin{align}\label{eq:WB}
\rho_{z}(\te) &= \tr \left( A^{\prime}_{(z)}\, \rho^{\prime}(\te) \right) = 
\tr \left( A^{\prime}_{(z)}\, S(t)\, \rho^{\prime}(t)\, S^{-1}(t) \right) \nn\\
&= \tr \left( S^{-1}(t)\, A^{\prime}_{(z)}\, S(t)\, \rho^{\prime}(t) \right) =
\tr \left( B^{\prime}_{(z)}\rho^{\prime} \right) \, .
\end{align}
Here we have used eq.~\eqref{eq:Z1} and
\begin{equation}\label{eq:WC}
B^{\prime}_{(z)}=S^{-1}(t)\, A^{\prime}_{(z)}\, S(t)\, .
\end{equation}
In general, the operators $B^{\prime}_{(z)}$ are not diagonal even for diagonal $A^{\prime}_{(z)}$, such that off-diagonal elements of the classical density matrix $\rho^{\prime}(t)$ are needed for a determination of $\rho_{z}(\te)$. This clearly demonstrates the need to go beyond the local probabilities $p_{\tau}(t)$ and underlines the key role of the classical density matrix.

\section{Probabilistic computing}\label{sec:Probabilistic_Computing} 

In this section we propose a general formalism for probabilistic computing. It is based on the concept of the classical density matrix. This formalism goes beyond the usual procedure of performing different possible deterministic operations with certain probabilities, e.g. realized by random number generators, as described by the random operations or Markov chains in eq.~\eqref{eq:PCB}. Deterministic computing and quantum computing correspond to particular limits of our formalism. More general cases may provide models for important aspects of neural networks, neuromorphic computing or computing by biological processes as for the brain. A first central ingredient for the formalism are sequences of computational steps, either in time or in space or in some abstract space as layers of neural networks. The second central ingredient is a description of individual computational steps by a linear equation for the classical density matrix which contains the probabilistic information. Systems with these ingredients cover a large space of computational possibilities. They are not the most general ones, however. Non-linear transformations of the classical density matrix are not included in the approach of this section.

\subsection{Formalism for probabilistic computing}

The basic quantity for our formalism of probabilistic computing is the classical density matrix $\rho^{\prime}(t)$. It contains the probabilistic information available at a given value of the variable $t$ which orders the sequence of computational steps. The initial information is encoded in $\rho^{\prime}(t_{in})$, and we label the steps by $t=t_{in}+n\epsilon$ with positive integers $n$. We may set $\epsilon=1$, or equal to a basic time interval or position spacing if we want to use other units for $t$. A computational step maps $\rho^{\prime}(t)$ to $\rho^{\prime}(\te)$. After a certain number of steps the probabilistic information is read out from $\rho^{\prime}(t)$ at some particular $t$. We work with $M$ classical bits or Ising spins at each location $t$. The classical density matrix is then a real $2^{M}\times 2^{M}$-matrix, with matrix elements $\rho^{\prime}_{\tau\rho}$, $\tau$, $\rho=1,\ldots ,2^{M}$. We associate the basis states $\tau$ with the $2^{M}$ configurations of Ising spins, but the use of some other basis is possible as well \cite{CWQF}. The local probabilities $p_{\tau}(t)$ are given by the diagonal elements $\rho^{\prime}_{\tau\tau}(t)$,
\begin{equation}\label{eq:P1} 
p_{\tau}(t)=\rho^{\prime}_{\tau\tau}(t)\geq 0\com \tr\left( \rho^{\prime}(t)\right) =
\sum_{\tau}\, p_{\tau}(t)=1 \, .
\end{equation}
The positivity of the diagonal elements and the normalization \eqref{eq:P1} have to hold for all $t$. We have seen in the preceding section how the classical density matrix arises naturally in generalized Ising models. Using the occupation number basis for the generalized Ising models all elements of $\rho^{\prime}(t)$ obey $\rho^{\prime}_{\tau\rho}\geq 0$. In this section we take a more general approach and only require the property \eqref{eq:P1}.

A computational step is a linear map from $\rho^{\prime}(t)$ to $\rho^{\prime}(\te)$ as specified by the step evolution operator $S(t)$,
\begin{equation}\label{eq:P2} 
\rho^{\prime}(\te)=S(t)\rho^{\prime}(t)\, S^{-1}(t)\, .
\end{equation}
We consider here regular matrices $S(t)$, taking singular matrices as limiting cases. Different computational steps or ``gates'' are realized by different $S(t)$. For convenience, the step evolution operator is normalized such that the leading eigenvalues obey $|\lambda_{i}|=1$. (The overall multiplicative normalization of $S$ drops out in eq.~\eqref{eq:P2}.) We observe that $\tr(\rho^{\prime}(t))=1$ is preserved for arbitrary regular $S$, while the conditions $\rho^{\prime}_{\tau\tau}(t)\geq 0$, $\rho^{\prime}_{\tau\tau}(\te)\geq 0$ impose restrictions on the possible form of $S$. These restrictions guarantee that a local probability distribution $p_{\tau}(t)$ is mapped to another local probability distribution $p_{\tau}(\te)$, as characteristic for a typical step in probabilistic computing. For generalized Ising models in the occupation number basis the evolution equation \eqref{eq:P2} arises naturally, with $S(t)$ a nonnegative matrix. We will be here more general and admit arbitrary linear maps of probability distributions \eqref{eq:P2}, only requiring compatibility of $S(t)$ with the condition \eqref{eq:P1} for all $t$.
 
While generalized Ising models constitute simple realizations of probabilistic computing, there is no need to assume this setting for the general case. For example, there is no good reason why in a neural network the transport of information from one layer to the next should follow the restrictions of a generalized Ising models. In particular, for time dependent processes many forms of maps of probability distributions can be imagined. 

Within the general class of step evolution operators that respect the condition \eqref{eq:P1} we can identify simple limiting cases. Deterministic computing is realized if $S(t)$ are unique jump operators. All unique jump operators are compatible with eq.~\eqref{eq:P1}. We also may consider the particular case where $\rho^{\prime}(t)$ is a positive matrix (all eigenvalues $\lambda\geq 0$). The eigenvalues of $\rho^{\prime}(t)$ are not changed by the evolution \eqref{eq:P2}, such that the condition \eqref{eq:P1} is maintained for arbitrary regular $S(t)$. The special case of orthogonal $S(t)$, $S^{T}S=1$, realizes generalized quantum computing. The familiar unitary transformations correspond to a unitary subgroup of $SO(2^{M})$ which is specified by some appropriate complex structure. Unique jump operators are orthogonal. Approximate unique jump operators can account for errors in a deterministic computation. In general, they will not be orthogonal anymore.

For a general step evolution operator $S(t)$ the map of the probabilistic information $\rho^{\prime}(t)\raw\rho^{\prime}(\te)$ cannot be expressed as a map of local probabilities $\lbrace p_{\tau}(t)\rbrace\raw\lbrace p_{\tau}(\te)\rbrace$. Let two different $\rho^{\prime}_{1}(t)$ and $\rho^{\prime}_{2}(t)$, which have the same diagonal elements $p_{\tau}(t)=\rho^{\prime}_{1\,\tau\tau}(t)=\rho^{\prime}_{2\,\tau\tau}(t)$, be mapped by eq.~\eqref{eq:P2} to $\rho_{1}^{\prime}(\te)$ and $\rho_{2}^{\prime}(\te)$, respectively. The diagonal elements of $\rho_{1}^{\prime}(\te)$ can differ from the diagonal elements of $\rho_{2}^{\prime}(\te)$. In general, the off-diagonal elements of $\rho^{\prime}(t)$ matter for the computation of the diagonal elements of $\rho^{\prime}(\te)$, excluding the existence of a map of local probabilities. Maps of local probabilities require step evolution operators for which the diagonal elements of $\rho^{\prime}(\te)$ only depend on the diagonal elements of $\rho^{\prime}(t)$. This is realized for (no index sums here)
\begin{equation}\label{eq:NP1} 
S_{\tau\rho}(t)S^{-1}_{\sigma\tau}(t)=W_{\tau\rho}(t)\delta_{\rho\sigma}\, ,
\end{equation}
such that
\begin{equation}\label{eq:NP2} 
p_{\tau}(\te)=W_{\tau\rho}(t)p_{\rho}(t)\, .
\end{equation}
The condition \eqref{eq:NP1} is sufficient for implementing the map \eqref{eq:NP2} for arbitrary $p_{\rho}(t)$. Summing eq.~\eqref{eq:NP1} over $\tau$ yields
\begin{equation}\label{eq:NP3} 
\sum_{\tau}\, W_{\tau\rho}=1\, ,
\end{equation}
which, in turn, ensures the normalization $\sum_{\tau}\, p_{\tau}(\te)~=~1$. All unique jump operators obey eq.~\eqref{eq:NP1}. For general step evolution operators $S(t)$ the matrices $A^{(\tau)}$, which have for fixed $\tau$ the elements
\begin{equation}\label{eq:NP4} 
A^{(\tau)}_{\sigma\rho}=S_{\tau\rho}S^{-1}_{\sigma\tau}\, ,
\end{equation}
are not diagonal. Then
\begin{equation}\label{eq:NP5} 
p_{\tau}(\te)=\rho^{\prime}_{\tau\tau}(\te)=\tr\left (A^{(\tau)}\rho^{\prime}(t)\right )
\end{equation}
involves the off-diagonal elements of $\rho^{\prime}(t)$ and cannot be written in the form \eqref{eq:PCB}.

\subsection{Subsystems}

One often encounters situations where the information processing in subsystems becomes largely independent of the ``environment''. Only partial information from $\rho^{\prime}(t)$ is used for the subsystem. This is typical for the computing in biological systems. Having as input a very large number of pixels from ``pictures taken by the eye'', the information used for decision making has to be highly concentrated. We model this concentration process by the density matrix $\rho$ for subsystems, generalizing the bit-quantum map \eqref{eq:WC1}. The possible density matrices for the subsystem are spanned by traceless $R\times R$-matrices $L_{z}$,
\begin{equation}\label{eq:P3} 
\rho=R^{-1}\left (1+\rho_{z}L_{z}\right ) \, ,
\end{equation}
where the set of $L_{z}$ is assumed to be complete in the sense that every density matrix of the subsystem can be written in the form \eqref{eq:P3} for suitable real coefficients $\rho_{z}$. (For quantum subsystems one has $R=2^{Q}$ and $L_{z}$ are the hermitian generators of $SU(2^{Q})$.) We require that for all $z$ the diagonal elements of $\rho$ can be associated with probabilities in the subsystem,
\begin{equation}\label{eq:P4} 
\bar{p}_{\alpha}(t)=\rho_{\alpha\alpha}(t)\geq 0\com\tr(\rho(t))=1\, .
\end{equation}

The map from $\rho^{\prime}(t)$ to $\rho(t)$ associates the coefficients $\rho_{z}(t)$ to expectation values of suitable classical operators $A^{\prime}_{(z)}$,
\begin{equation}\label{eq:P5}
\rho_{z}(t)=\tr\left \{A^{\prime}_{(z)}\rho^{\prime}(t)\right \}.
\end{equation}
The set of these expectation values specifies the subsystem. For a given set $A^{\prime}_{(z)}$ the condition \eqref{eq:P4} imposes constraints on the classical density matrices $\rho^{\prime}(t)$ that are compatible with the subsystem. They correspond to the quantum conditions discussed previously, but are weaker if we do not demand that $\rho$ is a positive matrix. It is sometimes convenient to extend the set of $L_{z}$, $\bar{L}_{z}=L_{z}/R$ for $z\neq 0$, $L_{0}=1/R$. With $\tr(\rho^{\prime})=1$, eq.~\eqref{eq:P3} can be written equivalently as
\begin{equation}\label{eq:P6}
\rho=\bar{\rho}_{z}\bar{L}_{z}\, ,
\end{equation}
where the sum includes now $z=0$, with $\bar{\rho}_{0}=1$, $\bar{\rho}_{z}=\rho_{z}$ for $z\neq 0$.

The construction \eqref{eq:P3}, \eqref{eq:P5} defines a wide class of possible subsystems. The often encountered partial traces of $\rho^{\prime}$ are a particular case. Partial tracing is possible if the index $\tau$ can be written as a double index, $\tau=(\alpha,\lambda)$, $\rho=(\beta,\sigma)$. A possible subsystem is defined by   a partial trace
\begin{equation}\label{eq:P7} 
\rho_{\alpha\beta}=\sum_{\sigma}\, \rho^{\prime}_{\alpha\sigma ,\beta\sigma}\, .
\end{equation}
This subsystem corresponds in eq.~\eqref{eq:P6} to
\begin{equation}\label{eq:P8} 
\bar{L}_{z}= \bar{L}_{ab}\com (\bar{L}_{ab})_{\alpha\beta}=\delta_{a\alpha}\delta_{b\beta}\, ,
\end{equation}
and
\begin{equation}\label{eq:P9} 
 (\bar{A}_{(ab)}^{\prime})_{\alpha\lambda,\beta\sigma}=\delta_{a\beta}\delta_{b\alpha}\delta_{\lambda\sigma}\, ,
\end{equation}
where the $R^{2}$ generators $\bar{L}_{z}$ are labeled by $\bar{A}_{ab}$, and similar for $\bar{A}^{\prime}_{(z)}$. One infers
\begin{align}\label{eq:P10} 
\bar{\rho}_{ab}&=\tr\left (\bar{A}^{\prime}_{(ab)}\rho^{\prime}\right )=\delta_{a\beta}\delta_{b\alpha}\delta_{\lambda\sigma}\, (\rho^{\prime})_{\beta\sigma ,\alpha\lambda}\nn\\
&=\rho^{\prime}_{a\sigma ,b\sigma}\, ,
\end{align}
such that eq.~\eqref{eq:P6} yields indeed the subtrace \eqref{eq:P7}. Subtraces correspond to summations over ``unobserved'' spin configurations. Our setting for subsystems is more general.

The step evolution transfers from $\rho^{\prime}$ to $\rho$, 
\begin{align}
\rho(\te)&=R^{-1}\left [1+L_{z}\, \tr\left \{A^{\prime}_{(z)}(t)\rho^{\prime}(\te)\right \}\right ]\nn\\
&=R^{-1}\left [1+L_{z}\, \tr\left \{A^{\prime}_{(z)}(t)S(t)\rho^{\prime}(t)S^{-1}(t)\right \}\right ]\nn\\
\label{eq:P11} &=R^{-1}\left [1+L_{z}\, \tr\left \{B^{\prime}_{(z)}(t)\rho^{\prime}(t)\right \}\right ]\\
&=R^{-1}\left [1+\rho_{z}(\te) L_{z}\right ]\, ,
\end{align}
with
\begin{equation}\label{eq:P12} 
B_{(z)}^{\prime}(t)=S^{-1}(t)A^{\prime}_{(z)}(t)\, S(t)\, ,
\end{equation}
and
\begin{equation}\label{eq:P13} 
\rho_{z}(\te)=\tr\left \{B^{\prime}_{(z)}(t)\, \rho^{\prime}(t)\right \}=\exval{B^{\prime}_{(z)}(t)}\, .
\end{equation}

The subsystem is closed, in the sense that its evolution needs no information from the environment, if $\rho_{z}(\te)$ can be expressed in terms of the coefficients $\rho_{z}(t)$. Step evolution operators compatible with a closed subsystem permit to express $\exval{B^{\prime}_{(z)}(t)}$ in terms of $\exval{A^{\prime}_{(z)}(t)}$. We may assume that this relation is linear
\begin{equation}\label{eq:P14} 
\exval{B^{\prime}_{(z)}(t)}=\sum_{y}\, c_{zy}(t)\exval{A^{\prime}_{(y)}(t)}\, .
\end{equation}
This translates directly to
\begin{equation}\label{eq:P15} 
\rho_{z}(\te)=c_{zy}(t)\, \rho_{y}(t)\, .
\end{equation}
If, furthermore, a regular matrix $D(t)$ exists with
\begin{equation}\label{eq:P16} 
L_{z}c_{zy}(t)=D(t)L_{y}D^{-1}(t)\, ,
\end{equation}
one can infer for the subsystem
\begin{equation}\label{eq:P17} 
\rho(\te)=D(t)\rho(t)D^{-1}(t)\, .
\end{equation}
Without loss of generality we can normalize $D(t)$ such that its leading eigenvalues are normalized as $|\lambda_{i}|=1$. In summary, this procedure maps the step evolution operator $S(t)$ to a step evolution operator $D(t)$ for the subsystem. The quantum evolution described in the preceding section is a special case where the $L_{z}$ are hermitian and $D(t)=U(t)$ are unitary. More generally, subsystems for which the absolute value of all eigenvalues of $D(t)$ equals one or is close to one are of particular interest. The information in these subsystems can be transported without loss.

We observe that the step evolution operators $D(t)$ for the subsystem need not to have the same properties as $S(t)$. For example, a nonnegative matrix $S(t)$ can be mapped to a real matrix $D(t)$ that has also negative elements. We have seen this in the previous sections for the quantum subsystems. The hermitian $2^{Q}\times 2^{Q}$-matrix $\rho$ can be written as a real symmetric $2^{Q+1}\times 2^{Q+1}$-matrix by combining the real and imaginary part of the complex $2^{Q}$-component wave function into a real $2^{Q+1}$-component vector, and accordingly for $\rho$. The unitary evolution matrix becomes in this basis an orthogonal $2^{Q+1}\times 2^{Q+1}$-matrix $D$. It has no definite sign for its elements. This is a simple way of avoiding in practice the constraints from the nonnegative character of $S$ in generalized Ising models. 

A wide class of models with interesting subsystems can be constructed along these lines. In the preceding section we have seen in a simple example that suitable step evolution operators can dynamically implement conditions for a closed subsystem, even if for initial steps of the evolution the subsystem is not closed, cf. eqs~\eqref{eq:Z18}-\eqref{eq:Z21}. The concentration of relevant information could proceed by a sequence of dynamically realized subsystems with decreasing dimension, following maps $S\raw D_{1}\raw D_{2}$ etc. This could reflect different layers of computing in neural networks. It seems likely that the subsystems are not obtained by simply omitting part of the configurations as for the subtracing procedure. Our general setting for subsystems permits to reshuffle the information at every concentration step. The learning of a neural network could be associated to the learning which observables $A^{\prime}_{(z)}$ define useful subsystems at different stages of the concentration of the relevant information.

We will not explore here further all the rich properties of general subsystems. In appendix \ref{app:CNOT-gate_in_probabilistic_computing} we discuss as a concrete example the two qubit quantum system realized by six classical bits with $\rho_{kl}$ given by correlation functions. We ask what are the properties of $S$ needed to realize the CNOT-gate for this subsystem.

\section{Quantum computing}\label{sec:Quantum_Computing}

The possibilities for universal quantum computing depend on the question which transformations of the classical statistical information can be employed. If arbitrary non-linear transformations of the probability distribution or arbitrary step evolution operators are available, universal quantum computing can be achieved for an arbitrary number of qubits by a finite number of classical bits. If we restrict ourselves to linear stochastic changes of probability distributions, positive step evolution operators or deterministic classical bit operations, the realization of universal quantum computation needs an infinite number of classical Ising spins. We demonstrate in the next section how this infinite number of macroscopic yes/no decisions can be realized by a finite number of real variables. If we want to associate a classical observable to each quantum observable with eigenvalues $\pm 1$ (e.g. quantum spins in every direction), such that the possible measurement values and expectation values of the quantum and classical observables coincide, one needs anyhow an infinite number of two-level observables. We also realize this aspect in the next section.

For a finite number of Ising spins only restricted classes of quantum operations can be realized. This is somewhat similar to the situation for Majorana wires \cite{AOR,SFN,AAHE}. The present section is devoted to the case of a finite number of classical bits, arguing that even for a restricted set of quantum operations rather rich new computational possibilities may open up.

The operations of universal quantum computing can be constructed from three elementary gates: the Hadamard gate, the CNOT gate and the $\pi/4$-gate. More precisely, there exist efficient approximations of any unitary operation in terms of products of such gates \cite{SLL,MNI}. The $\pi/4$-gate or $T$-gate (often called $\pi/8$-rotation) is represented by a unitary transformation of a single spin
\begin{equation}\label{eq:QC1} 
U_{T}=\left (\begin{array}{cc}
1 & 0 \\ 
0 & e^{\frac{i\pi}{4}}
\end{array} \right )\, ,
\end{equation}
according to $\rho^{\prime}=U\rho\,  U^{\dagger}$,
\begin{equation}\label{eq:QC2} 
\rho_{1}^{\prime}=\dfrac{1}{\sqrt{2}}(\rho_{1}-\rho_{2})\com \rho_{2}^{\prime}=\dfrac{1}{\sqrt{2}}(\rho_{1}+\rho_{2})\com \rho_{3}^{\prime}=\rho_{3}\, .
\end{equation}
In quantum mechanics it corresponds to a rotation of the spin in the $1$-$2$-plane by an angle $\pi/4$. It is the square root of the transformation $U_{Z}U_{12}$, $U_{T}^{2}=U_{Z}U_{12}$ and has period eight, $U^{8}_{T}=1$.

Following the discussion in sect.~\ref{sec:V}, the map of classical probabilities realizing the $T$-gate cannot be realized by random operations \eqref{eq:PCB} with positive transition probabilities \eqref{eq:PCA}. We can, however, perform the map by a linear transformation obeying eqs~\eqref{eq:40A}-\eqref{eq:40C}, if we admit some of the coefficients $W_{\tau\rho}$ to be negative. Let us define the linear combinations of the probabilities $p_{\tau}$ defined in sect.~\ref{sec:Quantum_jump_in_classical_statistical_systems} by
\begin{equation}\label{eq:L6A} 
p_{1278}=p_{1}+p_{2}-p_{7}-p_{8}\com p_{3456}=p_{3}+p_{4}-p_{5}-p_{6}\, ,
\end{equation}
with
\begin{equation}\label{eq:L6B} 
\rho_{1}=-p_{1278}-p_{3456}\com \rho_{2}=-p_{1278}+p_{3456}\, .
\end{equation}
The $\pi/4$-rotation \eqref{eq:QC2} is realized by
\begin{align}\label{eq:L6C} 
p_{1278}^{\prime} &= \dfrac{1}{\sqrt{2}}\left (p_{1278}+p_{3456}\right )\com \notag \\
p_{3456}^{\prime} &= \dfrac{1}{\sqrt{2}}\left (-p_{1278}+p_{3456}\right )\, ,
\end{align}
provided the linear combination $\rho_{3}^{\prime}=\rho_{3}$ in eq.~\eqref{eq:5A} is not changed. For example, we can employ transformations that leave the differences $p_{1}-p_{2}$, $p_{3}-p_{4}$, $p_{5}-p_{6}$, $p_{7}-p_{8}$ invariant. For linear transformations realizing eq.~\eqref{eq:L6C} some of the coefficients $W_{\tau\rho}$ are negative. (Nevertheless, all probabilities remain positive.) In this section we demonstrate how to realize negative coefficients $W_{\tau\rho}$ by adding additional Ising spins for the classical statistical system realizing a single qubit. If one wishes, the additional spins may at the end be integrated out in order to realize an effective system for the three classical bits $s_{k}$.

We will establish in section \ref{sec:Quantum_Mechanics} that all quantum operations can be realized by a static memory material for an infinite number of Ising spins. On the other hand, it is not possible to realize simultaneously the Hadamard gate and the $T$-gate precisely by a generalized Ising model for a finite number of classical bits. This finding is highly interesting from a conceptual point of view. Indeed, one finds that quantum systems for an arbitrary number of qubits can be obtained as subsystems of generalized Ising models with an infinite number of discrete two-level variables. For a practical realization of quantum operations, however, an infinite number of bits is not available and we have to investigate what is possible with a finite number of bits. The issue is somewhat analogous to the description of a continuous classical rotation by bits. An infinite number of bits is needed for the precise description of the location of a point on a circle. As for classical rotations, one is interested in possible approximations by a finite number of bits. We will see that for a single qubit a rather large number of quantum operations can be realized by a moderate number of Ising spins.

We concentrate in this paper on the possible realizations of quantum gates. Other important aspects of quantum computing, as the preparation of the initial state and the read out, are not covered here.

\subsection{Unitary transformations for a single qubit}

Both the Hadamard gate and the $T$-gate are operations for a single qubit. We therefore concentrate first on the possible realizations of a single qubit by $M$ Ising spins. By consecutive multiplications of $U_{H}$ and $U_{T}$ one can generate arbitrary unitary transformations. With $U_{T}^{2}=U_{Z}U_{12}$, $U_{T}^{4}=U_{Z}$, $U_{12}=U_{T}^{6}$, $U_{T}^{2}U_{H}U_{T}^{2}=U_{23}=(1+i\tau_{1})/\sqrt{2}$, $U_{23}^{2}=iU_{X}$ etc., we can construct the transformations \eqref{eq:11A}. New transformations can be obtained by sequences of $\pi/4$-rotations and $\pi/2$-rotations around different axes. Every unitary transformation can be approximated with a given precision by products of $U_{H}$ and $U_{T}$ with a sufficient number of factors. As a consequence, with step evolution operators $S_{H}$ and $S_{T}$ realizing $U_{H}$ and $U_{T}$, one can realize arbitrary unitary transformations by suitable products of $S_{H}$ and $S_{T}$. For static memory materials this is only possible for $M\raw\infty$.

The proof of the argument that $S_{H}$ and $S_{T}$ cannot be realized by finite $M$ relies on the fact that the step evolution operators for $M$ Ising spins are real nonnegative $N\times N$-matrices, $N=2^{M}$. Indeed, all matrix elements obey $S_{\tau\rho}\geq 0$ by virtue of eq.~\eqref{eq:Z11}. Step evolution operators can realize unitary transformations only in the sector of leading eigenvalues $|\lambda_{i}|=1$. For eigenvectors of subleading eigenvalues $|\lambda_{j}|<1$ the norm decreases $\sim|\lambda_{j}|^{P}$ if $S$ is applied $P$-times, contradicting the conservation of the norm for unitary transformations. The possible spectra of leading eigenvalues of nonnegative $N\times N$ matrices are limited by the Perron-Frobenius theorem. Possible leading eigenvalues $\lambda=e^{i\alpha}$ have to obey $\alpha=2\pi k_{i}/h_{i}$, with positive integers $h_{i}$ and $k_{i}$ in the range $0\leq k_{i}<h_{i}$, $\sum_{i}\, h_{i}\leq N$. This describes one or several sets, labeled by $i$, with respective period $h_{i}$. The upper bound on $\sum_{i}\, h_{i}$ simply reflects that the total number of leading eigenvalues cannot exceed $N$. If there is only a single period $h$ one concludes $S^{h}=1$. This generalizes to $S^{\bar{h}}=1$ for finite $\bar{h}$.

If both $S_{H}$ and $S_{T}$ are nonnegative $N\times N$ matrices, also any arbitrary product of factors $S_{H}$ and $S_{T}$ is a nonnegative $N\times N$ matrix with corresponding restriction on the spectrum of leading eigenvalues. One concludes that there exists a finite $\bar{h}$ with $S^{\bar{h}}=1$. This contradicts the realization of arbitrary unitary matrices as products of $S_{H}$ and $S_{T}$. For any finite $\bar{h}$ there exist unitary matrices with $U^{\bar{h}}\neq 1$.

\subsection{Two component single quantum spin}

A finite number $M$ of Ising spins can realize suitable finite subgroups of $U(2)$. Central issues can already be understood for a two-component quantum spin, with quantum spin operators $\tau_{1}$ and $\tau_{3}$. In this case $\rho$ is a real $2\times 2$ matrix, setting $\rho_{2}=0$ in eq.~\eqref{eq:4}. The unitary transformations are now $O(2)$-rotations in the plane of the two components. We focus on the continuous part $SO(2)$. We have already seen how to realize with two Ising spins $s_{1}$ and $s_{3}$ the $\pi/2$-rotations. They can be implemented by the unique jump operation leading to $U_{31}$ in eq.~\eqref{eq:11A}. We next want to realize $\pi/4$-rotations,
\begin{equation}\label{eq:QC3} 
\rho_{1}^{\prime}=\dfrac{1}{\sqrt{2}}(\rho_{1}-\rho_{3})\com 
\rho_{3}^{\prime}=\dfrac{1}{\sqrt{2}}(\rho_{1}+\rho_{3})\, .
\end{equation}
The corresponding unitary matrix $U_{\pi/4}^{(31)}$ obeys $\left (U_{\pi/4}^{(31)}\right )^{2}=U_{31}$, and therefore has period four, $\left (U_{\pi/4}^{(31)}\right )^{4}=1$, as given by
\begin{align}\label{eq:QC4}
U_{\pi/4}^{(31)}&=\left (\begin{array}{cc}
\cos(\pi/4) & \sin(\pi/4) \\ 
-\sin(\pi/4) & \cos(\pi/4)
\end{array} \right )\nn\\
&=\cos(\pi/4)+i\sin(\pi/4)\tau_{2}=e^{i\frac{\pi}{4}\tau_{2}}\, .
\end{align}
For $M=2$ the only matrices $S$ with period four involve necessarily the correlation $s_{1}s_{3}$. It cannot be expressed as a transformation acting only on $\rho_{1}=\exval{s_{1}}$ and $\rho_{3}=\exval{s_{3}}$, and therefore cannot realize $U_{\pi/4}^{(31)}$.

For four Ising spins $(M=4)$ we employ, in addition to $s_{1}$ and $s_{3}$, the spins $s_{13+}$ and $s_{13-}$. For these two Ising spins we require that their expectation values correspond to the diagonal directions of the quantum spin,
\begin{align}\label{eq:QC5}
\rho_{kl\pm}&=\exval{s_{kl\pm}}=\exval{\dfrac{1}{\sqrt{2}}(S_{k}\pm S_{l})}=\dfrac{1}{\sqrt{2}}\tr\left (\rho(\tau_{k}\pm\tau_{l})\right )\nn\\
&=\dfrac{1}{\sqrt{2}}\left (\rho_{k}\pm\rho_{l}\right ).
\end{align}
Together with eq.~\eqref{eq:3} this relates the expectation values of $s_{kl\pm}$ and $s_{k}$,
\begin{equation}\label{eq:QC6}
\exval{s_{kl\pm}}=\dfrac{1}{\sqrt{2}}\left (\exval{s_{k}}\pm \exval{s_{l}}\right ).
\end{equation}
This realization of a single qubit by four Ising spins extends the quantum constraint. Beyond $\exval{s_{1}}^{2}+\exval{s_{3}}^{2}\leq 1$ the probability distributions that can describe the quantum system have also to obey eq.~\eqref{eq:QC6}. 

With four bits obeying the constraint \eqref{eq:QC6} the realization of eq.~\eqref{eq:QC3} and therefore of the $\pi/4$-rotation $U_{\pi/4}^{(31)}$ is a simple unique jump operation
\begin{equation}\label{eq:QC7}
s_{1}\raw s_{13-}\raw -s_{3}\raw -s_{13+}\raw -s_{1}\, .
\end{equation}
The $16\times 16$ matrix $S$ is easily constructed by performing corresponding transformations on the correlation functions. The transformation $S_{31}$, which realizes the $\pi/2$-rotation, acts on $s_{13\pm}$ as
\begin{equation}\label{eq:QC8}
s_{13+}\leftrightarrow -s_{13-}\, .
\end{equation}

The generalization to rotations by smaller angles is straightforward. For the $\pi/8$ rotation one employs four additional Ising spins. They are ``situated'' at intermediate directions in the sense that the quantum constraint identifies their expectation values to the expectation values of the quantum spin in the corresponding directions. This can be continued - the finer the angular resolution, the more Ising spins and the more relations for the quantum constraint are needed. Arbitrary rotations require an infinite number of classical bits.

The whole construction is very parallel to the determination of an angle by yes/no decisions - the finer the angular resolution, the larger the number of yes/no decisions or bits needed. One may associate the Ising spins with yes/no decisions in which half of a circle a ``particle'' is situated, with ``angle of the half circle'' given by the ``direction of the spin''. This has an analogue by the binning of real numbers. Admitting instead of discrete Ising spins continuous real numbers as the classical variables over which a classical probability distribution is defined, one can realize arbitrary rotations of the two-component quantum spin by static memory materials. A simple example is given in appendix \ref{app:Continuous_classical_probability_distributions}.

\subsection{Three component single quantum spin}

A straightforward generalization to a three component quantum spin is not possible. This is related to restrictions limiting the finite subgroups of the rotation group $SO(3)$. For the realization of any $\pi/4$-rotation one needs to define the $\pi/4$-rotations around all three axes in the $1$, $2$ and $3$ directions. This is required by compatibility with the $\pi/2$-rotations around these three axes. If the $\pi/4$-rotation in the $1-3$ plane is defined, we can obtain $\pi/4$-rotations in the other planes by multiplication with suitable $\pi/2$-rotations. We could try to introduce four further Ising spins $s_{12\pm}$ and $s_{23\pm}$, in order to realize the $\pi/4$-rotations in the $1-2$ and $2-3$ planes similar to eq.~\eqref{eq:QC7}. However, in the $1-2$ plane at $\rho_{3}=1/\sqrt{2}$ we have now only two spins with the four directions $s_{13+}$, $-s_{13-}$, $s_{23+}$, $-s_{23-}$. A $\pi/4$-rotation in the $1-2$ plane at the level $\rho_{3}=1/\sqrt{2}$ needs four spins or eight directions. If we add bits at the missing positions, new bits would be needed at other positions in order to realize the other $\pi/4$-rotations. This process does not close. Suitable products of $\pi/4$-rotations around different axes can generate an arbitrary $SO(3)$-rotation with any wanted precision. For static memory materials this needs an infinite number of bits.

For a closed realization of rotations by a finite number of bits these rotations have to form a discrete subgroup of $SO(3)$. If rotations in different planes are included, a maximal set of six bits can realize the symmetry group of the icosahedron, which is the maximal non-abelian discrete subgroup of $SO(3)$. We denote these six spins by $s_{k\pm}$, and associate their expectation value to $\rho_{z}$ as
\begin{align}\label{eq:QC9}
\exval{s_{1\pm}}&=a\rho_{1}\pm b\rho_{3}\com\nn\\
\exval{s_{2\pm}}&=a\rho_{2}\pm b\rho_{1}\com\nn\\
\exval{s_{3\pm}}&=a\rho_{3}\pm b\rho_{2}\, ,
\end{align}
with
\begin{equation}\label{eq:QC10}
a=\left (\dfrac{1+\sqrt{5}}{2\sqrt{5}}\right )^{1/2}\com b=\left (\dfrac{2}{5+\sqrt{5}}\right )^{1/2}\, ,
\end{equation}
such that
\begin{equation}\label{eq:QC11}
b=\dfrac{2a}{1+\sqrt{5}}\com a^{2}+b^{2}=1\, .
\end{equation}

These expectation values are the same as for a quantum spin in the six directions $(a,0,\pm b)$, $(\pm b,a,0)$ and $(0,\pm b,a)$, that correspond to six of the corners of the icosahedron. (The other corners correspond to the negative of the six direction vectors.) The corresponding spin operators $S_{k\pm}$ are
\begin{equation}\label{eq:QC12}
S_{k\pm}=a\tau_{k}\pm b\tilde{\tau}_{k}\, ,
\end{equation}
with $\tilde{\tau}_{3}=\tau_{2}$, $\tilde{\tau}_{2}=\tau_{1}$, $\tilde{\tau}_{1}=\tau_{3}$. The reflections corresponding to $U_{X}$, $U_{Y}$ and $U_{Z}$ in eq.~\eqref{eq:11A} can still be realized by appropriate exchanges of Ising spins. For example, the transformation $U_{Y}$, $\rho_{1}\raw -\rho_{1}$, $\rho_{3}\raw -\rho_{3}$, $\rho_{2}\raw\rho_{2}$, is now realized by
\begin{equation}\label{eq:QC13}
U_{Y}:\, s_{1\pm}\raw -s_{1\pm}\com s_{2+}\leftrightarrow s_{2-}\com s_{3+}\leftrightarrow s_{3-}\, .
\end{equation}
The $\pi/2$-rotations $U_{12}$, $U_{31}$, or the Hadamard gate $U_{H}$, however, can no longer be realized by appropriate step evolution operators. They are replaced by $2\pi/5$-rotations around appropriate axes. This allows for a larger discrete subgroup of the $SO(3)$-rotations. The quantum condition places additional restrictions beyond $\rho_{z}\rho_{z}\leq 1$, since six expectation values are given by three numbers $\rho_{z}$ in eq.~\eqref{eq:QC9}. An example is
\begin{equation}\label{eq:QC14}
\rho_{1}=\dfrac{1}{2a}\left (\exval{s_{1+}}+\exval{s_{1-}}\right )=\dfrac{1}{2b}\left (\exval{s_{2+}}-\exval{s_{2-}}\right )\, ,
\end{equation}
which relates
\begin{equation}\label{eq:QC15}
\exval{s_{2+}}-\exval{s_{2-}}=\dfrac{2}{1+\sqrt{5}}\left( \exval{s_{1+}}+\exval{s_{1-}}\right)\, .
\end{equation}

\subsection{Two quantum spins}

We next ask which unitary operations for two qubits can be realized by a finite number of Ising spins or classical bits. We discuss here the particular question which unitary transformations can be realized by unique jump step evolution operators for a finite number of Ising spins. This is to a large extent an issue of group theory for finite groups. First of all, for a finite number of Ising spins we can only realize a discrete subgroup $D$ of $SU(4)$. The first step therefore consists in a classification of the finite subgroups of $SU(4)$ \cite{HH}. We are interested in discrete subgroups $D$ that contain a suitable subgroup $D_{1}\times D_{2}$, where each factor $D_{i}$ acts on a single qubit. The discrete group $D$ may or may not contain the CNOT-gate.

For classical Ising spins the unique jump operations form the group $P(2^{M})$ of all possible permutations between the $2^{M}$ states $\tau$. If $D$ is realized by unique jump operations, it has to be a subgroup of $P(2^{M})$. (If we want to realize $D$ by simple maps between the $M$ Ising spins, it has to be a subgroup of the much smaller permutation group $P(M)$.) Only part of the permutations $P(2^{M})$ constitute maps within the space of expectation values of spins and correlations employed for the definition of the coefficients $\rho_{z}$. We may call this subgroup $\bar{P}(2^{M})$. In other words, $\bar{P}(2^{M})$ should be a map of the coefficients $\rho_{z}\raw\rho^{\prime}_{z}$ for \emph{every} classical probability distribution that leads to a given set of $\rho_{z}$ obeying the quantum constraint. The determination of $\bar{P}(2^{M})$ does not seem to be a simple task. Obviously the permutations $P(M)$ of the $M$ spins belong to $\bar{P}(2^{M})$, but things get more complicated once maps between expectation values of spins and correlations are involved. The discrete transformation group $D$ has to be simultaneously a subgroup of $SU(4)$ and of $\bar{P}(2^{M})$.

We have not yet investigated the problem which is the minimal number of Ising spins needed to realize a given subgroup $D$ of $SU(4)$ by unique jump operations. On the other hand, for a sufficient number $M$ of classical bits any subgroup $D$ can be realized by static memory materials. One places spins at all points of the geometric figure related by the discrete transformations $D$. Then appropriate maps between spins at those points, belonging to $P(M)$, constitute unique jump operations realizing $D$ for the quantum subsystem. The quantum constraints identify the expectation values of all Ising spins with the expectation values of quantum operators $\hat{A}$, $\hat{A}^{2}=1$, $\tr(\hat{A})=0$, associated to the subgroup $D$ of $SU(4)$ by $D$-transformations of a given selected operator $\hat{A}_{0}$. The whole construction parallels the icosahedron for a single qubit, with $SU(2)$ replaced by $SU(4)$.

\section{Real classical variables}\label{sec:Real_classical_variables}

In this section we consider classical statistical systems for which the variables are real numbers instead of discrete Ising spins. The probability distribution is $p(x) \geq 0$, with $\int_x p(x) =1$, and $x$ denoting a (possibly multi-component) continuous variable. For applications to computing this may play a role for neuromorphic computing where the action potentials of neurons and similar quantities are given by real numbers. Also for traditional computers real numbers can actually be represented with high precision by a moderate number of bits. Any real number admits an infinite number of classical two-level observables or classical bits. They correspond to the infinite number of finer and finer binning, with yes/no decisions if the number is within a bin or outside.

It is rather obvious that arbitrary unitary transformations for $Q$ qubits can be performed by classical statistical systems with a sufficient number of real variables. For the example of a single qubit three classical real variables $x_k$, $k=1, 2, 3$, can clearly realize arbitrary $\rho_k$. The interesting question is rather if all quantum two-level observables can be associated to two-level classical observables, and if this association remains simple enough for practical purposes.

We demonstrate this issue by two examples. The first are Gaussian probability distributions over $x=(x_1, x_2, x_3)$. We can easily define three classical spins $s_k$ such that the quantum  system is determined by $\rho_k = \langle s_k \rangle$. Arbitrary unitary transformations are no problem and can be realized by deterministic unique jump operations. The quantum spins in directions different from the axes find, however, no correspondence in simple classical two-level observables. This shortcoming is remedied in our second example where we construct a family of classical probability distributions such that every quantum two-level observable can be associated with a corresponding classical two-level observable. Only for the second example all properties of quantum mechanics are realized by the classical statistical ensemble.

\subsection{Gaussian probability distribution}

Consider first a Gaussian probability distribution for three real variables $x_k \in \mathbb{R}$, $k = 1,2,3$, $a>0$,
\begin{equation}\label{eq:R1}
p(x) = (a\sqrt{\pi})^{-3} \exp \left\lbrace - \frac{1}{a^2} 
    \sum_k \left( x_k - \bar{x}_k \right)^2 \right\rbrace \, .
\end{equation}
It is normalized,
\begin{equation}\label{eq:R2}
\int \diff^3 x \, p(x) = 1\, ,
\end{equation} 
and depends on four parameters $\bar{x}_k$ and $a$. Three classical Ising spins are defined by
\begin{equation}\label{eq:R3}
s_k = \theta(x_k) - \theta(-x_k)\, .
\end{equation}
The values for these spin-observables are $s_k(x)=1$ for $x_k > 0$ and $s_k(x) = -1$ for $x_k < 0$. They have discrete values $\pm 1$ for all classical states defined by $\{ x_1, \, x_2, \, x_3 \}$ and correspond to yes/no decisions if $x_k$ is positive or not. The classical expectation value of $s_k$ is given by 
\begin{align}\label{eq:R4}
\langle s_k \rangle &= \int \diff^3 x \, p(x)  s_k(x) \notag \\
&= \frac{1}{a \sqrt{\pi}} \bigg\lbrace \int_0^\infty \diff x_k \, 
\exp \left( - \frac{(x_k - \bar{x}_k)^2}{a^2} \right) \notag \\
& \quad - \int_{-\infty}^0 \diff x_k \,
\exp \left( - \frac{(x_k - \bar{x}_k)^2}{a^2} \right) \bigg\rbrace \, .
\end{align}
Using
\begin{equation}\label{eq:R5}
y_k = \frac{x_k - \bar{x}_k}{a}
\end{equation}
one finds
\begin{align}\label{eq:R6}
\langle s_k \rangle &= \frac{1}{\sqrt{\pi}} \bigg\lbrace \int_{-\bar{x}_k/a}^{\infty} 
\diff y_k \, \eul^{- y_k^2} - \int_{-\infty}^{-\bar{x}_k/a} \diff y_k \, \eul^{- y_k^2} 
\bigg\rbrace \notag \\
&= \frac{2}{\sqrt{\pi}} \int_0^{\bar{x}_k/a} \diff y_k \, \eul^{-y_k^2} \, .
\end{align}
The result is the error function
\begin{equation}\label{eq:R6A}
\langle s_k \rangle = \erf\left( \frac{\bar{x}_k}{a} \right)\, ,
\end{equation}
and therefore $\langle s_k \rangle = \pm 1$ for $\bar{x}_k/a \to \pm \infty$ and $\langle s_k \rangle = 0$ for  $\bar{x}_k = 0$.

One can obviously realize arbitrary $\rho_k = \langle s_k \rangle$ by probability distributions with suitable $\bar{x}_k$. In general, these probability distributions do not constitute a quantum subsystem, however. For quantum subsystems an additional quantum constraint has to be obeyed. Defining the ``purity''
\begin{equation}\label{eq:R7}
P = \sum_k \langle s_k \rangle^2 = \sum_k \left( \text{erf}\left( \frac{\bar{x}_k}{a} \right)
\right) ^2\, ,
\end{equation}
the quantum condition $P \leq 1$ restricts $a$ as function of $\bar{x}_k$. In particular, pure quantum states are realized by $P=1$, fixing uniquely $a(\bar{x})$. For example, the pure state $\rho_1=\rho_2=0$, $\rho_3 = 1$ corresponds to $\bar{x}=(0,0,1)$ and the limit distribution characterized by $a(0,0,1)=0$. Another pure state with $\rho_1 = \rho_2 = 1/\sqrt{2}$, $\rho_3 = 0$ can be realized by $\bar{x} = (1/\sqrt{2},\, 1/\sqrt{2},\, 0)$ with $a(\bar{x})$ obeying
\begin{equation}\label{eq:R5}
\text{erf} \left( \frac{1}{\sqrt{2} a} \right) = \frac{1}{\sqrt{2}} \, .
\end{equation}

The two states are related by a rotation of the vectors $(\rho_1, \rho_2, \rho_3)$ or $(\bar{x}_1, \bar{x}_2, \bar{x}_3)$. The function $a(\bar{x})$ is, however, not rotation invariant since its value depends on the direction of $\bar{x}$. Inserting $a(\bar{x})$ the probability distribution \eqref{eq:R1} depends only on the three variables $\bar{x}$. Due to the lack of rotation invariance of $a(\bar{x})$ it is not invariant under a simultaneous rotation of $x$ and $\bar{x}$. We observe a clash between rotation invariance of $p(x)$ for constant $a$ and the pure state condition with $a(\bar{x})$. Nevertheless, arbitrary unitary $SU(2)$-transformations of the quantum subsystem can be achieved by rotations of $\bar{x}$ accompanied by corresponding changes of $a(\bar{x})$. The drawback of the lack of rotation invariance of the classical probability distribution is that rotated classical spins do not correspond to rotated quantum spins.

\subsection{Real classical variables and infinitely many classical spins}

We can define a continuum (infinitely many) of classical spin observables $s(\bm{e}) = s(\Omega)$ which depend on the direction of a unit vector $\bm{e}$ or the corresponding angle $\Omega$ on the sphere. For a given $\bm{e}$ we divide the space of all classical states $x$, e.g. $\mathbb{R}^3$, into two halves, $\mathcal{I}_+(\bm{e})$ and $\mathcal{I}_-(\bm{e})$, with $x \in \mathcal{I}_+(\bm{e})$ if $\bm{x\cdot e} > 0$, and $x \in \mathcal{I}_-(\bm{e})$ for $\bm{x \cdot e} < 0$. The spin observable $s(\bm{e})$ takes the value one for $x \in \mathcal{I}_+(\bm{e})$ and minus one for $x \in \mathcal{I}_-(\bm{e})$,
\begin{equation}\label{eq:R6}
s(\bm{e}; x) = 
\begin{cases} 
1 \quad &\text{ for } \bm{x \cdot e} > 0 \\
-1 \quad &\text{ for } \bm{x \cdot e} < 0
\end{cases}\, ,
\end{equation}
e.g.
\begin{equation}\label{eq:R7}
s(\bm{e}; x) = \theta(\bm{x\cdot e}) - \theta(- \bm{x\cdot e})\, .
\end{equation}
The three spins $s_k$ in eq.~\eqref{eq:R3} are particular examples for $\bm{e} = (1,0,0)$, $(0,1,0)$ and $(0,0,1)$. For continuous classical states the definition of infinitely many different classical spin observables or classical Ising spins is rather straightforward. 

With classical states characterized by real numbers the need of infinitely many classical spin observables for the realization of a quantum spin finds a natural setting. For the quantum spin an infinite number of yes/no decisions can be taken by arbitrary directions $\bm{e}$ of the spin operator. The same holds for classical spin variables if the classical states correspond to real numbers. One should not be scared by the formal appearance of an infinite number of spins. Real numbers can be characterized by an infinite number of possibilities of binning. Different binnings lead to different possibilities for realizing macroscopic spin variables or yes/no decisions. One such binning is the bit representation of real numbers in conventional computers. This demonstrates that for efficient precision  computations a rather modest number of bits is sufficient, even though formally an infinite number of bits is needed for an arbitrarily accurate real number. Most likely the situation is similar for those classical systems that realize quantum subsystems. 

Arbitrary unitary transformations of the quantum density matrix \eqref{eq:4} can be performed by changes of the probability distribution \eqref{eq:R1}, i.e. by rotating the vector $\bar{x}$ accompanied by a change of $a(\bar{x})$. Nevertheless, for an arbitrary direction $\bm{e}$ the quantum spin in direction $\bm{e}$ does not correspond to the classical spin observable $s(\bm{e})$ given by eq.~\eqref{eq:R7}. For both the quantum spin and the classical spin the possible measurement values are $\pm 1$. The expectation values, and therefore the associated probabilities to find one or minus one, differ, however. 

This is easily demonstrated for a pure quantum state with $\rho_1 = \rho_2 = 0$, $\rho_3=1$, as realized by the classical probability distribution
\begin{align}\label{eq:R8}
p(x) &= \lim_{a\to 0} (a\sqrt{\pi})^{-3} \exp\left\lbrace - \frac{1}{a^2} \left( x_1^2 + x_2^2 + (x_3 - 1)^2 \right) \right\rbrace \notag \\
&= \delta(x_1) \delta(x_2) \delta(x_3 - 1)\, . 
\end{align}
For the quantum spin in the direction $\bm{e}$, given by the operator
\begin{equation}\label{eq:R9}
\hat{S}(\bm{e}) = e_k \hat{S}_k = e_k \tau_k\, ,
\end{equation}
the expectation value is given by 
\begin{equation}\label{eq:R10}
\langle \hat{S}(\bm{e}) \rangle = e_k \rho_k\, .
\end{equation}
For our particular state one has 
\begin{equation}\label{eq:R11}
 \langle \hat{S}(\bm{e}) \rangle = e_3\, .
\end{equation} 
In contrast, for the classical spin it only matters if the point $(0,0,1)$ lies in $\mathcal{I}_+ (\bm{e})$ or in $\mathcal{I}_- (\bm{e})$. It lies in $\mathcal{I}_+ (\bm{e})$ if $e_3 > 0$, and otherwise in $\mathcal{I}_- (\bm{e})$. For this particular classical state the classical expectation values are
\begin{equation}\label{eq:R12}
\langle s(\bm{e}) \rangle = \theta(e_3) - \theta(- e_3)\, .
\end{equation}
For a probability distribution of the type \eqref{eq:R8} it is not surprising that classical spins have sharp values. An exception are the directions for which $e_3 = 0$. In this case the limiting procedure yields $\langle s(\bm{e})\rangle = 0$. We conclude that only for the particular directions $\bm{e}= (0,0,1)$ or $\bm{e} = (\cos\phi, \, \sin\phi , \, 0)$ the expectation values of the classical spins coincide with the ones of the quantum spin in the same direction. 

The case of a point-like probability distribution with sharp values of the classical spins may seem somewhat particular. The clash between quantum and classical expectation values persists, however, for other states for which $p(x)$ is a smooth distribution. For the pure quantum state $\rho_1 = \rho_2 = 1/\sqrt{2}$, $\rho_3 = 0$, realized by eq.~\eqref{eq:R1} with $\bar{x}_1 = \bar{x}_2 = 1/\sqrt{2}$, $\bar{x}_3=0$ and $a$ obeying eq.~\eqref{eq:R5}, the expectation value of the quantum spin $\hat{S}(\bm{e})$ reads
\begin{equation}\label{eq:R13}
\langle \hat{S}(\bm{e}) \rangle = \frac{1}{\sqrt{2}} (e_1 + e_2) \, .
\end{equation}
We may first look at the direction $\bm{e} = (1,0,0)$ with $\langle \hat{S} (\bm{e}) \rangle = 1/\sqrt{2}$, which corresponds to the particular classical spin $s_1$ in eq.~\eqref{eq:R3}, and therefore
\begin{equation}\label{eq:R14}
\langle s(\bm{e}) \rangle = \erf \left( \frac{1}{\sqrt{2} a} \right) = \frac{1}{\sqrt{2}} \, .
\end{equation} 
As it should be by construction, the quantum and classical expectation values agree. Consider next the quantum spin with sharp value, $e_1 = e_2 = 1/\sqrt{2}$, where $\langle \hat{S} (\bm{e}) \rangle = 1$. For the classical spin the region $\mathcal{I}_+$ is given by $x_1 + x_2 > 0$, while all points with $x_1 + x_2 < 0$ lie in $\mathcal{I}_-$. Since $a$ differs from zero the probability distribution $p(x)$ does not vanish in $\mathcal{I}_-$. This implies for the expectation value of the classical spin $s(\bm{e})$ that it has to be smaller than one. Indeed, one has for the difference from one
\begin{equation}\label{eq:R15}
1 - \langle s(\bm{e}) \rangle = 2 \int_{\mathcal{I}_-(\bm{e})} \diff x \, p(x)\, .
\end{equation}

In conclusion, the Gaussian probability distributions \eqref{eq:R1} can realize arbitrary unitary transformations of a quantum subsystem characterized by the expectation values of three particular classical spins $s_1, s_2, s_3$. For these particular spins the expectation values of the quantum spins agree with the expectation values of the classical spins in these particular directions. Because of the $\bar{x}$-dependence of $a(\bar{x})$ this agreement does not hold for arbitrary directions.

If one wants to achieve the identification of possible outcomes of quantum measurements of the quantum spin $\hat{S}_k(\bm{e})$ with the measurements of classical spin observables $s(\bm{e})$ given by eq.~\eqref{eq:R7} one has to realize classical probability distributions for which
\begin{equation}\label{eq:R16}
\langle s(\bm{e}) \rangle = e_k \rho_k = e_k \langle s_k \rangle
\end{equation} 
for arbitrary $e_k$. Here $\langle s_k \rangle$ are the expectation values of three classical spins in the particular directions $\bm{e} = (1,0,0)$, $(0,1,0)$, $(0,0,1)$. The relation \eqref{eq:R16} relates the expectation values of classical spin observables in arbitrary directions to the expectation values of three particular classical spins.

\subsection{Classical probability distribution for one-qubit quantum state}

Consider the classical Ising spins $s(\bm{e}, x)$ defined by eq.~\eqref{eq:R7}. We want to find a family of probability distributions $p(\bm{\rho}, x)$ that is characterized by a vector $\bm{\rho}$, such that the expectation values of the classical spins obey
\begin{equation}\label{eq:R17}
\langle s(\bm{e}) \rangle = \int \diff^3 x \, p(\bm{\rho}, x) s(\bm{e}, x) = \bm{\rho}\cdot \bm{e}\, . 
\end{equation}
In particular, the three spins $s_1$, $s_2$, $s_3$ corresponding to $\bm{e_1} = (1, 0, 0)$, $\bm{e_2}=(0, 1, 0)$, $\bm{e_3} = (0, 0, 1)$ obey $\langle s_k \rangle = \rho_k$. They define the density matrix for the quantum subsystem by eq.~\eqref{eq:4}. We require the quantum condition $\sum_k \rho_k^2 \leq 1$. For probability distributions obeying eq.~\eqref{eq:R17} the quantum expectation values of the quantum spins $\hat{S}(\bm{e})$ in arbitrary directions $\bm{e}$ coincide with the classical expectation values of the classical spins $s(\bm{e})$ in the same directions. This holds for arbitrary $\bm{\rho}$ obeying $|\bm{\rho}| \leq 1$. We will concentrate on pure quantum states for which $\bm{\rho}$ is a unit vector and $\bm{\rho}\cdot \bm{e} = \cos\vartheta$ involves the relative angle $\vartheta$ between the unit vectors $\bm{\rho}$ and $\bm{e}$. Mixed quantum states pose no additional problem. 

Let us parameterize the variable $\bm{x}$ by its length $r$ and direction given by a unit vector $\bm{f}$, 
\begin{equation}\label{eq:R18}
\bm{x} = r\bm{f}\, , \quad |\bm{f}| = 1\, .
\end{equation} 
We find a family of probability distributions realizing the quantum condition \eqref{eq:R17}, given by 
\begin{equation}\label{eq:R19}
p(\bm{\rho}, x) = \tilde{p}(r)\, (\bm{\rho}\cdot \bm{f}) \, \theta(\bm{\rho}\cdot \bm{f})\, .
\end{equation}
The detailed form of $\tilde{p}(r) \geq 0$ is not important, being only restricted by the normalization condition $\int_x p(x) = 1$. The probability distribution \eqref{eq:R19} is positive for all $x$ since it vanishes whenever the scalar product $\bm{\rho}\cdot\bm{f}$ is negative. It differs from zero only in one half-space determined by $\bm{\rho}$. We can equivalently write it as
\begin{equation}\label{eq:R20}
p(\bm{\rho}, x) = \frac{\tilde{p}(r)}{r} (\bm{\rho}\cdot\bm{x})\, \theta(\bm{\rho}\cdot\bm{x})\, .
\end{equation}
A rather smooth behavior at $r=0$ may be obtained by $\tilde{p}(r\to 0) \sim r^3$ or similar.

For a proof of the relation \eqref{eq:R17} we employ the fact that the probability distribution is invariant under simultaneous rotations of $\bm{\rho}$ and $\bm{x}$ or $\bm{\rho}$ and $\bm{f}$. This contrasts with the Gaussian probability distribution \eqref{eq:R1}, for which $\bar{x}_k$ plays a similar role as $\rho_k$. For the present case there is no longer a clash between the quantum constraint and combined rotation invariance. Without loss of generality we can therefore take a pure state characterized by $\bm{\rho} = (0, 0, 1)$ for which the probability distribution is given by
\begin{equation}\label{eq:R23}
p = \tilde{p}(r)\, f_3 \, \theta(f_3) \, .
\end{equation}
We can further choose an arbitrary direction of $\bm{e}$ in the $(x_1, x_2)$ plane and we take $(e_1 > 0)$
\begin{equation}\label{eq:R25}
\bm{e} = (e_1, 0, e_3)\, .
\end{equation}
In consequence, the classical expectation value of $s(\bm{e})$ reads 
\begin{align}\label{eq:R26}
\langle s(\bm{e}) \rangle &= \int \diff^3 x\, \tilde{p}(r) \frac{x_3\, \theta(x_3)}{r} \notag \\
& \quad \times\left[ \theta(e_1 x_1 + e_3 x_3) - \theta(-e_1 x_1 - e_3 x_3) \right]\, .
\end{align}
With $R^2 = x_1^2 + x_3^2 = r^2 - x_2^2$ we can perform the $x_2$-integration
\begin{align}\label{eq:R27}
\langle s(\bm{e})\rangle &= \int \diff x_1\, \diff x_3\, h(R)\, x_3\, \theta(x_3) \notag \\
& \quad \times 
\left[ \theta(e_1 x_1 + e_3 x_3) - \theta ( -e_1 x_1 - e_3 x_3 ) \right] \, ,
\end{align}
with $r^2 = R^2 + x_2^2$ and
\begin{equation}\label{eq:R28}
h(R) = \int \diff x_2\, \frac{\tilde{p}(r)}{r}\, .
\end{equation}
We next choose
\begin{align}\label{eq:R29}
x_1 = R\sin\alpha\, &, \quad x_3=R\cos\alpha\, , \notag \\
e_1 = \sin\varphi\, &, \quad e_3 = \cos\varphi\, ,
\end{align}
such that 
\begin{align}\label{eq:R30}
\langle s(\bm{e}) \rangle &= \int_0^\infty \diff R\, h(R) R^2\, \int_{-\pi/2}^{\pi/2}
\diff\alpha \cos\alpha \notag \\
&\quad \times
\left[ \theta \left( \cos(\alpha - \varphi)\right) - \theta \left( - \cos(\alpha - \varphi) \right) \right] \notag \\
&= \frac{1}{2} \left\lbrace \int_{\varphi - \pi/2}^{\pi/2} \diff\alpha \cos\alpha
- \int_{-\pi/2}^{\varphi - \pi/2} \diff\alpha \cos\alpha \right\rbrace\, .
\end{align}
Here we have employed the normalization condition
\begin{equation}\label{eq:R31}
\int_0^\infty \diff R\, h(R) R^2 = \frac{1}{2}\, ,
\end{equation}
which follows from 
\begin{equation}\label{eq:R32}
\int_x p(x)=\int \diff x_1\, \diff x_3\, h(R) x_3 \, \theta(x_3) = 1\, .
\end{equation}
The trivial angular integral \eqref{eq:R30} yields indeed
\begin{equation}\label{eq:R33}
\langle s(\bm{e})\rangle = \cos\varphi = e_3 = \bm{\rho}\cdot \bm{e}\, .
\end{equation}

We can rotate the vector $\bm{e}$ around the $x_3$-axis without changing $s(\bm{e})$. The choice \eqref{eq:R25} for $\bm{e}$ is therefore without loss of generality. For an integration of states characterized by a different unit vector $\bm{\rho}$ we employ invariance under simultaneous rotations of $\bm{\rho}$, $\bm{f}$ and $\bm{e}$, as manifest by the formulation of $p$ and $s$ in terms of scalar products. This implies eq.~\eqref{eq:R17} for arbitrary unit vectors $\bm{\rho}$ and $\bm{e}$ and concludes the proof. The family of normalized classical probability distributions \eqref{eq:R19} describes completely all aspects of pure states for a single quantum spin. 

It is rather obvious that the state of a qubit can be described by three real numbers $\rho_k$, further reduced to two for pure states by the condition $|\bm{\rho}|=1$. All unitary transformations can be achieved by manipulations of these three numbers, e.g. rotations of the vector $\bm{\rho}$. What our construction in terms of probability distributions $p(\bm{\rho})$ and classical spin observables $s(\bm{e})$ brings in addition is the crucial property that individual measurements of the quantum spin in an arbitrary direction yields only the discrete values $\pm 1$, and the possibility to perform this measurement by measuring classical two-level observables. The continuous wave aspect of quantum mechanics only involves $\rho$, while the discrete particle side needs a probabilistic setting similar to the one given here. 

\section{Quantum mechanics}\label{sec:Quantum_Mechanics}

In this section we discuss the realization of full quantum mechanics by suitable classical statistical probability distributions for Ising spins. We consider an arbitrary number $Q$ of quantum spins and want to realize arbitrary unitary transformations \eqref{eq:I1} for the hermitian $2^{Q}\times 2^{Q}$-density matrix $\rho$. As we have argued before, an infinite number of Ising spins is needed in order to realize for suitable quantum subsystems all unitary transformations by static memory materials. We have seen in the preceding section how infinitely many classical spins can arise from continuous real variables. In the present section we work on a more abstract level for which it does not matter how the infinitely many classical bits are realized by a given system. Quantum subsystems of classical statistical systems can be realized on different levels, depending on the observables for which a direct correspondence between quantum and classical observables exists, and depending on the way the classical probability distributions have to change for the realization of arbitrary unitary transformations \eqref{eq:I1}.

On the first level we only demand that an arbitrary positive density matrix $\rho$, $\rho^{\dagger}=\rho$, $\tr(\rho)=1$, can be realized by the expectation values of a suitable number of Ising spins according to eqs~\eqref{eq:30}, \eqref{eq:31}. The density matrix defines the quantum subsystem. Since $\rho$ is specified by $2^{Q}-1$ real quantities in the range $-1\leq \rho_{\mu_{1}\mu_{2}\ldots\mu_{Q}}\leq 1$, one needs the expectation values $\exval{\sigma_{\mu_{1}\mu_{2}\ldots\mu_{Q}}}$ of $2^{Q}-1$ classical two-level observables with possible measurement values $\pm 1$. Those classical two-level observables need not be independent Ising spins. One also can use suitable classical correlation functions of Ising spins. In this case some of the two-level observables $\sigma_{\mu_{1}\mu_{2}\ldots\mu_{Q}}$ are products of ``fundamental'' Ising spins. A minimal number of fundamental Ising spins needed to realize arbitrary $\rho$ is given by $M=3Q$. At this first level only the quantum observables that correspond to the generators $L_{\mu_{1}\mu_{2}\ldots\mu_{Q}}$ have a direct correspondence to classical observables $\sigma_{\mu_{1}\mu_{2}\ldots\mu_{Q}}$. They have the same possible measurement values $\pm 1$ and the same expectation values, and therefore also the same probabilities to find a given possible measurement value. 

At the second level we impose the positivity of $\rho(t)$. This defines a first ``quantum constraint'' for the classical probability distributions that are compatible with such a quantum subsystem. The positivity of $\rho(t)$ guarantees that arbitrary hermitian operators $\hat{A}(t)$ can be associated with quantum observables $A(t)$. The possible measurement values of the quantum observables are given by the eigenvalues of $\hat{A}(t)$, and the probabilities to find these values can be computed from $\rho(t)$. Since an arbitrary positive $\rho(t)$ can be realized by some suitable classical probability distribution $p_{\tau}(t)$, and the same holds for $\te$, one can realize an arbitrary map $\rho(t)\raw\rho(\te)$ by some suitable (non-unique) map of classical probability distributions $p_{\tau}(t)\raw p_{\tau}(\te)$. This includes arbitrary unitary transformations \eqref{eq:I1}. At this level there is no guaranty that a suitable map can be realized by a static memory material or a deterministic unique jump operation. At the second level a general quantum observable $A(t)$ has no direct correspondence to a classical observable. This direct correspondence remains restricted to the quantum observables corresponding to the generators $L_{\mu_{1}\mu_{2}\ldots\mu_{Q}}$, or to linear combinations of commuting generators.

At the third level the correspondence between quantum observables and classical observables is extended to a larger set of observables. We will require that arbitrary quantum observables corresponding to operators $\hat{A}$ with $\hat{A}^{2}=1$, $\tr(\hat{A})=0$, can be identified with suitable two-level classical observables. The requirement that the expectation values of these quantum observables, as computed from the density matrix by the usual quantum rule, coincide with the classical expectation values, computed from the classical probability distribution, imposes a second set of quantum constraints on the classical probability distributions realizing this type of subsystem. On the third level continuous symmetries as rotations can be realized simultaneously on the quantum and classical level. This requires an infinite number of Ising spins, or ``continuous Ising spins'', corresponding to the infinite number of bits needed for a precise localization of a point on a sphere. The third level greatly enhances the possibilities to realize arbitrary unitary transformations by static memory materials or unique jump operations.

In the present section we are mainly concerned with the third level. We construct quantum subsystems for which all unitary transformations can be realized by unique jump operations. In appendix \ref{app:Two-qubit_quantum_subsystem_with_classical_correlation_functions} the discussion is extended to the case of ``composite'' classical two-level observables where part of the classical observables is expressed by products of ``fundamental'' Ising spins. We pay attention to the question under which circumstances the quantum constraint on the third level is sufficient to ensure the positivity of the density matrix.

\subsection{Continuous family of Ising spins}

For a representation of arbitrary unitary operations on a single qubit by static memory materials or unique jump operations one needs an infinite number of classical bits. We may consider a continuous family of Ising spins, or ``continuous Ising spins'',
\begin{equation}\label{eq:W1}
s(\Omega)=s(\bm{e})=s(e_{k})\com e_{k}e_{k}=1\com s^{2}(\Omega)=1\, ,
\end{equation}
one for each point $\bm{e}=(e_{1},e_{2},e_{3})$ on a unit sphere, or for each direction characterized by the angle $\Omega$. (More precisely, we take only half of the unit sphere, identifying $s(-\bm{e})=-s(\bm{e})$.) 

We start with a two-component qubit where $\Omega$ is replaced by the angle $\varphi$ on a circle, $\bm{e}=(e_{1},e_{3})$, $e_{1}^{2}+e_{3}^{2}=1$. The quantum condition is imposed such that the expectation value of the Ising spin $s(\varphi)$ equals the one of the quantum spin $S(\varphi)$ in the direction characterized by $\varphi$
\begin{equation}\label{eq:W2}
\exval{s(\varphi)}=\exval{S(\varphi)}=\tr\left \{\rho(\cos(\varphi)\tau_{1}+\sin(\varphi)\tau_{3})\right \}.
\end{equation}
Here we assume (for a given $t$) a probability distribution $p[s(\varphi)]$ for the Ising spins with direction $\varphi$, 
\begin{equation}\label{eq:W3}
\exval{s(\varphi)}=\int\, \cD s(\varphi)\, p[s(\varphi)]\, s(\varphi)
\end{equation}
where
\begin{equation}
\int\, \cD s(\varphi)=\prod_{\varphi}\sum_{s(\varphi)=\pm 1}.
\end{equation}
(More generally, the expectation value $\exval{s(\varphi)}$ is computed from the ($t$-local) classical density matrix. In the occupation number basis $s(\varphi)$ is a diagonal classical operator such that the probabilities $p[s(\varphi)]$ correspond to the diagonal elements of the classical density matrix.) The quantum constraint expresses the expectation values of all Ising spins $s(\varphi)$ in terms of the two numbers $\rho_{1}$ and $\rho_{3}$,
\begin{equation}\label{eq:W4}
\exval{s(\varphi)}=\cos(\varphi)\rho_{1}+\sin(\varphi)\rho_{3}\, .
\end{equation}

This is a strong constraint on the probability distributions $p[s(\varphi)]$ that are compatible with this quantum subsystem. Infinity many expectation values are given in terms of two real numbers. Let us first concentrate on a pure state with $\rho_{1}^{2}+\rho_{3}^{2}=1$,
\begin{equation}\label{eq:W5}
\rho_{1}=\cos(\psi)\com \rho_{3}=\sin(\psi)\, .
\end{equation}
The quantum condition takes the simple form
\begin{equation}\label{eq:W6}
\exval{s(\varphi)}=\cos(\varphi-\psi)\, .
\end{equation}
The expectation values $\exval{s(\varphi)}$ have to follow a simple harmonic function, being invariant under a simultaneous shift of $\varphi$ and $\psi$. This generalizes to arbitrary mixed quantum states for which we can always write ($r\geq 0$)
\begin{align}\label{eq:W7}
& \rho_{1} = r\cos(\psi)\com \rho_{3}=r\sin(\psi)\, ,\nn\\
& \langle s(\varphi) \rangle = r\cos(\varphi-\psi)\, .
\end{align}
The quantum condition requires $0\leq r\leq 1$, such that $|\exval{s(\varphi)}|\leq 1$ holds for arbitrary $\varphi$. 

The quantum conditions for the finite number of Ising spins discussed previously obtain for particular choices of $\varphi$. The spins $s_{1}$ and $s_{3}$ correspond to $\varphi=0$ and $\varphi=\pi/2$, respectively, reproducing eq.~\eqref{eq:3}. Similarly, the spins $s_{13+}$ and $s_{13-}$ are realized for $\varphi=\pi/4$ and $\varphi=-\pi/4$, such that eq.~\eqref{eq:W4} implements the quantum conditions \eqref{eq:QC5}, \eqref{eq:QC6}. 

Rotations on the circle by an arbitrary angle $\alpha$ can now be  realized by unique jump operations $s(\varphi)\raw s(\varphi-\alpha)$,
\begin{align}\label{eq:W8}
\exval{s^{\prime}(\varphi)}&=\exval{s(\varphi-\alpha)} =r\cos(\varphi-\alpha-\psi)\nn\\
&=\cos(\varphi)\rho_{1}^{\prime}+\sin(\varphi)\rho_{3}^{\prime}\, ,
\end{align}
with
\begin{equation}\label{eq:W9}
\rho_{1}^{\prime}=r\cos(\psi+\alpha)\com \rho_{3}^{\prime}=r\sin(\psi+\alpha)\, .
\end{equation}
In particular, for $\alpha=\epsilon \omega$ we can take the continuum limit $\epsilon\raw 0$. This is one of the simplest classical systems that can realize a continuous unitary evolution for a quantum subsystem \cite{CWIT}.

For infinitely many Ising spins the generalization to a three-component qubit is straightforward. Characterizing the direction of the Ising spin by a unit vector $(e_{1},e_{2},e_{3})$, and the direction of the quantum spin by the vector $(\rho_{1},\rho_{2},\rho_{3})$, the quantum condition reads \cite{CWPO}
\begin{equation}\label{eq:W10}
\exval{s(\bm{e})}=e_{k}\rho_{k}\, .
\end{equation}
Since $|\exval{s(\bm{e})}|\leq 1$ for arbitrary $\bm{e}$ we infer, in particular, the constraint $\rho_{k}\rho_{k}\leq 1$, cf. eq.~\eqref{eq:7}. The setting with three classical bits corresponds to the particular unit vectors $(1,0,0)$, $(0,1,0)$ and $(0,0,1)$, while for the discrete setting with six Ising spins the unit vectors $\bm{e}$ correspond to corners of the icosahedron. Arbitrary $SO(3)$-rotations on the unit sphere can now be realized by unique jump operations
\begin{equation}\label{eq:W11}
s^{\prime}(\bm{e})=s(\bm{e}^{\prime})\com e_{k}^{\prime}=e_{l}O_{lk}\com O^{T}O=1 \ .
\end{equation}
With
\begin{equation}\label{eq:W12}
\exval{s^{\prime}(\bm{e})}=e_{k}^{\prime}\rho_{k}=e_{k}\rho_{k}^{\prime}\, ,
\end{equation}
this transfers to
\begin{equation}\label{eq:W13}
\rho_{k}^{\prime}=O_{kl}\rho_{l}\, .
\end{equation}
The unitary transformation of the quantum density matrix,
\begin{align}\label{eq:W14}
\rho^{\prime}&=\dfrac{1}{2}\left (1+\rho_{k}^{\prime}\tau_{k}\right )=\dfrac{1}{2}\left (1+\rho_{l}\, O^{T}_{lk}\tau_{k}\right )\nn\\
&=\dfrac{1}{2}\left (1+\rho_{l}\, U\tau_{l}\, U^{\dagger}\right )=U\rho\,  U^{\dagger}\, ,
\end{align}
reflects that $SO(3)$ is a subgroup of $U(2)$, with appropriate unitary matrices $U$ associated to $O$ according to
\begin{equation}\label{eq:W15}
U\tau_{l}\, U^{\dagger}=O^{T}_{lk}\tau_{k}\, .
\end{equation}
The overall phase of $U$, corresponding to the abelian $U(1)$-factor of $U(2)$, plays no role for $\rho$.

Our system of a continuous family of Ising spins $s(\bm{e})$ is set up as a probabilistic ensemble in classical statistics. The transfer matrices or the step evolution operators $S$ are even deterministic unique jump operators, similar to cellular automata. Depending on the choice of $S(t)$ an arbitrary unitary evolution of a single quantum spin can be described, both for pure and mixed states. If for an infinitesimal step from $t$ to $t+\epsilon$ the rotation $s(\bm{e})\raw s(\bm{e}^{\prime})$ is infinitesimal, the unitary transformation of the density matrix is also infinitesimal
\begin{equation}\label{eq:W16}
\rho(t+\epsilon)=U(t)\rho(t)\, U^{\dagger}(t)\com U(t)=\exp\left (-i\epsilon H(t)\right )\, ,
\end{equation}
with hermitian Hamiltonian $H(t)$. The dependence of expectation values on the location on the chain $t$ is mapped to the time evolution in quantum mechanics. This includes $t$-dependent Hamiltonians. The possible measurement values $\pm 1$ of the quantum spin in arbitrary directions correspond to the possible measurement values of the classical Ising spin in the same direction. Expectation values of the quantum spin in an arbitrary direction are computable from the standard rules of classical statistics. The issue of the correct choice of a correlation function for a sequence of ideal measurements is discussed in detail in refs.~\cite{CWA,CWB,CWPO}. This correlation function is obtained by the multiplication of quantum operators. Our discussion establishes the realization of an arbitrary evolution of a single quantum spin as a particular classical statistical system.

\subsection{Arbitrary unitary transformations for two qubits}

Arbitrary unitary transformations for two qubits can be realized by a continuous family of Ising spins $s(e_{\mn})$, with $e_{\mn}=e_{z}$ the components of a $15$-dimensional unit vector which obeys additional constraints. Here we employ $\mu,\nu=(0,1,2,3)$ and omit $e_{00}$. The quantum condition reads
\begin{equation}\label{eq:W17}
\exval{s(\bm{e})}=\exval{s(e_{\mn})}=e_{\mn}\rho_{\mn}=e_{z}\rho_{z}\, ,
\end{equation}
with $\rho_{\mn}=\rho_{z}$ the coefficients defining the complex $4\times 4$ density matrix $\rho$ by eq.~\eqref{eq:13}. We identify $s(-e_{\mn})=-s(e_{\mn})$ and use $s(e_{z})$, $s(e_{\mn})$ and $s(\bm{e})$ synonymously. If we only impose the condition $e_{z}e_{z}=1$, the vectors $\bm{e}$ span half the unit sphere $S^{14}$. Arbitrary $SO(15)$-rotations can be realized as unique jump operations in this case. This includes the $SU(4)$ subgroup of $SO(15)$ which is embedded in $SO(15)$ such that the $15$-dimensional fundamental representation of $SO(15)$, that can be associated to $e_{\mn}$ and $\rho_{\mn}$, transforms as the $15$-dimensional adjoint representation of $SU(4)$. The embedding is precisely provided by eq.~\eqref{eq:13}, with
\begin{equation}\label{eq:W18}
U\rho\, U^{\dagger}=\dfrac{1}{4}\left (1+O_{\mn,\tau\sigma}\rho_{\tau\sigma}L_{\mn}\right )\, ,
\end{equation}
and $O$ an orthogonal $15\times 15$-matrix.

Imposing only the constraint $e_{z}e_{z}=1$, $e_{z}=e_{\mn}$, and demanding the condition \eqref{eq:W17} for all spins $s(e_{z})$, realizes a $15$-component single qubit rather than two three-component qubits. Indeed, the conditions $|\exval{s(e_{\mn})}|\leq 1$ for all $e_{\mn}$ on $S^{14}$ would require $\rho_{z}\rho_{z}\leq 1$, not compatible with the relation $\rho_{z}\rho_{z}=3$ for a pure quantum state of two qubits. For a description of two qubits further constraints on $e_{z}$ have to be imposed. They will be of a form such that the subgroup of $SU(4)$-transformations can still be obtained as unique jump operations, corresponding to transformations on the manifold spanned by $e_{\mn}$.

The basic principle for the selection of the Ising spins $s(e_{z})$ requires that a classical bit is associated to each traceless two-level observable $\hat{A}$ in the quantum system. Such two-level observables obey $\hat{A}^{2}=1$, $\tr(\hat{A})=0$, and $\hat{A}$ has therefore an equal number of eigenvalues $+1$ and $-1$. As for any hermitian traceless operator, $\hat{A}$ can be written as a linear combination of the generators $L_{z}$
\begin{equation}\label{eq:W18A}
\hat{A}(e_{z})=e_{z}L_{z}\, .
\end{equation}
Our classical statistical system of Ising spins is chosen such that every $\hat{A}(e_{z})$ corresponds to a classical bit $s(e_{z})$, and vice versa, with
\begin{equation}\label{eq:W18B}
\exval{s(e_{z})}=\exval{\hat{A}(e_{z})}=\tr\left (\rho\hat{A}(e_{z})\right )=e_{z}\rho_{z}\, .
\end{equation}
Here $\exval{s(e_{z})}$ is determined by the classical statistical ensemble, while $\exval{\hat{A}(e_{z})}$ is the expectation value in the quantum system. (We identify $\hat{A}(-e_{z})=-\hat{A}(e_{z})$, $s(-e_{z})=-s(e_{z})$.) The condition $\hat{A}^{2}=1$ requires
\begin{equation}\label{eq:W18C}
e_{z}e_{z}=1\com d_{zyw}e_{z}e_{y}=0\, ,
\end{equation}
where $d_{zyw}$ are the symmetric coefficients defined by
\begin{equation}\label{eq:W18D}
\left \{L_{z},L_{y}\right \} = 2\delta_{zy}+2d_{zyw}L_{w}\, .
\end{equation}
This follows from the condition
\begin{equation}\label{eq:W18E}
\hat{A}^{2}=\dfrac{1}{2}e_{z}e_{y}\left \{L_{z},L_{y}\right \}=e_{z}e_{z}+d_{zyw}e_{z}e_{y}L_{w}=1\, .
\end{equation}
The second part of the constraint \eqref{eq:W18C} constitutes the additional conditions on the unit vectors $e_{z}$ alluded to above.

The manifold of vectors $\bm{e}$ obeying the condition \eqref{eq:W18C} is closely related to the homogeneous space $SU(4)/SU(2)\times SU(2)\times U(1)$. This space has eight real dimensions, such that the second condition \eqref{eq:W18C} amounts to six constraints selecting an eight-dimensional subspace of $S^{14}$. The connection to $SU(4)$ becomes apparent from the fact that any hermitian $4\times 4$-matrix with two eigenvalues $+1$ and two eigenvalues $-1$ can be diagonalized by some particular $SU(4)$ transformation $U_{(30)}(e_{z})$
\begin{equation}\label{eq:W18F}
\hat{A}(e_{z})=U_{(30)}(e_{z})\, L_{30}\, U_{(30)}^{\dagger}(e_{z})\, .
\end{equation}
On the other hand, $L_{30}$ is left invariant by unitary transformations of the subgroup $SU(2)\times SU(2)\times U(1)$, implying that $U_{(30)}(e_{z})$ is specified only modulo $SU(2)\times SU(2)\times U(1)$-transformations. 

The $SU(4)$-transformation acting on the space of vectors $\bm{e}$ is uniquely determined if we specify the transformation of a sufficient number of independent vectors $\bm{e}$. Its embedding into $SO(15)$ is analogous to eq.~\eqref{eq:W18}. A unique jump operation acts on the Ising spins as $s(e_{z})\raw s(e_{z}^{\prime})$, $e_{z}^{\prime}=e_{y}O_{yz}$, $O^{T}O=1$. Here the orthogonal matrices $O$ belong to the $SU(4)$ subgroup according to
\begin{equation}\label{eq:W18G}
\hat{A}^{\prime}=e_{z}^{\prime}L_{z}=e_{y}\, O_{yz}L_{z}=e_{y}\, U^{\dagger}L_{y}\, U\, .
\end{equation}
For the quantum subsystem this corresponds to a transformation of the density matrix according to
\begin{equation}\label{eq:W18H}
\exval{s(e_{z}^{\prime})}=e_{z}^{\prime}\rho_{z}=e_{z}\rho_{z}^{\prime}\com \rho_{z}^{\prime}=O_{zy}\rho_{y}\, ,
\end{equation}
or
\begin{equation}\label{eq:W18GG}
\rho^{\prime}=\dfrac{1}{4}\left (1+\rho_{z}^{\prime}L_{z}\right )=U\rho\, U^{\dagger}\, .
\end{equation}
Correspondingly, one has
\begin{align}\label{eq:W18HH}
\exval{\hat{A}^{\prime}}&=\tr(\rho\hat{A}^{\prime})=
e_{z}\, \tr \left( U\rho\, U^{\dagger}L_{z}\right) \nn \\
&= \tr (\rho^{\prime}\hat{A}) = e_{z}\rho_{z}^{\prime} = \exval{s(e_{z}^{\prime})}\, .
\end{align}

We conclude that a classical statistical system of continuous Ising spins $s(e_{z})$, with $e_{z}$ obeying eq.~\eqref{eq:W18C}, admits a quantum subsystem provided the quantum constraint \eqref{eq:W17} holds for all $s(e_{z})$ and the density matrix \eqref{eq:13} is positive. If the spins $s(e_{z})$ are independent, arbitrary unitary transformations for the quantum subsystem can be realized by unique jump operations on the classical spins. It is also conceivable to construct quantum subsystems for two qubits if some of the $s(e_{z})$ are composite, given by products of other spins. We discuss this in appendix \ref{app:Two-qubit_quantum_subsystem_with_classical_correlation_functions}.

\subsection{Quantum constraints for two qubits}

We next investigate the remaining part of the quantum constraint for $\rho$ resulting from the requirement of positivity. While for a single qubit the positivity of $\rho$ follows automatically if eq.~\eqref{eq:W10} holds for all $\bm{e}$ on $S^{2}$, for two or more qubits it may need additional restrictions on the space of $\rho_{z}$ and therefore on the classical probability distributions that can realize the quantum subsystem. We will see that the constraint \eqref{eq:W17} is indeed not sufficient to ensure the positivity of $\rho$. Thus the requirement of positive eigenvalues of $\rho$ imposes additional constraints on the allowed expectation values $\rho_{\mn}$. 

We concentrate on the quantum condition for a pure state, $\rho^{2}=\rho$. This requires
\begin{align}\label{eq:W18I}
\dfrac{1}{16}\left (1+\rho_{z}L_{z}\right )^{2}&=\dfrac{1}{16}\left (1+\rho_{z}\rho_{z}+2\rho_{z}L_{z}+d_{zyw}\rho_{z}\rho_{y}L_{w}\right )\nn\\
&=\dfrac{1}{4}\left (1+\rho_{z}L_{z}\right )\, ,
\end{align}
and therefore imposes on the allowed $\rho_{z}$ the conditions
\begin{equation}\label{eq:W18J}
\rho_{z}\rho_{z}=3\com d_{zyw}\rho_{z}\rho_{y}=2\rho_{w}\, .
\end{equation}
The space of $\rho_{z}$ obeying the condition \eqref{eq:W18J} is closely related to the projective space $\bC\bP^{4}$ in four complex dimensions, or to the homogeneous space $SU(4)/ SU(3)\times U(1)$ \cite{CWA,CWB,CWPT,CWPO,BH}. A pure state density matrix has one eigenvalue one, and all other eigenvalues zero. It can therefore be obtained from $\bar{\rho}=\diag(1,0,0,0)$ by a suitable unitary $SU(4)$-transformation $\bar{U}$,
\begin{equation}\label{eq:W18K}
\rho=\bar{U}\bar{\rho}\, \bar{U}^{\dagger}\, .
\end{equation}
Since $\bar{\rho}$ is invariant under a subgroup $SU(3)\times U(1)$, this specifies the homogeneous space $SU(4)/SU(3)\times U(1)$. Thus the values of $\rho_{z}$ corresponding to a pure quantum state belong to a manifold with six real dimensions. It is a submanifold of $S^{14}$, where the second condition \eqref{eq:W18J} imposes eight additional constraints. We observe that the vectors $e_{z}$ and $\rho_{z}$ belong to different manifolds.

The pure state condition \eqref{eq:W18J} guarantees the positivity of $\rho$. This implies for all traceless two-level observables $\hat{A}(e_{z})$, $\hat{A}^{2}=1$, the bound
\begin{equation}\label{eq:W18L}
|\exval{\hat{A}(e_{z})}|\leq 1\, .
\end{equation}
One can diagonalize $\hat{A}=\diag(1,1,-1,-1)$ such that in this basis $\exval{\hat{A}}=\rho_{11}+\rho_{22}-\rho_{33}-\rho_{44}$. Since all diagonal elements of $\rho$ obey $0\leq \rho_{\alpha\alpha}\leq 1$, $\sum_{\alpha}\, \rho_{\alpha\alpha}=1$, eq.~\eqref{eq:W18L} follows. The bound \eqref{eq:W18L} guarantees that pure quantum states can indeed be realized by Ising spins obeying eq.~\eqref{eq:W18B}. This statement extends easily to mixed quantum states.

In the opposite direction, the quantum constraint \eqref{eq:W17}, \eqref{eq:W18C} implies eq.~\eqref{eq:W18B} and therefore the bound \eqref{eq:W18L} for all traceless two-level observables $\hat{A}$, $\tr(\hat{A})=0$, $\hat{A}^{2}=1$. This bound is not strong enough to guarantee the positivity of $\rho$, however. Without loss of generality we can take $\rho$ diagonal (by an appropriate choice of basis), $\rho=\diag(\rho_{11},\rho_{22},\rho_{33},\rho_{44})$. For $\hat{A}=e_{\mn}L_{\mn}$ only $e_{30}$, $e_{03}$ and $e_{33}$ contribute to 
\begin{align}\label{eq:W18M}
\exval{\hat{A}}&=\tr(\rho\hat{A})=\rho_{11}(e_{30}+e_{03}+e_{33})+\rho_{22}(e_{30}-e_{03}-e_{33})\nn\\
&\quad +\rho_{33}(-e_{30}+e_{03}-e_{33})+\rho_{44}(-e_{30}-e_{03}+e_{33})\, .
\end{align}
For the example $\rho_{11}=\rho_{22}=\rho_{33}=7/20$, $\rho_{44}=-1/20$ one finds 
\begin{equation}\label{eq:W18N}
\exval{\hat{A}} = \dfrac{2}{5} \left( e_{30}+e_{03}-e_{33}\right) \, .
\end{equation}
For arbitrary $e_{\mn}$ obeying $e_{\mn}e_{\mn}\leq 1$ this implies $|\exval{\hat{A}}|\leq 2\sqrt{3}/5\leq 1$, in accordance with eq.~\eqref{eq:W18L}. On the other hand the density matrix of our example is not positive. We have therefore found Ising spins $e_{z}$ obeying eq.~\eqref{eq:W18C} with expectation values obeying the constraint \eqref{eq:W17}, that lead to a non-positive $\rho$. We conclude that the positivity of $\rho$ has to be imposed as a quantum constraint on the classical statistical probabilities which can realize the quantum subsystem, in addition to the constraint \eqref{eq:W17}.

Classical statistical probability distributions obeying both quantum constraints \eqref{eq:W17} and \eqref{eq:W18J} realize a quantum subsystem for two qubits. One can implement arbitrary unitary evolutions of the density matrix by suitable unique jump operations. These unique jump operations preserve the quantum constraints. We have therefore established a classical statistical representation of quantum systems with two spins. This includes entanglement \cite{CWA}.

\subsection{Quantum mechanics for an arbitrary number of spins}

The generalization to an arbitrary number $Q$ of qubits is straightforward. The continuous family of Ising spins depends on an $(2^{2Q}-1)$-dimensional unit vector $s(e_{\mu_{1}\mu_{2}\ldots \mu_{Q}})$, $\mu_{i}=0,1,2,3$, again with $e_{00\ldots 0}$ omitted. The quantum condition for the expectation values reads
\begin{equation}\label{eq:W28}
\exval{s(\bm{e})}=\exval{s(e_{\mu_{1}\mu_{2}\ldots \mu_{Q}})}=e_{\mu_{1}\mu_{2}\ldots \mu_{Q}}\rho_{\mu_{1}\mu_{2}\ldots \mu_{Q}}\, ,
\end{equation}
with $\rho_{\mu_{1}\mu_{2}\ldots \mu_{Q}}$ defining the hermitian $2^{Q}\times 2^{Q}$ density matrix by eq.~\eqref{eq:30}. Arbitrary $SO(2^{2Q}-1)$-rotations of the unit vector $e_{\mu_{1}\mu_{2}\ldots \mu_{Q}}$ can be realized as unique jump operations. They can equivalently be described by $SO(2^{2Q}-1)$-rotations of the coefficients $\rho_{\mu_{1}\mu_{2}\ldots \mu_{Q}}$. The subgroup of $SU(2^{Q})$-rotations, under which the fundamental vector representation of $SO(2^{2Q}-1)$ transforms as the adjoint representation of $SU(2^{Q})$, performs unitary transformations of the density matrix. Since the overall phase, corresponding to the abelian $U(1)$-factor of $U(2^{Q})$, does not matter, this classical statistical system realizes an arbitrary unitary evolution of $Q$ quantum spins.

Similar to the case of two qubits we have to constrain the vectors $e_{z}=e_{\mu_{1}\mu_{2}\ldots \mu_{Q}}$ such that all density matrices compatible with the positivity constraint can be realized by eq.~\eqref{eq:W28}. The construction \eqref{eq:W18A}-\eqref{eq:W18E} is valid for arbitrary $Q$. The manifold of $e_{z}$ is $SU(2^{Q})/SU(2^{Q-1})\times SU(2^{Q-1})\times U(1)$. We also impose the positivity constraint for $\rho$. This construction demonstrates that quantum mechanics corresponds to a particular subsystem of a classical statistical ensemble - in our case specified by the quantum conditions - and a particular unitary evolution law - in our case given by the unique jump step evolution operators.

\section{Discussion}\label{sec:Discussion} 

In this paper we discuss the bit-quantum map which defines the embedding of a quantum subsystem within a wider classical statistical system. Both classical statistical systems and quantum subsystems can be defined ``locally'' for every (discrete) point $t$ on a chain corresponding to time, space, or layers in neural networks. For arbitrary quantum operations the bit-quantum map should be complete in the sense that every quantum density matrix for the subsystem can be realized by a suitable classical probability distribution or classical density matrix. If arbitrary transformations of the classical statistical information from $t$ to $\te$ are allowed, universal quantum computing for an arbitrary number of qubits can be realized by a finite number of macroscopic classical two-level observables. We have discussed a correlation map that represents $Q$ qubits in terms of $3Q$ classical bits. If this map can be shown to be a complete bit-quantum map, general probabilistic operations on $3Q$ classical bits are sufficient to realize, in principle, arbitrary unitary operations on $Q$ qubits.

We have worked out several examples how quantum gates can be implemented within the classical statistical setting of generalized Ising models or static memory materials. For an infinite number of Ising spins, or continuous Ising spins, arbitrary quantum operations for an arbitrary number of qubits can be realized. In particular, this infinite number of classical bits can arise from a finite number of continuous real variables. For generalized Ising models a finite number of classical bits can only implement a discrete subgroup of the most general unitary evolution of a quantum system. These subgroups can be rather dense, however, allowing for a rich spectrum of quantum computational steps to be executed by classical statistical systems. For a given finite number of classical bits further quantum operations can be performed by a more general transport of probabilistic information between layers of a classical statistical system. 

We advocate that a general theoretical framework for computing is given by the quantum formalism for classical statistics, which is based on a classical density matrix instead of a local probability distribution. This covers classical deterministic computing, classical probabilistic computing and quantum computing. In particular, we believe that interesting aspects of neural networks can be implemented.

A general computational step is associated to the evolution of the classical density matrix \eqref{eq:I2} between neighboring $t$-layers. The ordering variable $t$ may be time, position of points or hypersurfaces in space, or the layers in a neural network, perhaps even in the brain. The evolution law \eqref{eq:I2} is linear, such that the superposition principle for solutions, familiar from quantum mechanics, applies. 

Consecutive steps of a computation correspond to the non-commutative matrix multiplication of step evolution operators $S(\te)S(t)$. Different realizations of computing correspond to different properties of $S$. For generalized Ising models in the occupation number basis $S$ is a nonnegative matrix. For steps of quantum computing $S$ is orthogonal such that no information is lost. In the presence of a complex structure orthogonal step evolution operators $S$ are mapped to unitary operators. For deterministic computing or cellular automata $S$ is a unique jump operator. These properties need not to hold for the evolution steps of the full density matrix $\rho^{\prime}(t)$. It is sufficient that they  are realized for suitable subsystems $\rho(\rho^{\prime})$, as for the quantum subsystem.

For deterministic classical computing or cellular automata the ``gates'' $S(t)$ are unique jump operators. Also the initial data given by $\rho^{\prime}(t_{in})$ single out one particular spin configuration. Its processing by a sequence of gates $S(t)S(t-\epsilon)\ldots S(t_{in}+\epsilon)S(t_{in})$ results in a single spin configuration at $t$, which is the output of the computation. For probabilistic computing either $S$ is not a unique jump operation, or $\rho^{\prime}(t_{in})$ is truly a probabilistic initial state, given by a probability distribution for the initial spin configurations that differs from zero for more than one spin configuration. For quantum computing the initial density matrix is probabilistic, such that quantum computing is always probabilistic computing. This holds even if $S$ is a unique jump operation. More precisely, the probabilistic information contained in $\rho'(t)$, or in the local probability distribution $\{p_\tau(t)\}$, has to obey quantum conditions in order to realize a suitable quantum subsystem. These quantum conditions enforce a truly probabilistic state for every $t$.

Deterministic computation is a special limiting case of probabilistic computation. Already the errors in deterministic computations are described by probabilistic computation. In this case $S$ is not precisely a unique jump operation - the entries accounting for the ``wrong output'' differ from zero. The noncommutativity familiar from quantum mechanics is characteristic for all forms of computing. Neural networks are typical cases of probabilistic computing. Only part of the initial information is relevant for layers at larger $t$ - the essential information is concentrated in appropriate subsystems. The statistical information available for the subsystem is necessarily truly probabilistic. 

Our findings and formalism open interesting directions for future research. The first concerns new computing architectures. One avenue are static memory materials for which $t$ is realized as a position in space. First simulations \cite{SEW} show that the information loss for suitable generalized Ising models may be moderate enough to permit even almost deterministic computations. Interesting classical interference structures for input information at two different boundaries may be exploited. Another avenue may employ the discrete unitary transformations that can be realized by a finite number of Ising spins. Even though this is not full quantum computing, powerful algorithms using new non-commutative structures may be developed.

A second direction of research may ask if the new non-commutative structures and probabilistic computing are already realized by neural networks or neuromorphic computing, perhaps even in our brain. Probabilistic aspects seem to be genuine for this type of computing. This is certainly the case for biological systems that cannot realize error-free deterministic computation steps. Even if the overall steps for neural networks or neuromorphic computing are realized in a deterministic manner, the effect on relevant subsystems will typically become probabilistic. As we have discussed, the probabilistic nature of the computation, or the noise or ``errors'' in deterministic steps, may even be an important ingredient for the achievement of computational goals.

The third direction addresses questions of the foundations of quantum mechanics. Once one realizes that quantum systems can be implemented as suitable subsystems of classical statistical systems \cite{CWQM}, several interesting questions emerge. Is time an ordering concept \cite{CWPT} in a general probabilistic ensemble covering the universe from the ``infinite past'' to the ``infinite future''? Why and how do the quantum conditions necessary for the realization of quantum subsystems arise in the ubiquitous way observed in nature? What is the precise nature of measurements and the associated measurement correlations?

The perhaps most important outcome of this work is the notion that quantum properties are not restricted to the microscopic, to very well isolated systems or to low temperature. They may play a role for macroscopic probabilistic systems. Perhaps they can even influence open systems as life.

\medskip\noindent
{\em Acknowledgment:} The author would like to thank Karlheinz Meier and Christian Pehle for stimulating discussions. This work is part of and supported by the DFG Collaborative Research Center ``SFB 1225 (ISOQUANT)''. 

\medskip\noindent
{\em Note added:} After the original version of this paper some important aspects of quantum computation have been implemented on small artificial neural networks \cite{PEME}. My colleague and friend Karlheinz Meier has passed away and I would like to dedicate this work to his memory.


\begin{appendices}

\section{Correlation map for two qubits}\label{app:Correlation_map_for_two_qubits}

In this appendix we discuss details of the correlation map from six classical Ising spins to two qubits. We present explicit classical probability distributions that realize families of entangled states for the quantum system. This is one by use of the normalized classical wave function. We employ the wave function in order to discuss how unitary transformations in the quantum system are realized by transformations of classical probability distributions.

We consider here pure quantum states for two qubits ($Q=2$). We concentrate on the correlation map for which the classical system employs $M=3Q=6$ Ising spins $s_{k}^{(1)}$, $s_{k}^{(2)}$. There are $2^{6}=64$ classical states $\tau$ with associated probabilities $p_{\tau}$. The task is to find the probability distributions $\lbrace p_{\tau}\rbrace$ that induce the wanted pure state quantum density matrix by virtue of the correlation map \eqref{eq:33}.

\subsection{Normalized classical wave function} 

A useful tool is the normalized classical wave function $q$ \cite{CWQP}. It is a real $64$-component vector with components $q_{\tau}$ obeying
\begin{equation}\label{eq:CC1} 
p_{\tau}=q_{\tau}^{2}\, .
\end{equation}
This wave function can be seen as a classical probability amplitude. It is defined by the probabilities $p_{\tau}$ up to signs, $q_{\tau}=\sigma_{\tau}\sqrt{p_{\tau}}$, $\sigma_{\tau}=\pm 1$. From $\sum_{\tau}\, p_{\tau}=1$ one infers that $q$ is a vector on the unit sphere $S^{63}$,
\begin{equation}\label{eq:CC2}
q_{\tau}q_{\tau}=1\, .
\end{equation}
(The normalized classical wave function $q$ can be constructed from the classical wave function $\tilde{q}$ and $\bar{q}$ of sect.~\ref{sec:Static_Memory_Materials} by a non-linear transformation \cite{CWIT}.)

The classical Ising spins $s_{k}^{(i)}$ can be represented as commuting classical operators $\hat{s}_{k}^{(i)}$, such that
\begin{equation}\label{eq:CC3}
\exval{s_{k}^{(i)}}=q_{\tau}\left (\hat{s}_{k}^{(i)}\right )_{\tau\rho}q_{\rho}=\langle q|\hat{s}_{k}^{(i)}|q\rangle\, .
\end{equation}
For diagonal operators $(\hat{s}_k^{(i)})_{\tau\rho} = (\tilde{s}_k^{(i)})_\tau\, \delta_{\tau\rho}$ one has
\begin{equation}\label{eq:194A}
\langle s_k^{(i)}\rangle = \sum_\tau q_\tau^2\, (\tilde{s}_k^{(i)})_\tau =
\sum_\tau p_\tau\, (\tilde{s}_k^{(i)})_\tau\, ,
\end{equation}
and therefore associates $(\tilde{s}_k^{(i)})_\tau$ with the value that the classical spin $s_k^{(i)}$ takes in a given classical state $\tau$.

We may use a direct product representation,
\begin{equation}\label{eq:CC4}
\hat{s}_{1}^{(i)}=t^{(i)}\otimes 1\otimes 1\, ,\,  \hat{s}_{2}^{(i)}=1\otimes t^{(i)}\otimes 1\, ,\, \hat{s}_{3}^{(i)}= 1\otimes 1\otimes t^{(i)}\, ,
\end{equation}
with $4\times 4$-matrices
\begin{equation}\label{eq:CC5}
t^{(1)}=\left (\begin{array}{cccc}
1 &  &  &  \\ 
 & 1 &  &  \\ 
 &  & -1 &  \\ 
 &  &  & -1
\end{array} \right )\, ,\,  t^{(2)}=\left (\begin{array}{cccc}
1 &  &  &  \\ 
 & -1 &  &  \\ 
 &  & 1 &  \\ 
 &  &  & -1
\end{array} \right )\, .
\end{equation}
An arbitrary vector $q$ can be constructed as a linear combination of basis states that we take as direct products of four component vectors, $\alpha,\beta,\gamma,\bar{\alpha},\bar{\beta},\bar{\gamma}=1,\ldots ,4$, $\tau=(\alpha,\beta,\gamma)$,
\begin{equation}\label{eq:CC6}
b_{\tau}^{(\bar{\alpha},\bar{\beta},\bar{\gamma})}=b^{(\bar{\alpha},\bar{\beta},\bar{\gamma})}_{\alpha,\beta,\gamma}=b_{\alpha}^{(1,\bar{\alpha})}b_{\beta}^{(2,\bar{\beta})}b_{\gamma}^{(3,\bar{\gamma})}
\end{equation}
with
\begin{equation}\label{eq:CC7}
b_{\alpha}^{(k,\bar{\alpha})}=\delta_{\alpha}^{\bar{\alpha}}\, .
\end{equation}
In this basis $\hat{s}_{1}^{(i)}$ acts on the first factor, $\hat{s}_{2}^{(i)}$ on the second and $\hat{s}_{3}^{(i)}$ on the third. The basis states are eigenstates of the classical spin operators $\hat{s}_k^{(i)}$ with eigenvalues $\pm 1$.

An arbitrary classical wave function is expanded in this basis with coefficients $q_{\bar{\alpha}, \bar{\beta}, \bar{\gamma}}$,
\begin{equation}\label{eq:198A}
q_\tau = q_{\alpha, \beta, \gamma} = q_{\bar{\alpha}, \bar{\beta}, \bar{\gamma}}\, 
b ^{\bar{\alpha}, \bar{\beta}, \bar{\gamma}}_{\alpha, \beta, \gamma}\, .
\end{equation}
The classical spin operators act as
\begin{align}\label{eq:198B}
(\hat{s}_1^{(i)})_{\alpha\beta\gamma}^{\quad \enspace \alpha'\beta'\gamma'}\, 
q_{\alpha'\beta'\gamma'} &= (t^{(i)})^{\alpha'}_\alpha\, q_{\alpha' \beta \gamma}\, ,
\notag \\
(\hat{s}_2^{(i)})_{\alpha\beta\gamma}^{\quad \enspace \alpha'\beta'\gamma'}\, 
q_{\alpha'\beta'\gamma'} &= (t^{(i)})^{\beta'}_\beta\, q_{\alpha \beta' \gamma}\, ,
\notag \\
(\hat{s}_3^{(i)})_{\alpha\beta\gamma}^{\quad \enspace \alpha'\beta'\gamma'}\, 
q_{\alpha'\beta'\gamma'} &= (t^{(i)})^{\gamma'}_\gamma\, q_{\alpha \beta \gamma'}\, ,
\end{align}
with diagonal $(t^{(i)})_{\alpha}^{\alpha'} = t_\alpha^{(i)}\, \delta_\alpha^{\alpha'}$. Expectation values of the classical spins and therefore the quantum density matrix can be computed directly from the classical wave function. For example, one has
\begin{equation}\label{eq:198C}
\langle s_2^{(2)} \rangle = \rho_{02} = \sum_{\alpha \gamma}
\left( q_{\alpha 1 \gamma}^2 - q_{\alpha 2 \gamma}^2 + q_{\alpha 3 \gamma}^2 - q_{\alpha 4 \gamma}^2 \right)
\end{equation}
or
\begin{equation}\label{eq:198D}
\langle s_1^{(1)} s_3^{(2)} \rangle = \rho_{13} = \sum_{\alpha \beta \gamma}
t_\alpha^{(1)} t_\gamma^{(2)} q_{\alpha \beta \gamma}^2\, .
\end{equation}

Classical wave functions for some simple quantum states can easily be established. For the example of the pure quantum state with $\exval{s_{3}^{(1)}}=\exval{s_{3}^{(2)}}=1$, and therefore $\exval{s_{3}^{(1)}s_{3}^{(2)}}=1$, $\exval{s_{k}^{(1)}s_{l}^{(2)}}=\exval{s_{k}^{(1)}s_{3}^{(2)}}=\exval{s_{3}^{(1)}s_{l}^{(2)}}=0$, $\exval{s_{k}^{(1)}}=\exval{s_{l}^{(2)}}=0$ for $k,l=1,2$, a possible classical wave function obeys
\begin{equation}\label{eq:CC8} 
q_{0}=q_{0}^{(1)}\otimes q_{0}^{(2)}\otimes q_{0}^{(3)}\, ,
\end{equation}
with
\begin{equation}\label{eq:CC9}
t^{(1)}q_{0}^{(3)}=t^{(2)}q_{0}^{(3)}=q_{0}^{(3)}\, ,
\end{equation}
and
\begin{equation}\label{eq:CC9A}
\langle q_{0}^{(1)}| t^{(i)}|q_{0}^{(1)}\rangle =\langle q_{0}^{(2)}| t^{(i)}|q_{0}^{(2)}\rangle=0\, .
\end{equation}
This is realized by
\begin{align}\label{eq:CC10} 
&q_{0}^{(3)}=\left (\begin{array}{c}
1 \\ 
0 \\ 
0 \\ 
0
\end{array} \right )\, ,\, q_{0}^{(1)}=\dfrac{1}{\sqrt{2(a^{2}+b^{2})}}\left (\begin{array}{c}
a \\ 
b \\ 
\pm b \\ 
\pm a
\end{array} \right )\, ,\nn\\
&q_{0}^{(2)}=\dfrac{1}{\sqrt{2(c^{2}+d^{2})}}\left (\begin{array}{c}
c \\ 
d \\ 
\pm d \\ 
\pm c
\end{array} \right )\, .
\end{align}
Due to the different choices of $a,b,c,d$ the wave function is not unique. Further solutions of eq.~\eqref{eq:CC9} may take the more general form $q_{0}=b_{0}\otimes q_{0}^{(3)}$, with $16\times 16$ matrix $b_{0}$ not taking the direct product form $q_{0}^{(1)}\otimes q_{0}^{(2)}$. As a result, we obtain a whole family of probability distributions that realize this quantum state.

\subsection{Classical probability distributions for pure entangled quantum states}

It is instructive to construct explicitly the classical probability distributions for selected entangled quantum states. The realization of entangled quantum states poses no particular problem. The maximally entangled state \eqref{eq:25} obeys $\exval{s_{1}^{(1)}s_{1}^{(2)}}=\exval{s_{2}^{(1)}s_{2}^{(2)}}=\exval{s_{3}^{(1)}s_{3}^{(2)}}=-1$, while $\exval{s_{k}^{(i)}}=0$, $\exval{s_{k}^{(1)}s_{l}^{(2)}}=0$ for $k\neq l$. It can be realized by a classical wave function in direct product form,
\begin{equation}\label{eq:CC11} 
 q_{en}= q_{en}^{(1)}\otimes  q_{en}^{(2)}\otimes  q_{en}^{(3)}\, ,
\end{equation} 
obeying
\begin{equation}\label{eq:CC12} 
\hat{s}^{(1)}_{k}\hat{s}^{(2)}_{k}q_{en}^{(k)}=-q_{en}^{(k)}\, ,
\end{equation}
and
\begin{equation}\label{eq:CC13} 
\langle q^{(k)}_{en}|\hat{s}^{(i)}_{k}|q_{en}^{(k)}\rangle =0\, .
\end{equation}
The solution,
\begin{equation}\label{eq:CC14} 
q_{en}^{(k)}=\dfrac{1}{\sqrt{2}}\left (\begin{array}{c}
0 \\ 
1 \\ 
\pm 1\\
0
\end{array} \right )\, ,
\end{equation}
corresponds for every $k$-sector to equal probabilities $p_{+-}=p_{-+}=1/2$, while $p_{++}=p_{--}=0$. Taking things together, the eight probabilities $p_{\tau}$ for which $s_{k}^{(1)}$ and $s_{k}^{(2)}$ are opposite for all $k$ take all the value $1/8$. The other $56$ probabilities, for which at least for one $k$ the spins $s_{k}^{(1)}$ and $s_{k}^{(2)}$ take the same value, vanish.

Wave functions that can be written in the direct product form
\begin{equation}
q_{\vartheta}=q_{\vartheta}^{(1)}\otimes q_{\vartheta}^{(2)}\otimes q_{\vartheta}^{(3)}
\end{equation}
imply the relations
\begin{align}
\rho_{k0}&=\exval{s_{k}^{(1)}}= \langle q^{(k)}| t^{(1)}|q^{(k)}\rangle \, ,\nn\\
\rho_{0k}&=\exval{s_{k}^{(2)}}= \langle q^{(k)}| t^{(2)}|q^{(k)}\rangle \, ,\nn\\
\label{eq:CC16}\rho_{kk}&=\exval{s_{k}^{(1)}s_{k}^{(2)}}= \langle q^{(k)}| t^{(1)} t^{(2)}|q^{(k)}\rangle \, , \\
\end{align}
and, for $k\neq l$,
\begin{align}
\rho_{kl}&=\exval{s_{k}^{(1)}s_{l}^{(2)}}= \langle q^{(k)}| t^{(1)} |q^{(k)}\rangle\, \langle q^{(l)}| t^{(2)} |q^{(l)}\rangle\nn\\
\label{eq:CC17} &=\rho_{k0}\rho_{0l} \, .
\end{align}

A family of pure quantum states for which these conditions hold is given by the wave function
\begin{equation}\label{eq:CC18} 
\psi_{\vartheta}=\left (\begin{array}{c}
0 \\ 
\cos(\vartheta) \\ 
\sin(\vartheta) \\ 
0
\end{array} \right )\, .
\end{equation}
We will construct explicitly classical wave functions and the classical probability distribution for this family of pure quantum states. For these classical wave functions the non-vanishing coefficients $\rho_{\mu\nu}$ or expectation values of classical spins have to obey
\begin{align}\label{eq:CC19} 
\rho_{30}&=-\rho_{03}=\cos^{2}(\vartheta)-\sin^{2}(\vartheta)\com \rho_{33}=-1\, ,\nn\\
\rho_{11}&=\rho_{22}=2\cos(\vartheta)\sin(\vartheta)\, .
\end{align}
For $q^{(1)}$ one infers (with $\varepsilon_{1},\varepsilon_{2}=\pm 1$)
\begin{equation}\label{eq:CC20} 
q^{(1)}=\left (\begin{array}{c}
a \\ 
b \\ 
\varepsilon_{2} b \\ 
\varepsilon_{1} a
\end{array} \right )\com a^{2}+b^{2}=\dfrac{1}{2}\, ,
\end{equation}
and
\begin{equation}\label{eq:CC21} 
a^{2}-b^{2}=\cos(\vartheta)\sin(\vartheta)\, .
\end{equation}
Up to signs, this implies
\begin{equation}\label{eq:CC22} 
a=\dfrac{1}{2}\left (\cos(\vartheta)+\sin(\vartheta)\right )\com b=\dfrac{1}{2}(\cos(\vartheta)-\sin(\vartheta))\, .
\end{equation}
The same relation holds for $q^{(2)}$. In the sectors with $k=1,2$ one therefore as $p_{++}=p_{--}=a^{2}$, $p_{+-}=p_{-+}=b^{2}$. For $q^{(3)}$ one finds
\begin{equation}\label{eq:CC23} 
q^{(3)}=\left (\begin{array}{c}
0 \\ 
\pm\cos(\vartheta) \\ 
\pm\sin(\vartheta) \\ 
0
\end{array} \right )\, ,
\end{equation}
such that in the $k=3$ sector $p_{+-}=\cos^{2}(\vartheta)$, $p_{-+}=\sin^{2}(\vartheta)$, $p_{++}= p_{--} = 0$. For $\cos(\vartheta)=-\sin(\vartheta)=1/\sqrt{2}$ we recover the maximally entangled state \eqref{eq:25}. The probabilities in a given sector $k$ are independent of the probabilities in the other sectors. All probabilities $p_\tau$ are therefore specified. We have established a family of classical probability distributions which realize the entangled pure quantum state \eqref{eq:CC18}.

\subsection{Representation of unitary transformations by changes of classical probability distributions}

The realization of arbitrary pure states by the correlation map for two qubits comes in pair with the possibility to realize arbitrary unitary $SU(4)$-transformations on a given pure state. It is instructive to understand how unitary transformations in the quantum subsystem are related to transformations of the classical wave function for the classical system. The quantum wave function of an arbitrary pure quantum state obtains (up to an irrelevant overall phase) by a unitary $SU(4)$-transformation $U$ from some arbitrary fixed pure state $\psi_{0}$,
\begin{equation}\label{eq:CC24} 
\psi=U\psi_{0}\, .
\end{equation}
Similarly, an arbitrary classical wave function can be expressed as an orthogonal $SO(64)$ transformation $O$ from an arbitrary fixed wave function $q_{0}$,
\begin{equation}\label{eq:CC25} 
q=O\, q_{0}\, .
\end{equation}
The simple relation \eqref{eq:CC25} is an important advantage of the use of classical wave functions, since the expression of a general transformation between probabilities is a complicated object, while the rotations \eqref{eq:CC25} realize directly the normalization condition $\sum_\tau p_\tau = \sum_\tau q_\tau^2 = 1$ and obey a group structure.

Let us now consider a pair $(\psi_{0},q_{0})$ such that the pure state density matrix $\rho^{(0)}$ constructed from $\psi_{0}$ is the correlation map of the probability distribution associated to $q_{0}$. The coefficients defining the density matrix $\rho_{0}$ obey
\begin{equation}\label{eq:CC26} 
\rho_{\mn}^{(0)}=\langle \psi_{0}|L_{\mn}|\psi_{0}\rangle\, ,
\end{equation}
with $L_{\mn}$ the fifteen generators of $SU(4)$. By our choice they coincide with the expectation values of the fifteen diagonal classical operators $M_{\mn}$,
\begin{equation}\label{eq:CC27} 
\rho_{\mn}^{(0)}=\langle q_{0}|M_{\mn}|q_{0}\rangle\, ,
\end{equation}
where
\begin{equation}\label{eq:CC28} 
M_{k0}=\hat{s}_{k}^{(1)}\com M_{0k}=\hat{s}_{k}^{(2)}\com M_{kl}=\hat{s}_{k}^{(1)}\hat{s}_{l}^{(2)}\, .
\end{equation}
Completeness of the correlation map is established if for every $U$ one can find $O$ such that
\begin{align}\label{eq:CC29} 
\rho_{\mn}&=\langle \psi_{0}|U^{\dagger}L_{\mn}\, U|\psi_{0}\rangle
=\langle q_{0}|O^{T}M_{\mn}O|q_{0}\rangle\nn\\
&=\langle \psi |L_{\mn}|\psi \rangle=\langle q|M_{\mn}|q\rangle\, .
\end{align}

While the unitary quantum transformations $U$ are defined as linear transformations on $\psi_0$, this linearity needs not to hold for the orthogonal transformations $O$ acting on $q$. We will see that the transformations of the classical wave functions realizing the correlation map are necessarily non-linear. The matrices $O$ needed to realize a given $U$ will depend on the wave function $q$. Assume that two quantum wave functions $\psi_1$ and $\psi_2$ are transformed by the same $U$. This does not hold for the associated classical wave functions $q_1$ and $q_2$. On the classical level the orthogonal transformations $O_1$ and $O_2$, that realize a given unitary transformation of the quantum subsystem by transformations of $q_1$ and $q_2$, are different.

\subsection{Transformations of classical wave functions for a single qubit}

Let us first discuss the simpler problem for a single qubit, $Q=1$, where $\psi$ has two components, $q$ has eight components, $L_{\mn}$ is replaced by $\tau_{k}$ and $M_{\mn}$ by $\hat{s}_{k}$, 
\begin{equation}\label{eq:CC36} 
\hat{s}_{1}=\tau_{3}\otimes 1\otimes 1\com \hat{s}_{2}=1\otimes \tau_{3}\otimes 1\com \hat{s}_{3}=1\otimes 1\otimes \tau_{3}\, .
\end{equation}
In this case we have already established that arbitrary quantum states can be realized by appropriate classical probability distributions. Therefore orthogonal transformations of the classical wave function exist for every arbitrary unitary transformation in the quantum subsystem. We will see that already for a simple qubit the transformation of the classical wave function is a non-linear orthogonal transformation. The non-linearity is genuine for the correlation map and extends to an arbitrary number of qubits. 

An arbitrary pure quantum state,
\begin{equation}\label{eq:CC37} 
\rho_{k}=\exval{s_{k}}=\langle q|\hat{s}_{k}|q\rangle\com \rho_{k}\rho_{k}=1\, ,
\end{equation}
can be realized by
\begin{equation}\label{eq:CC38} 
q=q^{(1)}\otimes q^{(2)}\otimes q^{(3)}\, ,
\end{equation}
with $q^{(k)}$ being two-component real unit vectors. With the representation \eqref{eq:CC36} the expectation values of the classical spins have a simple form, 
\begin{equation}\label{eq:CC39} 
\rho_k = \exval{q^{(k)}\tau_{3}q^{(k)}}\, .
\end{equation}
We can parameterize $q^{(k)}$ as
\begin{equation}\label{eq:CC40} 
q^{(k)}=\left (\begin{array}{c}
a_{k} \\ 
b_{k}
\end{array} \right )\com a_{k}^{2}=\dfrac{1+\rho_{k}}{2}\com b_{k}^{2}=\dfrac{1-\rho_{k}}{2}\, .
\end{equation}
For suitable $a_k$, $b_k$ any pure quantum state with $\sum_k \rho_k^2 = 1$ can be realized.

We first demonstrate explicitly that the classical wave functions realizing a given initial quantum state can be transformed suitably in order to realize any arbitrary unitary transformation of this quantum state. We choose an initial state with $\exval{s_{3}}=1$, $\exval{s_{1}}=\exval{s_{2}}=0$,
\begin{equation}\label{eq:CC41} 
q_{3}=\dfrac{1}{2}\left (\begin{array}{c}
1 \\ 
1
\end{array} \right )\otimes \left (\begin{array}{c}
1 \\ 
1
\end{array} \right )\otimes \left (\begin{array}{c}
1 \\ 
0
\end{array} \right )\com \psi_{3}=\left (\begin{array}{c}
1 \\ 
0
\end{array} \right )\, .
\end{equation}
We want to show that an orthogonal matrix exists that rotates $q_{3}$ to $q$ as given by eqs~\eqref{eq:CC38}, \eqref{eq:CC40}. We can take a direct product
\begin{align}\label{eq:CC42} 
O&=O^{(1)}\otimes O^{(2)}\otimes O^{(3)}\, ,\nn\\
O^{(k)}&=\left (\begin{array}{cc}
\cos(\vartheta_{k}) & \sin(\vartheta_{k}) \\ 
-\sin(\vartheta_{k}) & \cos(\vartheta_{k})
\end{array} \right )\, ,
\end{align}
with
\begin{align}\label{eq:CC43} 
&\cos(\vartheta_{1})\sin(\vartheta_{1})=\dfrac{\rho_{1}}{2}\com \cos(\vartheta_{2})\sin(\vartheta_{2})=\dfrac{\rho_{2}}{2}\, ,\\
&\cos(\vartheta_{3})=a_{3}=\sqrt{\dfrac{1+\rho_{3}}{2}}\com \sin(\vartheta_{3})=-b_{3}=\sqrt{\dfrac{1-\rho_{3}}{2}}\nn\, .
\end{align}
One can verify by explicit computation that arbitrary $\rho_k$ for a pure quantum state can be obtained by this orthogonal transformation. Since every initial and final pure quantum states admits associated normalized classical wave functions \eqref{eq:CC40}, it is not surprising that orthogonal transformation between the classical wave functions can be found.

We next show that the transformations of classical wave functions that realize unitary quantum transformations cannot be linear. For this purpose we consider two different quantum states $\psi_1$ and $\psi_2$ and show that the matrix $O$ that is associated to a given unitary transformation $U$ of the quantum system depends on the associated wave functions $q_1$ and $q_2$. Consider first the quantum state $\psi_1$ which is an eigenstate of $s_3$, and the unitary transformation $U=\tau_1$, 
\begin{equation}\label{eq:CC44} 
\psi_1 = \left( \begin{array}{c}
1 \\
0
\end{array} \right)
\com U_{1}=\tau_{1}\com U_{1}\psi_{1}=\left (\begin{array}{c}
0 \\ 
1
\end{array} \right )\, .
\end{equation}
This corresponds in eq.~\eqref{eq:CC42} to $\vartheta_{1}=\vartheta_{2}=0$, $\vartheta_{3}=-\pi/2$, or
\begin{equation}\label{eq:CC45} 
O_{1}=1\otimes 1\otimes \left (\begin{array}{cc}
0 & -1 \\ 
1 & 0
\end{array} \right )\, .
\end{equation}
Applying $O_{1}$ to a direct product wave function,
\begin{equation}\label{eq:CC46} 
q=q^{(1)}\otimes q^{(2)}\otimes \left (\begin{array}{c}
a \\ 
b
\end{array} \right )\, ,
\end{equation}
leaves $q^{(1)}$ and $q^{(2)}$ invariant,
\begin{equation}\label{eq:CC47} 
O_{1}q=q^{(1)}\otimes q^{(2)}\otimes \left (\begin{array}{c}
-b \\ 
a
\end{array} \right )\, .
\end{equation}
After the transformation $\rho_{1}$ and $\rho_{2}$ have not changed. 

In contrast, if we next apply $U_{1}$ to an eigenstate of $\tau_{2}$,
\begin{equation}\label{eq:CC48} 
\psi_{2}=\dfrac{1}{\sqrt{2}}\left (\begin{array}{c}
1 \\ 
i
\end{array} \right )\com U_{1}\psi_{2}=\dfrac{1}{\sqrt{2}}\left (\begin{array}{c}
i \\ 
1
\end{array} \right )\, ,
\end{equation}
the unitary transformation $U_1 = \tau_1$ reverses the sign of $\rho_{2}$, $\rho_{2}=1\raw\rho_{2}=-1$. This is not compatible with the transformation $O_1$ in eq.~\eqref{eq:CC47}. We therefore need a different matrix $\tilde{O}_{2}$ to represent the action of $U_{1}$ on a classical wave function representing $\psi_{2}$. This demonstrates the necessary non-linearity of the orthogonal transformations of the classical wave functions.

We conclude that the matrix $O$ realizing a given $U$ cannot be found by embedding $SU(2)$ as a subgroup of $SO(8)$. The relation between $U$ and $O$ cannot be realized on the operator level. For the latter it would be necessary to find for an arbitrary $SU(2)$-matrix $U$ an $SO(8)$-matrix $O$ such that
\begin{equation}\label{eq:CC49} 
U^{\dagger}\tau_{k}U=b_{kl}\tau_{l}
\end{equation}
implies for the same coefficients $b_{kl}$
\begin{equation}\label{eq:CC50} 
O^{T}\hat{s}_{k}O=b_{kl}\hat{s}_{l}\, .
\end{equation}
Then eqs~\eqref{eq:CC24},~\eqref{eq:CC25} would hold for every $(\psi_{0},q_{0})$ obeying eqs~\eqref{eq:CC26}, \eqref{eq:CC27}, in contrast to the findings above. The impossibility to realize the pair \eqref{eq:CC49}, \eqref{eq:CC50} can be seen easily for infinitesimal transformations,
\begin{equation}\label{eq:CC51} 
U=1+i\alpha_{m}\tau_{m}\, ,
\end{equation}
for which
\begin{equation}\label{eq:CC52} 
b_{kl}=\delta_{kl}+2\alpha_{m}\varepsilon_{mkl}\, .
\end{equation}
For
\begin{equation}\label{eq:CC53} 
O_{\tau\rho}=\delta_{\tau\rho}+\beta_{\tau\rho}\com \beta_{\tau\rho}=-\beta_{\rho\tau}\, ,
\end{equation}
one has
\begin{equation}\label{eq:CC54} 
O^{T}\hat{s}_{k}O=\hat{s}_{k}-[\beta ,\hat{s}_{k}]\, .
\end{equation}
The commutator of an antisymmetric matrix $\beta$ with a diagonal matrix $\hat{s}_{k}$ has all diagonal elements vanishing, making a relation $[\beta,\hat{s}_{k}]=-2\alpha_{m}\varepsilon_{mkl}\hat{s}_{l}$ impossible.

The lesson of this simple exercise is that no linear transformation of classical wave functions exists that realizes arbitrary unitary transformations for the quantum subsystem. Just as for the probabilities, arbitrary unitary transformations require non-linear transformations of the normalized classical wave functions. This statement holds for many quantum bit maps that realize quantum density matrices by expectation values of classical spins. It generalizes to $Q=2$ or a higher number of qubits.

\section{Unique jump realizations of the CNOT-gate}\label{app:Unique_jump_realizations_of_the_CNOT-gate} 
 
The possible realization of the CNOT-gate by a unique jump operation depends on the number of classical Ising spins or classical bits used. For two qubits ($Q=2$) and fifteen classical bits we have discussed a simple unique jump operation in sect.~\ref{sec:Two_entangled_quantum_spins}. In this appendix we concentrate on $Q=2$ and the correlation map with six classical bits $s_{k}^{(1)}$, $s_{k}^{(2)}$, as discussed in sect.~\ref{sec:Bit_quantum_maps}. We will first show that no unique jump operation exists that realizes the map \eqref{eq:25} for the corresponding correlation functions for arbitrary probability distributions $\lbrace p_{\tau}\rbrace$. It would be sufficient, however, to realize eq.~\eqref{eq:25} for those probability distributions that obey the quantum condition. This is a more complex question that has not yet found a complete answer. We describe the issue by the discussion of a concrete example of a particular unique jump operation.

\subsection{CNOT-gate with six Ising spins}

The transformation \eqref{eq:26} is no longer a simple transformation among Ising spins. For example, the exchange $\rho_{03}\leftrightarrow\rho_{33}$ exchanges the expectation values $\exval{s_{3}^{(2)}}$ and the correlation function $\exval{s_{3}^{(1)}s_{3}^{(2)}}$. As an example, the exchange
\begin{equation}\label{eq:43}
\exval{s_{3}^{(2)}}\leftrightarrow \exval{s_{3}^{(1)}s_{3}^{(2)}}
\end{equation}
can be realized by a conditional switch of $s_{3}^{(2)}$: it changes sign if $s_{3}^{(1)}$ is negative, and remains invariant if $s_{3}^{(1)}$ is positive. This conditional switch is, however, not the only possibility to realize $\rho_{03}\leftrightarrow\rho_{33}$. For an assessment which type of transformations of the classical probability distribution $\lbrace p_{\tau}\rbrace$ can realize the CNOT-gate we have to proceed to a more systematic study, for which we detail a typical step in the following.

We may label the $2^{6}=64$ states $\tau$ by $(++\tilde{\rho})$, $(+-\tilde{\rho})$, $(-+\tilde{\rho})$, $(--\tilde{\rho})$, where the two signs stand for the values of $s_{3}^{(1)}$ and $s_{3}^{(2)}$, and the collective index $\tilde{\rho}$ denotes the $16$ possibilities for the remaining spins $s_{1}^{(1)}$, $s_{2}^{(1)}$, $s_{1}^{(2)}$ and $s_{2}^{(2)}$. With similar notations for the probabilities $p_{\tau}$ this yields
\begin{align}\label{eq:44}
\exval{s_{3}^{(2)}}&=\sum_{\tilde{\rho}}\left (p_{++\tilde{\rho}}-p_{+-\tilde{\rho}}+p_{-+\tilde{\rho}}-p_{--\tilde{\rho}}\right )\, ,\nn\\
\exval{s_{3}^{(1)}s_{3}^{(2)}}&=\sum_{\tilde{\rho}}\left (p_{++\tilde{\rho}}-p_{+-\tilde{\rho}}-p_{-+\tilde{\rho}}+p_{--\tilde{\rho}}\right ).
\end{align}
The transformation \eqref{eq:34} therefore corresponds to
\begin{equation}\label{eq:45}
X+Y\leftrightarrow X-Y
\end{equation}
with
\begin{equation}\label{eq:46}
X=\sum_{\tilde{\rho}}\left (p_{++\tilde{\rho}}-p_{+-\tilde{\rho}}\right )\com Y=\sum_{\tilde{\rho}}\left (p_{-+\tilde{\rho}}-p_{--\tilde{\rho}}\right ).
\end{equation}
The conditional switch of $s_{3}^{(2)}$ if $s_{3}^{(1)}=-1$ leaves $X$ invariant while $Y$ changes sign. It is obvious that this is not the only transformation among the $p_{\tau}$ that realizes eq.~\eqref{eq:45}. 

If we combine eq.~\eqref{eq:43} with the invariance of $\rho_{30}=\exval{s_{3}^{(1)}}$, the condition \eqref{eq:45} is supplemented by the invariance of $Z$
\begin{equation}\label{eq:47}
Z=\sum_{\tilde{\rho}}\left (p_{++\tilde{\rho}}+p_{+-\tilde{\rho}}-p_{-+\tilde{\rho}}-p_{--\tilde{\rho}}\right ).
\end{equation}
Using the normalization condition,
\begin{equation}\label{eq:48}
\sum_{\tilde{\rho}}\left (p_{++\tilde{\rho}}+p_{+-\tilde{\rho}}+p_{-+\tilde{\rho}}+p_{--\tilde{\rho}}\right )=1\, ,
\end{equation}
we infer the invariance of 
\begin{equation}\label{eq:49}
b_{30}=A_{++}^{33}+A_{+-}^{33}\, ,
\end{equation}
where we denote
\begin{equation}\label{eq:50}
A_{\alpha\beta}^{33}=\sum_{\tilde{\rho}}\, p_{\alpha\beta\tilde{\rho}}\, .
\end{equation}
This leaves us with two variables $A_{++}^{33}$ and $A_{--}^{33}$, with
\begin{equation}\label{eq:51}
A_{+-}^{33}=b_{30}-A_{++}^{33}\com A_{-+}^{33}=1-b_{30}-A_{--}^{33}\, ,
\end{equation}
and
\begin{equation}\label{eq:52}
X=2A_{++}^{33}-b_{30}\com Y=1-b_{30}-2A_{--}^{33}\, .
\end{equation}
Insertion into eq.~\eqref{eq:36} yields
\begin{equation}\label{eq:53}
A_{++}^{33}+A_{--}^{33}\leftrightarrow A_{++}^{33}-A_{--}^{33}+1-b_{30}\, .
\end{equation}
Again, the conditional switch, which corresponds to invariant $A_{++}^{33}$ and $A_{--}^{33}\leftrightarrow A_{-+}^{33}=1-b_{30}-A_{--}^{33}$, is not the only possibility to realize eq.~\eqref{eq:53}.

This type of considerations can be continued in order to establish the conditions for the other relations in eq.~\eqref{eq:25}. Using arbitrary maps among the probabilities $p_{\tau}$ it is indeed possible to realize the CNOT-gate with six Ising spins. This follows from the fact that arbitrary values of $\rho_{\mn}$ obeying the quantum constraint can be obtained for suitable probability distributions $\lbrace p_{\tau}\rbrace$.

\subsection{Unique jump operation}

We will next show that the map \eqref{eq:25} underlying the CNOT-gate cannot be realized by a unique jump operation for arbitrary probability distributions. Unique jump operations map each state $\tau$ to precisely one other state $\tau^{\prime}$, with a corresponding map $p_{\tau}\raw p_{\tau^{\prime}}$. For this purpose we first observe that the $64$ states $\tau$ can be divided into four blocks of $16$ probabilities each, such that the CNOT-transformation acts within each block separately. Indeed, the invariance of $\rho_{30}$, $\rho_{01}$ and $\rho_{31}$ implies that the sum of probabilities $p_{\sigma_{1}\sigma_{2}++\rho_{2}\rho_{3}}$ for states with both $s_{3}^{(1)}=1$ and $s_{1}^{(2)}=1$ remains invariant under the transformation. Here $\sigma_{k}=\pm 1$ and $\rho_{k}=\pm 1$ denote the values of $s_{k}^{(1)}$ and $s_{k}^{(2)}$ in the corresponding state $\tau$. (We use shorthands $+,-$ for $+1$ and $-1$.) If unique jump operations are realized for arbitrary probabilities, the CNOT-transformation cannot exchange probabilities from the block $p_{\sigma_{1}\sigma_{2}++\rho_{2}\rho_{3}}$ with probabilities in the other blocks $p_{\sigma_{1}\sigma_{2}+-\rho_{2}\rho_{3}}$, $p_{\sigma_{1}\sigma_{2}-+\rho_{2}\rho_{3}}$ and $p_{\sigma_{1}\sigma_{2}--\rho_{2}\rho_{3}}$, for which $(s_{3}^{(1)},s_{1}^{(2)})$ takes values $(+,-)$, $(-,+)$ and $(-,-)$, respectively. A unique jump realization of the CNOT-gate can only exchange probabilities within each given block. 

As a consequence, each block can be discussed separately and we concentrate on the one with $(s_{3}^{(1)},s_{1}^{(2)})=(+,+)$. Realizing the transformation $\rho_{03}\leftrightarrow\rho_{33}$ within this block requires the relation \eqref{eq:43} now restricted to sums over probabilities within this block. Since $s_{3}^{(1)}=1$, both $\rho_{03}$ and $\rho_{33}$ are the same and therefore remain invariant under the transformation. A unique jump operation can therefore only exchange states within each subblock with $s_{3}^{(2)}=1$ or $s_{3}^{(2)}=-1$ separately. This defines two invariant subblocks with eight states each. Applying the same argument to the exchanges $\rho_{02}\leftrightarrow\rho_{32}$, $\rho_{10}\leftrightarrow\rho_{11}$, $\rho_{20}\leftrightarrow \rho_{21}$ the sixteen states are divided into subblocks in different ways. One ends with eight invariant subblocks with eight states each. The remaining transformations $\rho_{23}\leftrightarrow\rho_{12}$ and $\rho_{13}\leftrightarrow -\rho_{22}$ have to be realized by an exchange of states that respects the invariant subblocks. This is not possible, and we conclude that no transformation among the $64$ states $\tau$ is possible that realizes the map \eqref{eq:25} for arbitrary probability distributions.

\subsection{Conditional jumps}

The issue gets more complicated if we restrict the probability distributions $\lbrace p_{\tau}\rbrace$ to those that obey the quantum constraint. As an example we discuss a unique jump operation that realizes only part of the map \eqref{eq:25} for arbitrary $\lbrace p_{\tau}\rbrace$. The question arises if the missing relations can be enforced by the quantum constraints on $\lbrace p_{\tau}\rbrace$. We investigate the conditional jump $C$,
\begin{equation}\label{eq:N1} 
C:\,\begin{cases}
\textit{if }s_{3}^{(1)}=-1,& \text{flip sign of }s_{2}^{(2)}\text{ and }s_{3}^{(2)},\\
\textit{if }s_{1}^{(2)}=-1,& \text{flip sign of }s_{1}^{(1)}\text{ and }s_{2}^{(1)}.
\end{cases}
\end{equation}
This is a simple map between classical spins and their correlations. The spins $s_{3}^{(1)}$ and $s_{1}^{(2)}$ remain unchanged, while
\begin{align}\label{eq:N2} 
&s_{1}^{(1)}\leftrightarrow s_{1}^{(1)}s_{1}^{(2)}\com s_{2}^{(1)}\leftrightarrow s_{2}^{(1)}s_{1}^{(2)}\, ,\nn\\
&s_{2}^{(2)}\leftrightarrow s_{3}^{(1)}s_{2}^{(2)}\com s_{3}^{(2)}\leftrightarrow s_{3}^{(1)}s_{3}^{(2)}.
\end{align}
This ensures in eq.~\eqref{eq:25} the relations
\begin{align}\label{eq:N3}
&\rho_{10}\leftrightarrow\rho_{11}\com \rho_{20}\leftrightarrow\rho_{21}\com\rho_{02}\leftrightarrow\rho_{32}\com\rho_{03}\leftrightarrow\rho_{33}\, ,\nn\\
&\rho_{30},\rho_{01},\rho_{31}\text{ invariant.}
\end{align}

The situation is less simple for $\rho_{12}$, $\rho_{13}$, $\rho_{22}$ and $\rho_{23}$. The conditional jump $C$ transforms
\begin{align}\label{eq:N4}
&s_{1}^{(1)}s_{2}^{(2)}\leftrightarrow s_{1}^{(1)}s_{3}^{(1)}s_{1}^{(2)}s_{2}^{(2)}\com s_{1}^{(1)}s_{3}^{(2)}\leftrightarrow s_{1}^{(1)}s_{3}^{(1)}s_{1}^{(2)}s_{3}^{(2)}\, ,\nn\\
&s_{2}^{(1)}s_{2}^{(2)}\leftrightarrow s_{2}^{(1)}s_{3}^{(1)}s_{1}^{(2)}s_{2}^{(2)}\com s_{2}^{(1)}s_{3}^{(2)}\leftrightarrow s_{2}^{(1)}s_{3}^{(1)}s_{1}^{(2)}s_{3}^{(2)}.
\end{align}
Correspondingly, for a general probability distribution $\lbrace p_{\tau}\rbrace$ the two.point functions $\rho_{12}$, $\rho_{13}$, $\rho_{22}$ and $\rho_{23}$ are mapped to four-point functions. These four-point functions are not directly contained in the incomplete statistical information of the quantum subsystem. The relations \eqref{eq:25} do not need to hold, however, for arbitrary probability distributions. It is sufficient that $C$ maps a probability distribution obeying the quantum condition to another probability distribution obeying the quantum condition, and that eq.~\eqref{eq:25} holds for these particular pairs of probability distributions. 

A classical state $\tau$ can be characterized by the values of the six Ising spins, $\tau=(\sigma_{1},\sigma_{2},\sigma_{3},\rho_{1},\rho_{2},\rho_{3})$, with $\sigma_{1}=1$ for all states with $s_{1}^{(1)}=1$ while $\sigma_{1}=-1$ if $s_{1}^{(1)}=-1$, and similar for all $\sigma_{k}$, $\rho_{k}$. Correspondingly, we denote the classical probabilities $p_{\tau}$ by $p_{\sigma_{1}\sigma_{2}\sigma_{3}\rho_{1}\rho_{2}\rho_{3}}$. We often will use the shorthands $+$ or $-$ for a given $\sigma_{k}=+1$ or $\sigma_{k}=-1$. With this notation one has
\begin{align}\label{eq:N5}
\rho_{13}&=\sum_{\sigma_{2},\sigma_{3},\rho_{1},\rho_{2}}\, \biggl (p_{+\sigma_{2}\sigma_{3}\rho_{1}\rho_{2}+}+p_{-\sigma_{2}\sigma_{3}\rho_{1}\rho_{2}-}\nn\\
&\qquad-p_{+\sigma_{2}\sigma_{3}\rho_{1}\rho_{2}-}-p_{-\sigma_{2}\sigma_{3}\rho_{1}\rho_{2}+}\biggl )\, ,
\end{align}
or
\begin{align}\label{eq:N6}
\rho_{22}&=\sum_{\sigma_{1},\sigma_{3},\rho_{1},\rho_{3}}\, \biggl (p_{\sigma_{1}+\sigma_{3}\rho_{1}+\rho_{3}}+p_{\sigma_{1}-\sigma_{3}\rho_{1}-\rho_{3}}\nn\\
&\qquad-p_{\sigma_{1}+\sigma_{3}\rho_{1}-\rho_{3}}-p_{\sigma_{1}-\sigma_{3}\rho_{1}+\rho_{3}}\biggl ).
\end{align}
The map $C$ defines a map $p_{\sigma_{1}\sigma_{2}\sigma_{3}\rho_{1}\rho_{2}\rho_{3}}\raw p^{\prime}_{\sigma_{1}\sigma_{2}\sigma_{3}\rho_{1}\rho_{2}\rho_{3}}$ which associates to each set $(\sigma_{1},\sigma_{2},\sigma_{3},\rho_{1},\rho_{2},\rho_{3})$ a unique set $(\sigma_{1}^{\prime},\sigma_{2}^{\prime},\sigma_{3}^{\prime},\rho_{1}^{\prime},\rho_{2}^{\prime},\rho_{3}^{\prime})$. For a general classical probability distribution $\lbrace p_{\tau}\rbrace$ is is easy to see that $C$ does neither map $\rho_{13}$ to $\rho_{13}^{\prime}=-\rho_{22}$, nor $\rho_{12}$ to $\rho_{12}^{\prime}=\rho_{23}$. The question is if the map \eqref{eq:25} holds if we impose the quantum condition on the classical probability distribution.

Let us illustrate this issue for a set of four pure state density matrices with $\rho_{30}=\pm 1$, $\rho_{01}=\pm 1$. These states correspond to fixed values $s_{3}^{(1)}=\pm 1$, $s_{1}^{(2)}=\pm 1$. For these fixed values the map \eqref{eq:N4} reads
\begin{equation}\label{eq:NX}
\rho_{12}^{\prime}=\pm\rho_{12}\, ,\;\rho_{13}^{\prime}=\pm\rho_{13}\, ,\; \rho_{22}^{\prime}=\pm\rho_{22}\, ,\; \rho_{23}^{\prime}=\pm\rho_{23}\, ,
\end{equation}
with signs depending on the values of $s_{3}^{(1)}$ and $s_{1}^{(2)}$. For this state also $\rho_{31}=\pm 1$ assumes a fixed value. For $\rho_{30}^{2}+\rho_{01}^{2}+\rho_{31}^{2}=3$ we conclude that the quantum condition requires $\rho_{\mn}=0$ for all pairs $(\mu,\nu)$ except $(3,0)$, $(0,1)$ and $(3,1)$. The relations \eqref{eq:N3}, \eqref{eq:NX} ensure then $\rho_{\mn}^{\prime}=0$ except $\rho
_{30}^{\prime}=\rho_{30}$, $\rho^{\prime}_{01}=\rho_{01}$, $\rho_{31}^{\prime}=\rho_{31}$. This obeys the map \eqref{eq:25} and we conclude that for this particular quantum state eq.~\eqref{eq:25} is indeed realized by the map $C$.

Our next example considers pure state density matrices with $\rho_{30}=\pm 1$, $\rho_{03}=\pm 1$. Since the spins $s_{3}^{(1)}$ and $s_{3}^{(2)}$ assume fixed values for this quantum state, also $\rho_{33}=\rho_{30}\rho_{03}$ is fixed in this case. In consequence, the condition $\rho_{z}\rho_{z}=3$ is already saturated by $\rho_{30}$, $\rho_{03}$ and $\rho_{33}$, such that the quantum condition requires $\rho_{\mn}=0$ for all pairs $(\mu,\nu)$ except $(3,0)$, $(0,3)$ and $(3,3)$. For fixed values of $s_{3}^{(1)}, s_{3}^{(2)}=\pm 1$ the map \eqref{eq:N4} implies
\begin{equation}\label{eq:N6A}
\rho_{13}^{\prime}=\pm\rho_{11}\com\rho_{23}^{\prime}=\pm\rho_{21}
\end{equation}
and 
\begin{equation}\label{eq:N6B}
\rho_{12}^{\prime}=\pm\exval{s_{1}^{(1)}s_{1}^{(2)}s_{2}^{(1)}}\com \rho_{22}^{\prime}=\pm\exval{s_{2}^{(1)}s_{1}^{(2)}s_{2}^{(2)}}\, .
\end{equation}
The map \eqref{eq:25} is realized for this state if $\rho^{\prime}_{\mn}=0$ for all $(\mu,\nu)$ except $(3,0)$, $(0,3)$, $(3,3)$. This holds for $\rho_{13}^{\prime}$ and $\rho_{23}^{\prime}$ according to eq.~\eqref{eq:N6A}. The map \eqref{eq:25} therefore holds for this quantum state if the three-point functions in eq.~\eqref{eq:N6B} vanish.

Consider first the three conditions $\rho_{10}=0$, $\rho_{01}=0$, $\rho_{11}=0$,
\begin{align}\label{eq:N7}
\rho_{10}&=\sum_{\sigma_{2}\sigma_{3}\rho_{1}\rho_{2}\rho_{3}}\biggl (p_{+\sigma_{2}\sigma_{3}\rho_{1}\rho_{2}\rho_{3}}-p_{-\sigma_{2}\sigma_{3}\rho_{1}\rho_{2}\rho_{3}}\biggl )=0\, ,\nn\\
\rho_{01}&=\sum_{\sigma_{1}\sigma_{2}\sigma_{3}\rho_{2}\rho_{3}}\biggl (p_{\sigma_{1}\sigma_{2}\sigma_{3}+\rho_{2}\rho_{3}}-p_{\sigma_{1}\sigma_{2}\sigma_{3}-\rho_{2}\rho_{3}}\biggl )=0\, ,\nn\\
\rho_{11}&=\sum_{\sigma_{2}\sigma_{3}\rho_{2}\rho_{3}}\biggl (p_{+\sigma_{2}\sigma_{3}+\rho_{2}\rho_{3}}+p_{-\sigma_{2}\sigma_{3}-\rho_{2}\rho_{3}}\nn\\
&\qquad-p_{+\sigma_{2}\sigma_{3}-\rho_{2}\rho_{3}}-p_{-\sigma_{2}\sigma_{3}+\rho_{2}\rho_{3}}\biggl )=0\, .
\end{align}
Together with the normalization of the probability distribution this implies
\begin{align}\label{eq:N8}
&\sum_{\sigma_{2}\sigma_{3}\rho_{2}\rho_{3}}\, p_{+\sigma_{2}\sigma_{3}+\rho_{2}\rho_{3}}=\sum_{\sigma_{2}\sigma_{3}\rho_{2}\rho_{3}}\, p_{+\sigma_{2}\sigma_{3}-\rho_{2}\rho_{3}}\nn\\
&=\sum_{\sigma_{2}\sigma_{3}\rho_{2}\rho_{3}}\, p_{-\sigma_{2}\sigma_{3}+\rho_{2}\rho_{3}}=\sum_{\sigma_{2}\sigma_{3}\rho_{2}\rho_{3}}\, p_{-\sigma_{2}\sigma_{3}-\rho_{2}\rho_{3}}=\dfrac{1}{4}\, .
\end{align}
The quantities in eq.~\eqref{eq:N8} are the probabilities to find for $(s_{1}^{(1)}, s_{1}^{(2)})$ the values $(+,+)$, $(+,-)$, $(-,+)$ and $(-,-)$, respectively. For $\exval{s_{1}^{(1)}}=0$, $\exval{s_{1}^{(2)}}=0$, $\exval{s_{1}^{(1)}s_{1}^{(2)}}=0$ they have to be all equal.

For equal probabilities for the eight possible values of three spins the three-point correlation vanishes. This situation is, however, not necessarily realized for all probability distributions obeying the quantum condition. Let us impose first for three spins $s_{i}$ the conditions $\exval{s_{i}}=0$, $\exval{s_{i}s_{j}}=0$. The eight probabilities have to obey
\begin{align}\label{eq:N9}
&p_{+--}=p_{--+}=p_{-+-}=p_{++}\, ,\nn\\
&p_{---}=p_{++-}=p_{+-+}=p_{-++}=\dfrac{1}{4}-p_{+++}\, ,
\end{align}
with
\begin{equation}\label{eq:N10}
\exval{s_{1}s_{2}s_{3}}=4\left (p_{+++}-p_{---}\right ).
\end{equation}
Depending on the value of $p_{+++}$, $0\leq p_{+++}\leq 1/4$, the three point function $\exval{s_{1}s_{2}s_{3}}$ can assume all values between $-1$ and $+1$. We conclude that the conditions $\rho_{10}=\rho_{01}=\rho_{11}=\rho_{20}=\rho_{21}=0$ cannot be sufficient to enforce $\exval{s_{1}^{(1)}s_{1}^{(2)}s_{2}^{(1)}}=0$. They are even weaker than the conditions $\exval{s_{i}}=0$, $\exval{s_{i}s_{j}}=0$, since no relation for the correlation $\exval{s_{1}^{(1)}s_{2}^{(1)}}$ is imposed.

This situation does not change if we impose the additional relations $\rho_{02}=0$, $\rho_{22}=0$, $\rho_{12}=0$. Consistent with all these constraints one can realize non-zero values of $\exval{s_{1}^{(1)}s_{1}^{(2)}s_{2}^{(1)}}$. For example, one has $\exval{s_{1}^{(1)}s_{1}^{(2)}s_{2}^{(1)}}=1$ for the choice $p_{++++}=p_{+-++}=p_{++--}=p_{+---}=p_{-++-}=p_{--+-}=p_{-+-+}=p_{---+}=1/8$ and $p_{+++-}=p_{+-+-}=p_{++-+}=p_{+--+}=p_{-+++}=p_{--++}=p_{-+--}=p_{----}=0$, where $p_{\sigma_{1}\sigma_{2}\rho_{1}\rho_{2}}$ denotes probabilities for the spins $(s_{1}^{(1)}, s_{2}^{(1)}, s_{1}^{(2)}, s_{2}^{(2)})$. Since for the considered state the spins $s_{3}^{(1)}$ and $s_{3}^{(2)}$ take fixed values $\pm 1$ one has $\rho_{30}=\pm\rho_{01}$, $\rho_{32}=\pm\rho_{02}$, $\rho_{13}=\pm\rho_{10}$, $\rho_{23}=\pm\rho_{20}$. Thus $\rho_{31}$, $\rho_{32}$, $\rho_{13}$ and $\rho_{23}$ vanish automatically for vanishing $\rho_{01}$, $\rho_{02}$, $\rho_{10}$ and $\rho_{20}$. We conclude that the quantum condition is not sufficient for $C$ to realize the CNOT gate \eqref{eq:25} for arbitrary quantum states.

This discussion shows that the action of unique jump operations on quantum states needs a careful discussion of the role of the quantum constraint. So far we have not yet found a definite answer if a unique jump operation can realize the CNOT-gate for six classical bits.

We also recall that the question if the CNOT-gate can be realized by unique jump operations or not depends on the number of Ising spins used. For the $2^{2Q}-1$ Ising spins discussed at the beginning of this section a unique jump realization of the CNOT-gate is obvious. One may consider intermediate cases, as for even $Q$ a number of $15Q/2$ independent classical spins. They are ordered in pairs of neighboring quantum spins. For each pair one uses fifteen classical bits or Ising spins and realizes the CNOT-gate within a pair by the unique jump operation discussed in sect.~\eqref{sec:Two_entangled_quantum_spins}. While unitary transformations exchanging pairs of quantum spins are easy to realize as unique jump operations, it is now the exchange of a single quantum spin with another single quantum spin belonging to a different pair for which the unique jump operation poses a difficulty.

\bigskip

\section{CNOT-gate in probabilistic computing}\label{app:CNOT-gate_in_probabilistic_computing}

In this appendix we discuss the quantum subsystem for two qubits, realized by six Ising spins using the correlation map discussed in sect.~\ref{sec:Bit_quantum_maps}. We want to see what type of step evolution operators can realize the CNOT-gate. We employ the formalism for probabilistic computing in sect.~\ref{sec:Probabilistic_Computing}. There are fifteen classical observables $A^{\prime}_{(z)}=A^{\prime}_{(\mn)}$, $\mu,\nu=0,1,2,3$, $A^{\prime}_{00}$ omitted, whose expectation values determine the density matrix $\rho(t)$
\begin{equation}\label{eq:WD}
\rho_{\mn}(t)=\tr\left (A^{\prime}_{(\mn)}\rho^{\prime}(t)\right ).
\end{equation}
We label here the observables by the classical operators $A^{\prime}_{(\mn)}$, realized by diagonal $64\times 64$ matrices appropriate for $M=6$ classical bits. The density matrix at $\te$ obtains from the expectation values of transformed classical observables $B^{\prime}_{(\mn)}=S^{-1}A^{\prime}_{(\mn)}S$ according to
\begin{equation}\label{eq:WE}
\rho_{\mn}(\te)=\tr\left (B^{\prime}_{(\mn)}\rho^{\prime}(t)\right )=\tr\left (S^{-1}A^{\prime}_{(\mn)}S\rho^{\prime}(t)\right ).
\end{equation}
The CNOT-gate can be realized if one can find a suitable step evolution operator $S$ such that $\rho_{\mn}(\te)$ is related to $\rho_{\mn}(t)$ by eq.~\eqref{eq:26} for all $\rho^{\prime}(t)$ obeying the quantum constraint.

A sufficient condition for realizing the map $\rho_{\mn}\raw\rho^{\prime}_{\mn}=\pm\rho_{\rho\tau}$ is given by a relation on the operator level $B^{\prime}_{(\mn)}=S^{-1}A^{\prime}_{(\mn)}S=\pm A^{\prime}_{(\rho\tau)}$. For the weaker necessary condition, however, the relation of the type $B^{\prime}_{(\mn)}=\pm A^{\prime}_{(\rho\tau)}$ has to hold only once multiplied with arbitrary $\rho^{\prime}$ obeying the quantum constraint, and after performing the trace. This admits more possibilities than a map $A^{\prime}_{(\mn)}\raw B^{\prime}_{(\mn)}$ on the operator level.

It is straightforward to show that the CNOT-gate cannot be realized on the operator level. A realization on the operator level would require relations of the type \eqref{eq:26} on the level of the classical operators $B^{\prime}_{(\mn)}$ and $A^{\prime}_{(\mn)}$, which amounts to
\begin{widetext}
\begin{align}\label{eq:WF}
&[S,A^{\prime}_{(30)}]=[S,A^{\prime}_{(01)}]=[S,A^{\prime}_{(31)}]=0\, ,\nn\\
&SA^{\prime}_{(11)}=A^{\prime}_{(10)}S\com SA^{\prime}_{(10)}=A^{\prime}_{(11)}S\com SA^{\prime}_{(21)}=A^{\prime}_{(20)}S\com SA^{\prime}_{(20)}=A^{\prime}_{(21)}S\, ,\nn\\
&SA^{\prime}_{(33)}=A^{\prime}_{(03)}S\com SA^{\prime}_{(03)}=A^{\prime}_{(33)}S\com SA^{\prime}_{(32)}=A^{\prime}_{(02)}S\com SA^{\prime}_{(02)}=A^{\prime}_{(32)}S\, ,\nn\\
&SA^{\prime}_{(22)}=-A^{\prime}_{(13)}S\com SA^{\prime}_{(13)}=-A^{\prime}_{(22)}S\com SA^{\prime}_{(12)}=A^{\prime}_{(23)}S\com SA^{\prime}_{(23)}=A^{\prime}_{(12)}S\, .
\end{align}
\end{widetext}
Eq.~\eqref{eq:WF} requires that $S^{2}$ commutes with all $A^{\prime}_{(\mn)}$. The fifteen diagonal classical operators $(A^{\prime}_{(\mn)})_{\tau\rho}=(A^{\prime}_{(\mn)})_{\tau\tau}\delta_{\tau\rho}$ obey $(A^{\prime}_{(\mn)})_{\tau\tau}=1$ for states $\tau$ for which the product of Ising spins $s_{\mu}^{(1)}s_{\nu}^{(2)}$ takes the values one, and $(A^{\prime}_{(\mn)})_{\tau\tau}=-1$ for states where $s_{\mu}^{(1)}s_{\nu}^{(2)}=-1$. We emphasize the different product structure for classical operators $A^{\prime}_{(\mn)}$ and quantum operators $L_{\mn}$. While the classical operators $A^{\prime}_{(\mn)}$ all commute, this does not hold for the quantum operators. On the level of operator relations no step evolution obeying all relations \eqref{eq:WF} exists. The relations $[S,A^{\prime}_{(30)}]=0$, $[S,A^{\prime}_{(01)}]=0$ imply that $S$ must be block diagonal, with $16\times 16$ blocks acting separately in the sectors of fixed $s_{3}^{(1)}$ and $s_{1}^{(2)}$. The remaining relations have to hold for each sector separately. In the particular sector $s_{3}^{(1)}=1$, $s_{1}^{(1)}=1$ there is no difference between $A^{\prime}_{(11)}$ and $A^{\prime}_{(10)}$, $A^{\prime}_{(21)}$ and $A^{\prime}_{(20)}$, $A^{\prime}_{(33)}$ and $A^{\prime}_{(03)}$, or $A^{\prime}_{(32)}$ and $A^{\prime}_{(02)}$. In this sector $A^{\prime}_{(10)}$, $A^{\prime}_{(20)}$, $A^{\prime}_{(02)}$ and $A^{\prime}_{(03)}$ commute with $S$, such that $S$ has to be the unit matrix in this sector. This contradicts the last four relations \eqref{eq:WF}.

For a first exploration of the weaker condition \eqref{eq:WE} we discuss settings where eq.~\eqref{eq:WE} holds for particular states, corresponding to particular classical density matrices $\rho^{\prime}$ obeying the quantum constraint. These are ``basis states'' which will be discussed in appendix \ref{app:Maps_for_basis_states}. If eq.~\eqref{eq:WE} holds for all basis states discussed in appendix \ref{app:Maps_for_basis_states}, and for all $\rho^{\prime}$ realizing these basis states, the CNOT gate is realized by the step evolution operator $S$. Consider the pure quantum states $\rho_{30}=\rho_{01}=\rho_{31}=\pm 1$, for which the quantum condition requires for all $(\mu,\nu)$ except $(3,0)$, $(0,1)$ and $(3,1)$
\begin{equation}\label{eq:WG}
\rho_{\mn}=\tr\left (A^{\prime}_{(\mn)}\rho^{\prime}\right )=0\, .
\end{equation}
This restricts suitable sums of diagonal elements of $\rho^{\prime}$. We consider factorizing boundary conditions \eqref{eq:Z2} such that $\rho^{\prime}_{\tau\rho}=\tilde{q}_{\tau}\bar{q}_{\rho}$. One possible realization of these pure quantum states is obtained by partial equipartition in the Ising spins $s_{1}^{(1)}$, $s_{2}^{(1)}$, $s_{2}^{(2)}$ and $s_{2}^{(3)}$, 
\begin{equation}\label{eq:WH}
\tilde{q}_{\sigma_{1}\sigma_{2}\sigma_{3}\rho_{1}\rho_{2}\rho_{3}}=\dfrac{1}{4}\tilde{q}^{\prime}_{\sigma_{3}\rho_{1}}\com \bar{q}_{\sigma_{1}\sigma_{2}\sigma_{3}\rho_{1}\rho_{2}\rho_{3}}=\dfrac{1}{4}\bar{q}^{\prime}_{\sigma_{3}\rho_{1}}\, ,
\end{equation}
with $\tilde{q}^{\prime}_{\sigma_{3}\rho_{1}}$ and $\bar{q}^{\prime}_{\sigma_{3}\rho_{1}}$ equal to one for one particular combination $(\sigma_{3},\rho_{1})$, and zero otherwise. For example, $\tilde{q}^{\prime}_{+-}=\bar{q}^{\prime}_{+-}=1$ realizes the pure state $\rho_{30}=1$, $\rho_{01}=-1$, $\rho_{31}=-1$. The corresponding classical density matrix is block diagonal
\begin{equation}\label{eq:WI}
\rho^{\prime}=\dfrac{1}{16}E_{\sigma_{1}^{\prime}\sigma_{2}^{\prime}\rho_{2}^{\prime}\rho^{\prime}_{3}\sigma_{1}\sigma_{2}\rho_{3}}R_{\sigma^{\prime}_{3}\rho_{1}^{\prime}\sigma_{3}\rho_{1}}\, ,
\end{equation}
with partial equipartition matrix $E$ having all entries equal to one, and
\begin{equation}\label{eq:WJ}
R_{\sigma^{\prime}_{3}\rho_{1}^{\prime}\sigma_{3}\rho_{1}}=\delta_{\sigma_{3}^{\prime}\sigma_{3}}\delta_{\rho^{\prime}_{1}\rho_{1}}\tilde{q}_{\sigma_{3}\rho_{1}}.
\end{equation}

More generally, the equipartition matrix $E$ has all elements equal to one, $E_{\sigma\rho}=1$, where $\sigma$ and $\rho$ may stand for collective indices. The expectation values of classical operators $A^{\prime}$ vanish if the sum of all elements $A^{\prime}_{\rho\sigma}$ vanishes
\begin{equation}\label{eq:WK}
\exval{A}=\tr(A^{\prime}E)=\sum_{\rho\sigma}\, A^{\prime}_{\rho\sigma}=0\, .
\end{equation}
In particular, this holds if $A^{\prime}$ is diagonal and traceless, as the classical spin and correlation operators $A^{\prime}_{(\mn)}$. 

The equipartition matrix is left invariant by transformations with a matrix $T$,
\begin{equation}\label{eq:WL}
TE\, T^{-1}=E\, ,
\end{equation}
provided $T$ obeys for all $(\tau,\omega)$
\begin{equation}\label{eq:WM}
\sum_{\rho\sigma}\, T_{\tau\rho}T^{-1}_{\sigma\omega}=1\, .
\end{equation}
In particular, transformations obeying
\begin{equation}\label{eq:WN}
\sum_{\rho}\, T_{\tau\rho}=1\com \sum_{\sigma}\, (T^{-1})_{\sigma\omega}=1\, ,
\end{equation}
do not change the equipartition matrix. (In turn, the relations \eqref{eq:WN} imply also $\sum_{\tau}\, T_{\tau\rho}=1$, $\sum_{\omega}\, (T^{-1})_{\sigma\omega}=1$.) The $T$-matrices include all unique jump operations, but span a much larger space. 

By step evolution operators involving partial $T$-matrices one can easily realize the CNOT-gate for the particular pure states $\rho_{30}=\rho_{01}=\rho_{31}=\pm 1$. Subsequently, one could proceed to other basis states. As discussed already in appendix \ref{app:Unique_jump_realizations_of_the_CNOT-gate} the issue of a realization of the CNOT gate by suitable nonnegative step evolution operators remains open for the quantum subsystem defined by the use of classical correlations.

\section{Maps for basis states}\label{app:Maps_for_basis_states} 

On the level of the classical density matrices the bit-quantum map is linear, cf.~\eqref{eq:WC1}, \eqref{eq:WCA}. If one can realize this map for a suitable basis set of quantum density matrices, one may use this linearity in order to prove the bit-quantum map for arbitrary quantum density matrices. The basis set should permit to construct arbitrary $\rho$ as linear combinations of basis matrices.

\subsection{Basis set of pure state density matrices}

An arbitrary hermitian matrix $\rho$ can be expressed as a linear combination of pure state quantum density matrices $\rho_{k}^{\varepsilon}$. In particular, this holds for an arbitrary pure state density matrix $\rho$. The basis set $\lbrace \rho_{k}^{\varepsilon}\rbrace$ is typically redundant in the sense that it contains more basis matrices than linearly independent matrices $\rho$.

Consider first a single qubit, $Q=1$. The basis set contains six pure state density matrices
\begin{equation}\label{eq:B1}
\rho_{k}^{\varepsilon_{k}}=\dfrac{1}{2}\left (1+\varepsilon_{k}\tau_{k}\right )\com\varepsilon_{k}=\pm 1\, .
\end{equation}
An arbitrary hermitian $2\times 2$-matrix $\rho$ can be written as
\begin{equation}\label{eq:B2}
\rho=\dfrac{1}{2}\left (\rho_{0}+\rho_{z}\tau_{z}\right )=\sum_{k}\sum_{\varepsilon_{k}}\, \alpha_{k}^{\varepsilon_{k}}\rho_{k}^{\varepsilon_{k}}\, ,
\end{equation}
with
\begin{equation}\label{eq:B3}
\rho_{0}=\sum_{k}\sum_{\varepsilon_{k}}\, \alpha_{k}^{\varepsilon_{k}}\, ,
\end{equation}
and
\begin{equation}\label{eq:B4}
\rho_{z}=\alpha_{z}^{+}-\alpha_{z}^{-}.
\end{equation}
For $\tr(\rho)=1$ one requires $\sum_{k,\varepsilon_{k}}\, \alpha_{k}^{\varepsilon_{k}}=1$, and a pure state quantum density matrix obtains for
\begin{equation}\label{eq:B5}
\sum_{z}\,\left (\alpha_{z}^{+}-\alpha_{z}^{-}\right )^{2}=1\,  .
\end{equation}
We recall that a linear combination of two pure state density matrices $\rho_{1}$ and $\rho_{2}$, e.g., $\rho=\omega\rho_{1}+(1-\omega)\rho_{2}$, obeys the quantum condition only for $0\leq \omega\leq 1$. For $\omega\neq 0,1$ it is a mixed state density matrix. Expressing arbitrary pure state density matrices as a linear combination of $\rho_{k}^{\varepsilon}$ from the basis set therefore involves more than two coefficients $\alpha_{k}^{\varepsilon_{k}}$ to be different from zero (except trivial limiting cases). One can, of course, express eq.~\eqref{eq:B2} as a linear combination of two matrices $\rho_{1}$ and $\rho_{2}$ which are defined as partial sums. In this case $\rho_{1}$ and $\rho_{2}$ no longer are pure state density matrices. 

For $Q=2$ the basis set contains $36$ pure state density matrices, $\varepsilon_{k,l}=\pm 1$,
\begin{equation}\label{eq:B6}
\rho_{kl}^{\varepsilon_{k}\varepsilon_{l}}=\dfrac{1}{4}\left (1+\varepsilon_{k}L_{k0}+\varepsilon_{l}L_{0l}+\varepsilon_{k}\varepsilon_{l}L_{kl}\right ).
\end{equation}
They correspond to simultaneous eigenstates of the quantum spins $S_{k}^{(1)}$ and $S_{l}^{(2)}$ with eigenvalues $\varepsilon_{k}$ and $\varepsilon_{l}$, respectively. Since $S_{k}^{(1)}$ and $S_{l}^{(2)}$ commute, $L_{kl}=L_{k0}L_{0l}$, these states are also eigenstates of the product $S_{k}^{(1)}S_{l}^{(2)}$. An arbitrary hermitian $4\times 4$-matrix can be written as
\begin{equation}\label{eq:B7}
\rho=\dfrac{1}{4}\left (\rho_{0}+\rho_{\mn}L_{\mn}\right )=\sum_{k,l}\sum_{\varepsilon_{k},\varepsilon_{l}}\, \alpha_{kl}^{\varepsilon_{k}\varepsilon_{l}}\rho_{kl}^{\varepsilon_{k}\varepsilon_{l}}\, ,
\end{equation}
with
\begin{equation}\label{eq:B8}
\tr(\rho)=\rho_{0}=\sum_{k,l}\sum_{\varepsilon_{k},\varepsilon_{l}}\, \alpha_{kl}^{\varepsilon_{k}\varepsilon_{l}}.
\end{equation}
For the coefficients $\rho_{\mn}$ one has
\begin{align}\label{eq:B9}
\rho_{k0}&=\sum_{l}\sum_{\varepsilon_{l}}\, \left (\alpha_{kl}^{+\varepsilon_{l}}-\alpha_{kl}^{-\varepsilon_{l}}\right )\, ,\nn\\
\rho_{k0}&=\sum_{k}\sum_{\varepsilon_{k}}\, \left (\alpha_{kl}^{\varepsilon_{k}+}-\alpha_{kl}^{\varepsilon_{k}-}\right )\, ,\nn\\
\rho_{kl}&=\sum_{\varepsilon_{k},\varepsilon_{l}}\,\varepsilon_{k}\varepsilon_{l} \alpha_{kl}^{\varepsilon_{k}\varepsilon_{l}}.
\end{align}
For a given pair $(k,l)$ the linear combinations 
\begin{align}\label{eq:B9A}
\alpha_{kl}^{++}+\alpha_{kl}^{+-}+\alpha_{kl}^{-+}+\alpha_{kl}^{--}\, , \notag \\
\alpha_{kl}^{++}+\alpha_{kl}^{+-}-\alpha_{kl}^{-+}-\alpha_{kl}^{--}\, , \notag \\ 
\alpha_{kl}^{++}-\alpha_{kl}^{+-}+\alpha_{kl}^{-+}-\alpha_{kl}^{--}\, ,  \notag \\
\alpha_{kl}^{++}-\alpha_{kl}^{+-}-\alpha_{kl}^{-+}+\alpha_{kl}^{--}\, , 
\end{align}
that contribute to $\rho_{0}$, $\rho_{k0}$, $\rho_{0l}$ and $\rho_{kl}$, respectively, can be considered as independent coefficients, such that an arbitrary $\rho$ can indeed be realized. The choice of $\alpha_{kl}^{\varepsilon_{k}\varepsilon_{l}}$ realizing a given $\rho$ is not unique, reflecting the redundancy of the basis set. Generalizations to a higher number of qubits are straightforward.

\subsection{Bit-quantum map for basis set}

The linearity of the bit-quantum map \eqref{eq:WCA} can be exploited for the realization of quantum gates by suitable step evolution operators acting on classical density matrices. Assume that for two given classical density matrices $\rho_{1}^{\prime}$ and $\rho_{2}^{\prime}$ the evolution with some step evolution operator $S$ results in a unitary quantum evolution $U$ for the associated quantum density matrices,
\begin{equation}\label{eq:B10}
\rho(\rho_{a}^{\prime})=\rho_{a}\com \rho\left (S\rho_{a}^{\prime}S^{-1}\right )=U\rho_{a}U^{\dagger}.
\end{equation}
This entails the same unitary evolution for linear combinations,
\begin{align}\label{eq:B11}
\rho\left (S(\alpha_{1}\rho_{1}^{\prime}+\alpha_{2}\rho_{2}^{\prime})S^{-1}\right )&=\alpha_{1}\rho\left (S\rho_{1}^{\prime}S^{-1}\right )+\alpha_{2}\rho\left (S\rho_{2}^{\prime}S^{-1}\right )\nn \\
&=\alpha_{1} U\rho_{1}U^{\dagger} +\alpha_{2}U\rho_{2}U^{\dagger}\nn \\
&=U(\alpha_{1}\rho_{1}+\alpha_{2}\rho_{2})U^{\dagger}.
\end{align}
If for a suitable set of classical density matrices $\rho^{\prime\, \varepsilon_{k}\varepsilon_{l}}_{kl}$ the step evolution operator induces the unitary transformation for all associated quantum density matrices $\rho_{kl}^{\varepsilon_{k}\varepsilon_{l}}$ in the basis set, this set of classical density matrices can be used to realize the quantum operation $U$ for arbitrary quantum density matrices. If \textit{all} classical density matrices $\rho^{\prime}$ obeying $\rho(\rho^{\prime})=\rho_{kl}^{\varepsilon_{k}\varepsilon_{l}}$ obey eq.~\eqref{eq:B10}, the step evolution operator $S$ realizes the quantum operation $U$ for \textit{all} $\rho^{\prime}$ that obey the quantum condition.

\section{Continuous classical probability distributions for the two-component quantum spin}\label{app:Continuous_classical_probability_distributions}

The basic relation for the quantum subsystem is constituted by the identification of expectation values of classical two-level observables with quantum two-level observables by eqs~\eqref{eq:W10}, \eqref{eq:W17}. The precise realization of the classical expectation values is secondary. The two-level observables may be associated to yes-no-decisions which can be ``overlapping'' as we demonstrate by a simple example.

Consider a continuous probability distribution $p(\alpha)$ for the presence of a ``particle'' at a point on a circle parametrized by the angle $\alpha$. We define two-level observables $s(\varphi)$ or occupation numbers $n(\varphi)$ by a yes-no-decision if a particle is found in a half-circle with ``direction'' $\varphi$. This is like a trigger giving one if the particle is in the half-circle, and zero if not,
\begin{equation}\label{eq:CP1} 
n(\varphi,\alpha)=\begin{cases}
1 & \text{for }\alpha-\frac{\pi}{2}\leq \varphi\leq \alpha+\frac{\pi}{2}\\
0 &\text{else}\, .
\end{cases}
\end{equation}
The corresponding Ising spin is $s(\varphi)=2 n(\varphi)-1$. The expectation values reads (with periodic $\alpha$)
\begin{align}\label{eq:CP2} 
\exval{n(\varphi)}&=\int_{-\pi}^{\pi}\, p(\alpha)\, n(\varphi;\alpha)\, \mathrm{d}\alpha\nn\\
&=\int_{\varphi-\frac{\pi}{2}}^{\varphi+\frac{\pi}{2}}\, p(\alpha)\, \mathrm{d}\alpha\, .
\end{align}
For the probability distribution we take
\begin{equation}\label{eq:CP3} 
p(\alpha)=\dfrac{1}{2}\cos(\alpha-\psi)\,\hat{\theta}(\alpha-\psi)\, ,
\end{equation}
with
\begin{equation}\label{eq:CP4} 
\hat{\theta}(\alpha-\psi)=\begin{cases}
1 & \textit{for }-\frac{\pi}{2}\leq \alpha-\psi\leq \frac{\pi}{2}\\
0&\textit{else}.
\end{cases}
\end{equation}
Insertion into eq.~\eqref{eq:CP2} yields
\begin{equation}\label{eq:CP5} 
\exval{n(\varphi)}=\dfrac{1}{2}\left (1+\cos(\varphi-\psi)\right )
\end{equation}
or
\begin{equation}\label{eq:CP6} 
\exval{s(\varphi)}=\cos(\varphi-\psi)\, .
\end{equation}

The probability distribution \eqref{eq:CP3} obeys the quantum condition \eqref{eq:W6}. It therefore realizes a pure quantum state for a single two-component qubit. The quantum state is an eigenstate to the quantum spin in the $\psi$-direction
\begin{equation}\label{eq:CP7} 
\exval{s(\psi)}=\exval{S(\psi)}=1\, ,
\end{equation}
with $\psi$ determined by the classical statistical probability distribution $p(\psi;\alpha)$ according to eq.~\eqref{eq:CP3}. This normalized probability distribution depends on a single parameter $\psi$. Spin rotations by an angle $\gamma$ can be achieved for
\begin{equation}\label{eq:CP8} 
p(t)=p(\psi;\alpha)\com p(\te)=p(\psi+\gamma;\alpha)\, .
\end{equation}
This can be realized by a deterministic unique jump operation in the space of possible particle positions $\alpha(t)$. We may take $N$ discrete values $\alpha(t)$ on a circle and take the limit $N\raw\infty$ at the end. Then $\cL(t)$ in eq.~\eqref{eq:56} becomes a function of $\alpha(t)$ and $\alpha(\te)$. Correspondingly, the step evolution operator $S(t)$ is an $N\times N$-matrix. The rotation by $\gamma$ is achieved for $\beta\raw\infty$ with
\begin{equation}\label{eq:CP9} 
\cL(t)=-\beta\sum_{\alpha(\te)}\sum_{\alpha(t)}\, \left\{ \delta(\alpha(\te),\, \alpha(t)+\gamma)-1 \right\}\, .
\end{equation}
As an alternative to the particle on a circle, we can associate $N=2^{M}$ the $N$ different values of $\alpha(t)$ with $N$ different configurations of $M$ Ising spins.

\section{Two-qubit quantum subsystem with classical correlation functions}\label{app:Two-qubit_quantum_subsystem_with_classical_correlation_functions} 

It is an interesting question if arbitrary positive density matrices and unitary transformations for two quantum spins can be realized by two families of Ising spins $s^{(1)}(e_{k}^{(1)})$ and $s^{(2)}(e_{k}^{(2)})$. For the contributions to the density matrix proportional to $L_{k0}$ or $L_{0k}$ this parallels the discussion for a single qubit. Contributions $\sim L_{kl}$ have to be expressed by correlations of classical spins rather than be expectation values of additional Ising spins as in sect.~\ref{sec:Quantum_Mechanics}. 

\subsection{Two continuous Ising spins for two qubits}

The spins $s^{(1)}$ and $s^{(2)}$ are ``continuous Ising spins'' in the sense discussed in sect.~\ref{sec:Quantum_Mechanics}. Each one consists of a family of an infinite number of Ising spins, one for each point on a sphere. The quantum condition involves in this case the classical correlations of $s^{(1)}$ and $s^{(2)}$,
\begin{align}\label{eq:W19}
&\exval{s^{(1)}(e_{k}^{(1)})}=e_{k}^{(1)}\rho_{k0}\com \exval{s^{(2)}(e_{k}^{(2)})}=e_{k}^{(2)}\rho_{0k}\, ,\nn\\
&\exval{s^{(1)}(e_{k}^{(1)})s^{(2)}(e_{l}^{(2)})}=e_{k}^{(1)}e_{l}^{(2)}\rho_{kl}\, .
\end{align}
As a consequence, the expectation values of quantum spins and correlations between commuting quantum spins are related to expectation values of classical spins and appropriate correlations,
\begin{align}\label{eq:W20}
&\exval{e_{k}^{(1)}L_{k0}}=\exval{s^{(1)}(e_{k}^{(1)})}\com \exval{e_{k}^{(2)}L_{0k}}=\exval{s^{(2)}(e_{k}^{(2)})}\, ,\nn\\
&\exval{e_{k}^{(1)}e_{l}^{(2)}L_{kl}}=\exval{e_{k}^{(1)}L_{k0}\, e_{l}^{(2)}L_{0l}}=\exval{s^{(1)}(e_{k}^{(1)})\, s^{(2)}(e_{l}^{(2)})}\, .
\end{align}
No similar relations for correlations of non-commuting quantum operators exist, such that the quantum subsystem continues to be an incomplete statistical ensemble. The requirement that the quantum operators $e_{k}^{(1)}L_{k0}$ and $e_{k}^{(2)}L_{0k}$ have eigenvalues $\pm 1$ restricts $e_{k}^{(1)}$ and $e_{k}^{(2)}$ to be unit vectors of $S^{2}$,
\begin{equation}\label{eq:W21}
e_{k}^{(1)}e_{k}^{(1)}=1\com e_{k}^{(2)}e_{k}^{(2)}=1\, .
\end{equation}
With these constraints, the discrete possible measurement values as well as the expectation values of the quantum spins are directly inferred from the corresponding properties of the classical Ising spins.

The spins $s_{k}^{(1)}$ and $s_{k}^{(2)}$ in sect.~\ref{sec:Bit_quantum_maps} correspond to particular choices of $e_{k}^{(1)}$ and $e_{k}^{(2)}$, e.g.
\begin{equation}
s_{1}^{(a)}=s^{(a)}(1,0,0)\, ,\,  s_{2}^{(a)}=s^{(a)}(0,1,0)\, ,\,  s_{3}^{(a)}=s^{(a)}(0,0,1).
\end{equation}
Eq.~\eqref{eq:33} yields then simple expressions for the coefficients $\rho_{\mn}$.

\subsection{Quantum constraints and positivity of density matrix}

The condition \eqref{eq:W19} could be sufficient to ensure the positivity of the quantum density matrix, in contrast to the condition \eqref{eq:W17}. As compared to the setting \eqref{eq:W17} we now have only six unit vectors $e_{k}^{(1)},e_{k}^{(2)}$, instead of the fifteen unit vectors $e_{\mn}$ in sect.~\ref{sec:Quantum_Mechanics}. The spins $s(e_{kl})$ in sect.~\ref{sec:Quantum_Mechanics} are composite, expressed as products of the spins $s^{(1)}(e_{k}^{(1)})$ and $s^{(2)}(e_{l}^{(2)})$, 
\begin{equation}\label{eq:W22}
s(e_{kl})=s^{(1)}(e_{k}^{(1)})\, s^{(2)}(e_{l}^{(2)})\, .
\end{equation}
This leads to additional constraints on their expectation values of the type \eqref{eq:33}. They could ensure the positivity of $\rho$.

We first discuss families of density matrices for which positivity indeed follows from the quantum constraint \eqref{eq:W19}. We start by relating suitable quantum operators, as $M_{33}=\diag(1,1,1,-3)$, to the Ising spins, thus restricting the range for their expectation values through the range of expectation values of classical observables. In turn, this will yield restrictions on $\rho$. We exploit that linear combinations of two commuting quantum spin operators can be directly related to linear combinations of classical spins
\begin{equation}\label{eq:W23}
\alpha S_{1}+\beta S_{2}\; \hat{=}\; \alpha s_{1}+\beta s_{2}\, ,
\end{equation}
where $\exval{S_{1}}=\exval{s_{1}}$ and $\exval{S_{2}}=\exval{s_{2}}$. In our setting the associated quantum operators $\hat{S}_{1}$ and $\hat{S}_{2}$ have eigenvalues $\pm 1$, such that the combination $\alpha \hat{S}_{1}+\beta \hat{S}_{2}$ has the four eigenvalues $\pm\alpha\pm\beta$. These are also the possible measurement values of the classical combination $\alpha s_{1}+\beta s_{2}$. (Note that this property does not hold for non-commuting $\hat{S}_{1}$ and $\hat{S}_{2}$.) Consider now the commuting operators $L_{30}$, $L_{03}$ and $L_{33}$, and the linear combination
\begin{equation}\label{eq:W24}
M_{33}=L_{30}+L_{03}-L_{33} = \diag(1,1,1,-3)\, .
\end{equation}
By use of the quantum constraint \eqref{eq:W20} the expectation value is given by a sum of expectation values of Ising spins
\begin{equation}\label{eq:W25}
\exval{M_{33}}=\exval{s^{(1)}(0,0,1)+s^{(2)}(0,0,1)+s^{(1)}(0,0,1)s^{(2)}(0,0,1)}\, ,
\end{equation}
and therefore in the range
\begin{equation}\label{eq:W26}
-3\leq \exval{M_{33}}\leq 1\, .
\end{equation}
The expression of $\exval{M_{33}}$ in terms of the quantum density matrix involves the diagonal elements $\rho_{\alpha\alpha}$,
\begin{equation}\label{eq:W27}
\exval{M_{33}}=\rho_{11}+\rho_{22}+\rho_{33}-3\rho_{44}=1-4\rho_{44}\, .
\end{equation}
The quantum constraint \eqref{eq:W26} requires $\rho_{44}\geq 0$. By suitable other combinations this can be extended to all diagonal elements, $\rho_{\alpha\alpha}\geq 0$. We conclude that in a given basis where $L_{30}$, $L_{03}$ and $L_{33}$ are diagonal all diagonal elements of $\rho$ are positive. If, furthermore, $\rho$ is diagonal in this basis, it is positive.

Consider next the family of quantum density matrices that can be diagonalized by $SU(2)_{I}\times SU(2)_{II}$-transformations, where $SU(2)_{I}$ acts on the first quantum spin, and $SU(2)_{II}$ on the second. There exist classical Ising spins $s^{(1)}(e_{k}^{(1)})$, $s^{(2)}(e_{k}^{(2)})$ that can be brought by corresponding rotations of $e_{k}^{(1)}$ and $e_{k}^{(2)}$ to $s^{(1)}(0,0,1)$ and $s^{(2)}(0,0,1)$. One concludes that in the basis where $\rho$ is diagonal all $\rho_{\alpha\alpha}$ must be positive. For such density matrices the positivity follows from the constraint \eqref{eq:W19}. In other words, the constraint \eqref{eq:W19} ensures that all density matrices that can be diagonalized by $SU(2)_{I}\times SU(2)_{II}$-transformations are positive. 

There is no straightforward generalization to arbitrary $\rho$ for which $SU(4)$ transformations not belonging to $SU(2)_{I}\times SU(2)_{II}$ are needed for the diagonalization. At this stage it remains open if the quantum constraint \eqref{eq:W19} is sufficient, or if an additional constraint ensuring the positivity of $\rho$ is needed. If the quantum constraint \eqref{eq:W19} is not sufficient for guaranteeing positive $\rho$, we impose eq.~\eqref{eq:W18J} for pure states, or a corresponding generalization for mixed states, as a separate quantum constraint.

\subsection{Unitary transformations}

If the correlation map is complete, arbitrary $SU(4)$ transformations can be realized by suitable maps between classical probability distributions. We are further interested if arbitrary $SU(4)$-transformations can be realized by static memory materials. This restricts the maps between classical statistical probability distributions to those that can be realized by non-negative step evolution operators. We first consider unique jump operations transforming the Ising spins $s^{(1)}(e_{k}^{(1)})$ and $s^{(2)}(e_{k}^{(2)})$ into other Ising spins of this family. A continuous transformation group can be realized by rotations of $\bm{e}^{(1)}$ and $\bm{e}^{(2)}$, similar to the single qubit case. This $SO(3)\times SO(3)$-group can account for separate unitary transformations of each of the two quantum spins. It is not large enough, however, to perform arbitrary $SU(4)$-transformations.

We conclude that a realization of arbitrary $SU(4)$-rotations by unique jump operations requires for the setting \eqref{eq:W19} transformations involving the classical correlation functions. Since the Hadamard- and $T$-gates for single qubits are realized, the full $SU(4)$-transformations can be realized if the CNOT-gate can be realized. The large possibilities of unique jump operations involving both the Ising spins and suitable correlation functions remain to be investigated.

\end{appendices}

\vspace{2.0cm}\noindent



\vspace{2.0cm}\noindent

\bibliography{QCCB}

\end{document}